%% file: contact_paper.tex
\documentclass[10pt,a4paper]{article}         

\usepackage{geometry}
\usepackage{authblk}
\usepackage[numbers]{natbib}

\usepackage{lineno}
\modulolinenumbers[5]

\usepackage{microtype}
\usepackage[usenames,dvipsnames,svgnames,table]{xcolor} 

\usepackage[utf8]{inputenc}

\usepackage{pgfplots}
\pgfplotsset{compat=newest}

\pgfplotsset{table/search path = {figures/}}
\usepackage{amsmath,amsfonts,amssymb,amsthm}
\usepackage{mathtools}
\usepackage{xargs} 
\usepackage{bbold} 
\usepackage{scalerel} 
\usepackage{soulutf8}
\usepackage{tabto}
\usepackage{siunitx}
\usepackage{makecell}
\usepackage{tikz}

\usetikzlibrary{plotmarks, calc, spy, pgfplots.polar, external, patterns, pgfplots.groupplots, arrows.meta}
\usepgfplotslibrary{colorbrewer}

\usepackage{PTSansNarrow}
\usepackage[T1]{fontenc}
\usepackage{booktabs}
\usepackage{physics}
\usepackage[inline, shortlabels]{enumitem} 
\usepackage{comment}
\usepackage{tabularx}

\usepackage{pstool}
\usepackage{helvet}
\usepackage{standalone}

\usetikzlibrary{matrix}

\newcommand{\bs}[1] {\boldsymbol{#1}}
\newcommand{\mc}[1]{\mathcal{#1}}
\newcommand{\gv}[1]{\ensuremath{\mbox{\boldmath$ #1 $}}} 
\renewcommand{\div}[1]{\gv{\nabla} \cdot #1} 

\DeclareMathOperator*{\cond}{cond}

\DeclareMathOperator*{\assembly}{\scalerel*{\mathbb{A}}{\sum}}
\newcommand{\ie}[1] {\textit{i.e.}}
\newcommand{\eg}[1] {\textit{e.g.}}
\newcommand{\ea}[1] {\textit{et al.}}
\newcommand{\cf}[1] {\textit{cf.}}

\newcommand\enriched[2]{\bs{#1}_{#2}^\perp}
\newcommand\indexset{\iota_{h}^\perp}

\usepackage{algorithm}
\usepackage{algorithmicx}
\usepackage[noend]{algpseudocode}
\usepackage[normalem]{ulem}
\usepackage{subcaption}

\newcommand*{\algrule}[1][\algorithmicindent]{\hspace*{.5em}\vrule\vrule
	width 0pt height \baselineskip depth .25\baselineskip\hspace*{\dimexpr#1-.5em}}

\makeatletter
\newcount\ALG@printindent@tempcnta
\def\ALG@printindent{%
	\ifnum \theALG@nested>0
	\ifx\ALG@text\ALG@x@notext
	\else
	\unskip
	\ALG@printindent@tempcnta=1
	\loop
	\algrule[\csname ALG@ind@\the\ALG@printindent@tempcnta\endcsname]%
	\advance \ALG@printindent@tempcnta 1
	\ifnum \ALG@printindent@tempcnta<\numexpr\theALG@nested+1\relax
	\repeat
	\fi
	\fi
}%
\usepackage{etoolbox}
\patchcmd{\ALG@doentity}{\noindent\hskip\ALG@tlm}{\ALG@printindent}{}{\errmessage{failed to patch}}
\makeatother

\AtBeginEnvironment{algorithmic}{\lineskip0pt}

\algtext*{EndWhile}
\algtext*{EndFor}
\algtext*{EndIf}
\algdef{SE}[DOWHILE]{Do}{doWhile}{\algorithmicdo}[1]{\algorithmicwhile\ #1}%

\usepackage{etoolbox}
\makeatletter
\expandafter\patchcmd\csname\string\algorithmic\endcsname{\itemsep\z@}
{\itemsep=0.2ex plus2pt}{}{}
\makeatother

\makeatletter
\renewcommand{\ALG@beginalgorithmic}
{\tt}
\makeatother

\algnewcommand\algorithmicto{\textbf{to}}
\algrenewtext{While}{\textbf{while}\ }
\algrenewtext{EndWhile}{\textbf{end}}
\algrenewtext{Procedure}[2]{\textbf{function}\ \textproc{#1}\ifthenelse{\equal{#2}{}}{}{(#2)}}
\algrenewtext{EndProcedure}{\textbf{end}}

\newcommand*\Let[2]{\State #1 $\gets$ #2}
\algrenewcommand\algorithmicrequire{\textbf{Input:}}
\algrenewcommand\algorithmicensure{\textbf{Output:}}
\algrenewcommand\algorithmicfor{\textbf{for each}}

\usepackage{eqparbox}
\algrenewcommand{\algorithmiccomment}[1]{\hfill\eqparbox{COMMENT}{\color{gray} \it-- #1}}

\graphicspath{{./figures/}}

\makeatletter
\def\input@path{{./figures/}}
\makeatother

\usepackage{pdfcolfoot}

\begin{document}

\title{An interface-enriched generalized finite element formulation for locking-free coupling of non-conforming discretizations and contact}

\author[1,2]{Dongyu~Liu}
\author[1,3]{Sanne~J.~van~den~Boom}	
\author[4]{Angelo~Simone}	
\author[1]{Alejandro~M.~Arag\'{o}n}
	
\affil[1]{Faculty of Mechanical, Maritime and Materials Engineering, \break Delft University of Technology,
	Mekelweg 2, 2628 CD, Delft, The Netherlands}
\affil[2]{Faculty of Civil Engineering and Geosciences, Delft University of Technology Stevinweg 1, 2628 CN, Delft, The Netherlands}
\affil[3]{Department of Structural Dynamics, Netherlands Institute for Applied Scientific Research (TNO), Leeghwaterstraat 44-46, 2628 CA, Delft, The Netherlands}
\affil[4]{Department of Industrial Engineering, University of Padova, Padua, Italy}

\maketitle

\begin{abstract}
	We propose an enriched finite element formulation to address the computational modeling of contact problems and the coupling of non-conforming discretizations in the small deformation setting.
	The displacement field is augmented by enriched terms that are associated with generalized degrees of freedom collocated along non-conforming interfaces or contact surfaces.
	The enrichment strategy effectively produces an enriched node-to-node discretization that can be used with any constraint enforcement criterion; this is demonstrated with both multiple-point constraints and Lagrange multipliers, the latter in a generalized Newton implementation where both primal and Lagrange multiplier fields are updated simultaneously.
	The method's ability to ensure continuity of the displacement field---without locking---in mesh coupling problems, and to transfer fairly accurate tractions at contact interfaces---without the need for contact stabilization---is demonstrated by means of several examples.
	In addition, we show that the formulation is stable with respect to the condition number of the stiffness matrix by using a simple Jacobi-like diagonal preconditioner.
\end{abstract}

\section{Introduction}

The computational modeling of problems in contact mechanics requires careful considerations in order to prevent interpenetration between contacting bodies and ensure a proper transfer of contact tractions. A related problem arises in the coupling of non-conforming finite element discretizations, where the presence of hanging nodes, if not handled properly, yields a discontinuous displacement field. In this work these problems are addressed by means of an enriched finite element method that naturally leads to a simple computer implementation and inherently avoids the emergence of locking due to an over-constrained interface.

The numerical analysis of many engineering problems requires the coupling of meshes of different components.
These meshes are often non-conforming, \textit{i.e.}, the locations of their nodes do not coincide along the coupling interface, resulting in hanging nodes that call for special treatment. More importantly, the coupling procedure has to avoid over-constraining the coupling interface---to prevent locking---and ensure a proper transfer of tractions between subdomains.
Similar issues are shared by engineering applications that involve contact between bodies as, for instance, forging processes, gear systems, and impact problems~\cite{Wriggers2006}.
While coupling non-matching meshes is enforced by means of equality constraints, contact problems use inequality constraints, which makes modeling contact notoriously challenging because of its intrinsic highly nonlinear nature. In general, numerical methods used to resolve contact problems are either incapable of accurately transferring tractions between contacting bodies or require a very intricate computer implementation.

Coupling of non-conforming interfaces and contact is numerically handled by enforcing constraints, to guarantee continuity for the former and avoid interpenetration while allowing both bodies to slide or detach for the latter.
Although constraints can be enforced in several manners, their application often leads to locking due to an over-constrained interface.
In the multi-point constraint~(MPC) method, also known as master/slave approach, non-conformity is handled by constraining slave nodes on one side of the interface to those on the master side. This method, however, has been shown to be sensitive to the choice of master/slave surfaces insofar as guaranteeing continuity. A two-pass MPC approach, where both sides are taken as master and slave of each other~\cite{papadopoulos1992mixed,Haikal:2010}, ensures continuity but suffers from locking due to an over-constrained interface.
Dual space methods, such as finite element tearing and interconnecting (FETI)~\cite{farhat1991} and the mortar method~\cite{bernardi1989new, flemisch2005new}, use a Lagrange multiplier field to enforce compatibility at the interface. In both methods, the Lagrange multiplier field must be selected carefully to avoid locking by satisfying the inf-sup condition, also known as the Ladyzhenskaya--Babu{\v s}ka--Brezzi (LBB) condition~\cite{babuvska1973finite, brezzi2012mixed}.
A detailed account of several methodologies proposed in the literature to enforce contact constraints is given \S~\ref{sec:contact_review}.

Enriched finite element methods, whereby the primal field is enhanced or enriched by means of appropriate enrichment functions, provide an elegant approach to dealing with non-conforming meshes and contact.
For instance, Haikal and Hjelmstad~\cite{Haikal:2010} proposed an enriched stabilized discontinuous Galerkin formulation for coupling non-conforming meshes that recovers accurate tractions and is devoid of locking. This was accomplished by modified finite element (FE) shape functions and the addition of enriched nodes, effectively recovering an NTN type of constraint enforcement.

The eXtended/Generalized Finite Element Method has also been applied to solve contact and non-conforming interface problems~\cite{Duarte:2007, Dolbow2001, Hirmand2015, akula2019mortex, Khoei2007}.
Duarte \textit{et al.}~\cite{Duarte:2007} modified the FE partition of unity by means of clustering and used X/GFEM to deal with non-matching interfaces. For contact problems, Khoei and Nikbakht~\cite{Khoei2007} and Dolbow \textit{et al.}~\cite{Dolbow2001} used X/GFEM to model frictional contact, where contact surfaces were treated as embedded strong discontinuities. Hirmand \textit{et al.}~\cite{Hirmand2015} proposed an augmented Lagrangian-based stabilization technique to model frictional contact and thus obtain smooth contact tractions. Similarly, Akula \textit{et al.}~\cite{akula2019mortex} used the augmented Lagrange method (ALM) within the mortar method to model complex contact surfaces as embedded discontinuities in a background mesh using X/GFEM; a special stabilization technique was required when contacting bodies have high contrast in stiffness and/or mesh density.
Nevertheless, X/GFEM is in general unreliable insofar as obtaining accurate interface tractions or requires stabilization techniques. Arag\'{o}n and Simone~\cite{defem} reported the oscillatory nature of the traction field recovered by X/GFEM on a notched beam where a cohesive formulation was used to model perfect bonding. It has also been shown that X/GFEM cannot be used as an immerse boundary (or fictitious domain) method since recovered tractions in Dirichlet boundaries oscillate even when using the Barbosa-Hugues stabilization~\cite{Haslinger:2009}. The issue is caused by an over-constrained interface (or boundary), and thus the inf-sup condition is not satisfied~\cite{ramos2015}. Putting aside the issue of oscillatory tractions, X/GFEM may be unstable with respect to the condition number of system matrices, a problem that has prompted recent efforts in pursuit of a stable GFEM (SGFEM)~\cite{SGFEM, Babuska:2017}. Finally, X/GFEM may have poor accuracy in blending elements (elements containing both standard and enriched nodes), prescribing non-homogenous Dirichlet boundary conditions requires especial techniques, and the computer implementation is far from trivial~\cite{defem}.

Inspired by GFEM, another family of enriched FE formulations seeks to solve problems with discontinuities by placing enrichments to nodes created along the discontinuities---as opposed to X/GFEM, where enrichments are added to the nodes of the original mesh. The Interface-enriched Generalized Finite Element Method (IGFEM)~\cite{igfem} was first proposed to resolve weak discontinuities, \textit{i.e.}, situations where the gradient of the primal field is discontinuous, as in interface problems. The method later developed into the Hierarchical Interface-enriched Finite Element Method (HIFEM)~\cite{Soghrati2014}, whereby multiple interfaces within a single element are resolved via a hierarchical implementation of enrichment functions. The Discontinuity-Enriched Finite Element Method (DE-FEM)~\cite{defem} later proposed a generalization of IGFEM/HIFEM for the treatment of both weak and strong discontinuities---\textit{e.g.}, fracture---in a unified formulation.
Because enrichments are placed along discontinuities, this family of methods is devoid of many of the disadvantages of X/GFEM: the implementation is straightforward, the method is stable, \textit{i.e.}, the condition number grows with the problem size as in standard FEM~\cite{Aragon:2020a}, and there are no blending elements since enrichment functions are localized to cut elements and vanish at all original mesh nodes.
The latter therefore keep the Kronecker-$\delta$ property, which significantly simplifies the enforcement of non-homogeneous essential boundary conditions~\cite{ramos2015,defem}. Contrary to most non-standard FEMs in use today, in IGFEM/HIFEM/DE-FEM Dirichlet boundary conditions can be enforced strongly as in standard FEM. This is also true even along non-matching Dirichlet boundaries, where HIFEM/DE-FEM is used as an immerse boundary procedure with smooth recovered traction fields~\cite{vandenBoom2018,vanDenBoom:2019:cover}.
This is particularly interesting in the context of coupling non-conforming discretizations and contact and has largely inspired the present study.

In this paper we adopt the IGFEM paradigm for coupling non-conforming discretizations and solving contact problems. To this end, we place enriched nodes along non-conforming interfaces and contact surfaces, respectively, collocated at the locations of the non-conforming mesh nodes on the opposite surface. Their associated enriched functions are $C^0$-continuous and local to the enriched elements on one side of the non-conforming interface/contact surface. Notably, since the shape functions of the original mesh are kept intact, the partition of unity property is therefore lost in enriched elements, \textit{i.e.}, at any point in the element the sum of shape and enrichment functions does not add to unity. The formulation yields an enriched NTN contact discretization for which is straightforward to enforce constraints by any procedure; here we explore the enriched procedure with both MPC and ALM.
The proposed procedure is therefore a two-pass method and therefore shows no bias on the choice of master/slave surfaces.
The procedure is applied to several examples with linearized kinematics.
Results show that the method passes Taylor and Papodopoulos' contact patch test~\cite{Taylor:1991} and avoids locking due to an over-constrained interface; furthermore, traction transfer is fairly accurate without the need of stabilization techniques.
This is advantageous over other methods such as the two-pass mortar method which, although passes the patch test and shows no locking, requires interpolation of the pressure field that results in a complex formulation and corresponding computer implementation.
We also show the stability of the method insofar as the condition number of the stiffness matrix is concerned. Indeed, the condition number grows with mesh size $h$ as $\mathcal{O}\left( h^{-2}\right)$---\textit{i.e.}, at the same rate as that of the standard FEM---after the application of a simple diagonal preconditioner.

\section{Previous work on contact discretizations}\label{sec:contact_review}

We limit the scope of this survey to contact between deformable bodies, \textit{i.e.}, deformable-deformable contact.
Many contact discretization procedures have been proposed through the years.
Node-to-node (NTN)~\cite{oden1981exterior} and the node-to-segment (NTS)~\cite{Wriggers1985, BENSON1990141, Papadopoulos1993} discretizations are widely used, where constraints are enforced between a node pair or a node-segment pair, respectively.
Their applicability is, however, hindered by some intrinsic limitations:
Even though NTN can transfer tractions accurately, it can only be used with conforming discretizations along master/slave surfaces and only for infinitesimal sliding. The NTS approach overcomes this restriction but a one-pass approach---where nodes of one contacting surface are constrained to the segments of the other surface---is biased insofar as the choice of master/slave surfaces. A one-pass NTS approach not only fails to prevent interpenetration at times, it has been shown not to pass the contact patch test~\cite{Taylor:1991}, \textit{i.e.}, correct contact tractions cannot be recovered. Several works have attempted to address this issue. Papadopoulos and Taylor~\cite{papadopoulos1992mixed} proposed to enforce constraints in an average sense and integrate the contact pressure using Simpson's rule, thereby passing the patch test. Later, Zavarise and De~Lorenzis~\cite{zavarise2009modified} improved the one-pass NTS contact formulation to pass the contact patch test at the expense of losing symmetry in the contact contribution to the stiffness matrix.

A two-pass NTS approach could also be used to avoid interpenetration of contacting bodies, but this approach leads to an over-constrained contact interface and ultimately to locking. As a result, the traditional two-pass NTS contact discretization does not fulfill the LBB condition~\cite{Kikuchi1988, papadopoulos1992mixed} and therefore no convergence can be attained with mesh refinement. Papadopoulos \textit{et al.}~\cite{Papadopoulos1995} and Papadopoulos and Solberg~\cite{Papadopoulos1998} proposed a method that enforces continuity of tractions weakly, whereby nodes are divided into groups in which the gap is constrained and pressure continuity is enforced. Their approach avoids locking and passes the patch test when the same interpolation for geometry and traction field is used. The method was later extended to 3D by Jones and Papadopoulos~\cite{Jones2001}.

The idea of enforcing traction continuity or constraints in a weak sense can be understood as a segment-to-segment (STS) method, first proposed by Simo \textit{et al.}~\cite{SIMO1985163}, where displacement field is interpolated with linear shape functions, the pressure field is defined as piecewise constant over contact segment, so that the local variables (gaps) are evaluated in an ``average'' sense. Another STS-based procedure by Zavarise and Wriggers~\cite{Zavarise1998} also enforces contact constraints in a weak sense, whereby local variables are evaluated at integration points and then contact contributions to the stiffness matrix and force vector are calculated based on the integration of these local variables. El-Abbasi and Bathe~\cite{El-Abbasi2001} proposed yet another STS method in which integration points are projected onto the opposite contact surface. Although their formulation is capable of handling both linear and quadratic elements in contact, the LBB condition is satisfied only when the pressure is interpolated with linear continuous functions. In addition, a ``composite'' integration rule that combines Gaussian and Newton-Cotes rules is required to pass the patch test. All these formulations and their corresponding computer implementations are more complex than those of NTS contact.

The mortar method was first used to handle domain decomposition problems~\cite{bernardi1989new} and then contact~\cite{Belgacem1998, McDevitt2000, Wohlmuth2001}. The method also has single-pass and dual-pass versions. The traction field is defined on a \textit{mortar surface}, which is one of the two contacting surfaces or an intermediate surface that is introduced, and contact constraints are enforced in a weak sense. The single pass mortar method has been shown to pass the contact patch test~\cite{Puso2004}. This method was later extended to 3D and large deformations~\cite{Puso2004,Puso2004a, Fischer2005, Yang2005, Puso2008,Tur2009, Hueber2007, Laursen2012, Temizer2012a}. For large deformation kinematics, gap functions interpolated by shape functions in the mortar surface and nodally-averaged normal vectors  are used~\cite{Puso2008}. To avoid non-smooth normals, special techniques are also employed to obtain smooth surfaces, for instance by means of Hermite functions~\cite{Puso2004a} or by other averaging techniques~\cite{Popp2012}. Results of this method are generally much more accurate than those of traditional NTS procedures---but at the same the computer implementation is more involved and requires more computational resources~\cite{Farah2015}. Although some mortar-based methods satisfy the LBB condition, they are biased insofar as the choice of contact surfaces is concerned. Still, an appropriate choice of dual (Lagrange multiplier) shape functions should be made. Furthermore, the computer implementation is much more complex than standard NTS, which is also less demanding in terms of computational costs, especially in 3D. To avoid the contact surface bias issue, Solberg \textit{et al.}~\cite{Solberg2007} proposed a two-pass Mortar method with contact constraints enforced by means of Lagrange multipliers, thereby nodes are separated into \textit{active} and \textit{inactive} sets; however, a penalty-based stabilization is needed to avoid traction oscillations, and this method is not guaranteed to work for arbitrary geometries, particularly in 3D. Recently, this method was also extended to solve plasticity and self-contact by Puso and Solberg~\cite{Puso2020}. Park \textit{et al.}~\cite{Park2002} and Rebel \textit{et al.}~\cite{Rebel2002} proposed a method, different from the mortar method, where nodes from two contact surfaces are also projected into an intermediate surface, and an independent variable is defined to describe the motion of this surface. By enforcing the traction continuity weakly, constant stress can be transferred exactly for contact patch test. This method was also extended to handle frictional contact by Gonzalez~\cite{Gonzalez2006}.
On the basis of the mortar method, a dual Mortar method was proposed by Flemisch \textit{et al.}~\cite{flemisch2005new} to solve contact problems with curved interfaces in 2D, where discontinuous shape functions for dual variables are used because they are more stable and ensure more accurate solutions. This method was then extended to 3D contact problems with large deformation by Popp \textit{et al.}~\cite{Popp2010} and Popp and Wall~\cite{Popp2014}, and extended to frictional contact by Gitterle \textit{et al.}~\cite{Gitterle2010}. In general, the dual Mortar method offers the possibility of condensing the multipliers out of the system matrix (another advantage of using dual shape functions), resulting in a symmetric positive definite matrix. In their overview, Popp and Wall~\cite{Popp2014} highlight several advantages of the dual mortar method for contact over the traditional NTS approach. They show that, although smooth interpolations of the contact surfaces are needed (for instance using NURBS), especially for large deformation contact, solving problems with dual mortar can be more efficient than with the traditional single-pass mortar procedure.

Because the smoothness of contact surfaces influences the transfer of tractions greatly, several surface smoothing techniques have been proposed over the years: Belytschko \textit{et al.}~\cite{Belytschko2002} suggested to compute gap functions based on a least square fit of the original non-smooth surface, obtaining a smooth signed distance function which reduces traction oscillations; Padmanabhan and Laursen~\cite{Padmanabhan2001} and Sauer~\cite{ Sauer2013} proposed to use Hermite functions to interpolate contact surface and solution fields; El-Abbasi \textit{et al.}~\cite{El-Abbasi2001a} suggested to use cubic splines; Krstulovic-Opara \textit{et al.}~\cite{Krstulovic-Opara2002} proposed a Bézier formulation; Sauer~\cite{Sauer2011} used high-order Lagrange shape functions; and Stadler \textit{et al.}~\cite{Stadler2003} used a NURBS-based formulation.
These smoothing techniques aim at reducing traction oscillations and to improve convergence.

Contact constraints have also been enforced recently in the context of isogeometric analysis (IGA), whereby CAD features are used directly in the numerical analysis. Lu~\cite{Lu2011} and De Lorenzis \textit{et al.}~\cite{Lorenzis2011} proposed IGA formulations for frictionless and frictional contact, respectively; a similar approach to NTS was established in IGA, the so-called \textit{knot-to-surface}, where the main difference with respect to the traditional NTS lies in that the knot is used instead of mesh nodes and NURBS are used to describe contact surfaces. This method was later extended to 3D~\cite{Temizer2011, Temizer2012, DeLorenzis2012}, and a mortar-based IGA procedure was found to be more accurate than the standard knot-to-surface method. Corbett and Sauer~\cite{Corbett2014, Corbett2015} proposed to use a NURBS interpolation along contact surface, while the rest is discretized using linear elements. This modification makes the approach more efficient. For contact in large deformation, accurate results are still generally hard to obtain without mesh refinement. Dimitri \textit{et al.}~\cite{Dimitri2014, Dimitri2015} proposed to use T-splines to refine the discretization. Similar to the dual mortar method with FEM, a dual mortar method in IGA was also proposed by Seitz \textit{et al.}~\cite{Seitz2016} and offers better efficiency than the mortar-based IGA contact formulation~\cite{Dittmann2014}. An isogeometric collocation method was also used to solve contact problems~\cite{Kruse2015, DeLorenzis2015}, whereby contact forces are regarded as Neumann boundary conditions, and contact constraints are enforced as Dirichlet conditions. Since these conditions are fulfilled strongly, quadrature of Neumann BCs is eliminated and efficiency is improved greatly. Recently, Duong and Sauer~\cite{Duong2019} and Duong \textit{et al.}~\cite{Duong2019a} proposed an IGA contact formulation based on the surface potential method~\cite{Sauer2013a} and equipped with the refined boundary quadrature method~\cite{Duong2015}. In this procedure the ``two-half-pass'' method~\cite{Sauer2015} is used, \textit{i.e.}, only tractions on the slave surface are considered in each pass. XFEM is also used in this method to capture weak/strong discontinuities around contact boundaries~\cite{Duong2019a}. With a smoothing technique for post-processing, results are much more accurate than those obtained with traditional two-pass methods. However, IGA-based methods still suffer from locking, with the exception of the two-half-pass method, and the procedure requires a completely different discretization based on NURBS---which may not be straightforward to implement in general displacement-based FEM codes.

For most methods above, it is still not trivial to obtain accurate contact tractions for complex problems due to numerical artifacts~\cite{zavarise2009modified}. Even for mortar methods traction oscillations are generated for curved contact interfaces---which are relatively small and concentrated near the contact edge with the correct choice of mortar surface~\cite{akula2019mortex}. These numerical errors do not necessarily reduce with mesh refinement. Some STS and mortar based methods~\cite{Papadopoulos1998, Jones2001,Solberg2007,Park2002,Temizer2013, Puso2020} are quite accurate in the definition of contact tractions, pass the contact patch test, and avoid locking. Their formulations and implementations are, however, much more complicated than those of traditional NTN or NTS procedures.

There are also procedures that aim at transforming the contact problem into an equivalent NTN contact constraint enforcement.
For instance, the virtual element method (VEM) has recently been explored to solve problems in contact mechanics~\cite{Wriggers2016}. In VEM non-conforming contacting nodes are projected onto the opposite contact surface, and new nodes are generated at those locations, transforming the problem into a VEM-conforming mesh for which NTN contact constraints can be enforced straightforwardly. Besides, adding contact constraints to a general VEM code is straightforward. Results show that VEM passes the patch test. However, a penalty-based stabilization is needed to avoid traction oscillations. This method was later developed to solve frictional contact in large deformation kinematics by Wriggers and Rust~\cite{Wriggers2019a}, and solve contact with curved contact surfaces by Aldakheel \textit{et al.}~\cite{Bashir-ahmed2020}.

Enriched formulations have also been used to convert non-conforming contact discretizations into an equivalent NTN enforcement.
For instance, the enriched discontinuous Galerkin formulation for coupling non-conforming discretizations by Haikal and Hjelmstad~\cite{Haikal:2009} was later extended to solve frictionless contact with finite deformation kinematics~\cite{Haikal:2010}.
Masud~\textit{et al.}~\cite{Masud:2012} further combined the ideas put forward by Haikal~\textit{et al.}~\cite{Haikal:2009,Haikal:2010} with Nitsche's method in a variational multiscale framework that could be used not only for coupling non-conforming meshes but also for solving frictional contact problems.
The methodology proposed in this paper follows the enrichment strategy proposed by Haikal and Hjelmstad~\cite{Haikal:2010} in which, instead of modifying the shape functions of elements in contact, we keep the standard basis intact. Both strategies could be understood as a form of $h$-hierarchical refinement along contact surfaces and thus share some similarities with the non-hierarchical $h$-refinement approach used to handle non-conforming meshes in 3D proposed by Jiao and Heath~\cite{Jiao2004,Jiao2004a}. Their approach is based on the concept of common refinement of two meshes which is defined as ``the intersections of the elements of the input meshes''. The new discretization contains therefore nodes that can be found in both non-matching meshes. Our approach follows a similar strategy in that all standard nodes from one contact surface are projected to the other and vice versa.

\section{Problem description and formulation}

\subsection{Governing equations}

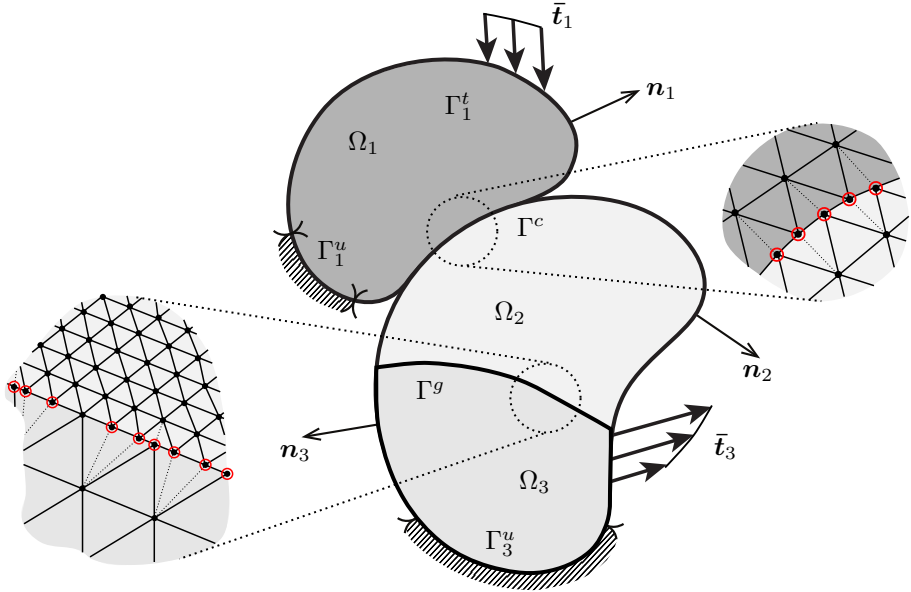
\begin{figure}
	\centering
	\def\svgwidth{0.7\textwidth}
	\input{potatos_contact_nonmatch.pdf_tex}
	\caption{Problem schematic showing the contact between two domains $\Omega_1$ and $\Omega_2$ along their contacting surface~$\Gamma^c_{12}$. The coupling between two domains $\Omega_2$ and $\Omega_3$ (perfect bonding) is also illustrated.
	Traction $\bs{\bar{t}}_1$ and $\bs{\bar{t}}_3$ are applied on boundaries $\Gamma_1^t$ and $\Gamma_3^t$, and boundaries $\Gamma_1^u$ and $\Gamma_3^u$ are subject to a prescribed displacement. Outer surface normals are denoted $\bs n_i, i=1\ldots3$.
	The insets show non-matching meshes at the contacting interface, with original standard FEM nodes shown in black, and enriched nodes, on the opposite surface, shown as red circles.}\label{fig:coupling_schematic}
\end{figure}

Figure~\ref{fig:coupling_schematic} shows domain $\Omega \subset \mathbb{R}^d$ composed of three $d$-dimensional subdomains $\Omega_i \subset \Omega$ such that $\overline{\Omega} = \cup_i \overline {\Omega}_i$. The subdomain boundaries are denoted $\partial \Omega_i \equiv \Gamma_i = \smash{\overline{\Omega}}_i \setminus \Omega_i$ and their outer normals $\bs n_i$.   Domains $\Omega_1$ and $\Omega_2$ are in frictionless contact along the surface $\Gamma_{12} = \Gamma_1 \cap \Gamma_2 \equiv \Gamma^c$, and domains $\Omega_2$ and $\Omega_3$ are perfectly bonded along the interface $\Gamma_{23} = \Gamma_2 \cap \Gamma_3 \equiv \Gamma^g$. Such interface could be physical (\textit{e.g.}, the interface between two different materials), numerical (\textit{e.g.}, a non-conforming discretization with hanging nodes), or a combination thereof.
Boundaries $\Gamma_i^u$ and  $\Gamma_i^t$ denote regions with prescribed Dirichlet and Neumann boundary conditions, respectively. These regions are disjoint, \textit{i.e.}, $\Gamma_i^u \cap \Gamma_i^t = \emptyset$.
The displacement (primal) field is denoted by $\bs{u}$ and the traction (dual) field by $\bs{t}$. These are composed by considering the fields within all domains $\Omega_i$. We therefore denote $\bs{u}_i$ the restriction of $\bs{u}$  to the $i$th domain, \textit{i.e.}, $\bs{u}_i = \left. \bs{u} \right|_ {\Omega_i}$. The same notation applies to the traction field $\bs{t}_i$ and to other subscripted quantities.

We are interested in solving the elastostatics frictionless contact boundary value problem whose strong form is expressed as: Given body force $\bs b_i : \Omega_i \to \mathbb{R}^d$, prescribed displacement $\bar{\bs{u}}_i : \Gamma_i^u \to \mathbb{R}^d$, and traction $\bar{\bs{t}}_i : \Gamma_i^t \to \mathbb{R}^d$ fields, find $\bs{u}_i$  such that
\begin{alignat}{2}
	\div \bs\sigma_i + \bs b_i & = \bs 0 \quad                                               &  & \text{in }\Omega_i, \label{eq:equi}             \\
	\intertext{with boundary conditions}
	\bs u_i                    & = \bar{\bs u}_i \quad                                       &  & \text{on }  \Gamma_i^u,                         \\
	\bs \sigma_i \bs n_i       & = \bar{\bs t}_i \quad                                       &  & \text{on } \Gamma_i^t, \label{eq:equi_traction} \\
	\intertext{interface conditions for perfect bonding}
	\bs u_2                    & = \bs u_3 \quad                                             &  & \text{on }  \Gamma^g \label{eq:equality},       \\
	\bs \sigma_2 \bs n_2       & = \bs \sigma_3 \bs n_3 \quad                                &  & \text{on } \Gamma^g,
	\intertext{and contact conditions}
	p_n                        & = \bs t_1^c \cdot \bs n_1  \leq 0 \quad                     &  & \text{on }\Gamma^c ,                            \\
	g_n                        & = \left(\bs x_2 - \bs x_1\right) \cdot \bs n_1 \geq 0 \quad &  & \text{on } \Gamma^c ,                           \\
	p_n g_n                    & = 0 \quad                                                   &  & \text{on } \Gamma^c. \label{eq:gappressure}
\end{alignat}
In Equation~\eqref{eq:equi}, $\bs \sigma_i = \bs{\mathcal{D}}_i : \bs \varepsilon_i$ is the Cauchy stress tensor which obeys Hooke's law, $\bs{\mathcal{D}}_i$ is a fourth-order constitutive tensor, and  $\bs \varepsilon_i = \frac{1}{2} \left( \grad \bs u_i + \grad \bs u_i^{\intercal} \right)$ the linearized strain tensor (small deformation theory).
For the perfect bonding between domains $\Omega_2$ and $\Omega_3$, equal displacements and tractions are enforced through interface $\Gamma^g$. Contact between domains $\Omega_1$ and $\Omega_2$ is enforced through the classical Hertz--Signorini--Moreau conditions, also known as Karush--Kuhn--Tucker conditions in the theory of optimization~\cite{Wriggers2006}; these consist of a contact pressure $p_n$, which can only be in compression, and a gap function $g_n$,  which ensures that the contact surfaces cannot penetrate one other. Finally, Equation~\eqref{eq:gappressure} ensures that pressure vanishes when the gap between the two contact surfaces is open, and the gap vanishes when they are in contact.

Equations~\eqref{eq:equi}-\eqref{eq:gappressure} can be seen as the solution to a constrained optimization problem on the functional given by
\begin{equation} \label{eq:functional}
	\Phi \left( \bs{u}, \bs{\lambda}^g, \bs{\lambda}^c \right) = \sum_{i=1}^3 \Pi_i \left( \bs{u}_i \right) + \Pi^g \left( \bs{u}_2, \bs{u}_3, \bs{\lambda}^g   \right) + \Pi^c \left( \bs{u}_1, \bs{u}_2, \bs{\lambda}^c \right),
\end{equation}
where the first term represents the potential energy in all domains. For the $i$th domain, the potential energy is
\begin{equation} \label{eq:potential_energy}
	\Pi_i \left( \bs{u}_i \right) = \frac{1}{2} \int_{\Omega_i} \bs{\varepsilon} \left( \bs{u}_i \right) : \bs{\sigma} \left( \bs{u}_i \right)  \, \dd{\Omega} - \int_{\Omega_i} \bs{b} \cdot  \bs{u}_i \, \dd{\Omega} - \int_{\Gamma_i^t} \bar{\bs{t}} \cdot  \bs{u}_i \, \dd{\Omega}.
\end{equation}
The other two terms in~\eqref{eq:functional} are associated with constraints at the perfectly bonded interface $\Gamma^g$ and at the contact interface $\Gamma^c$, enforced with Lagrange multipliers $\bs{\lambda}^g$ and $\bs{\lambda}^c$, respectively; their form varies depending on how the constraints are enforced. In this work we use the standard Lagrange multiplier method for the perfectly bonded interface, and an augmented Lagrangian type of enforcement for contact (which has been shown to outperform both Lagrange and penalty methods). Their forms are therefore respectively written as
\begin{align}
	\Pi^g \left( \bs{u}_2, \bs{u}_3, \bs{\lambda}^g \right) & = \int_{\Gamma^g} \bs{\lambda}^g \cdot \left( \bs{u}_2 - \bs{u}_3 \right) \, \dd{\Gamma},                                                                  \\
	\Pi^c \left( \bs{u}_1, \bs{u}_2, \bs{\lambda}^c \right) & = \int_{\Gamma^c} \frac{1}{2\epsilon_n} \left[ \big< \hat{\lambda}_n \big>^2 - \lambda_n^2\right] \, \dd{\Gamma},\label{eq:augmented_lagrangian_potential}
\end{align}
where $\left< \bullet \right>$ denotes the Macaulay brackets, $\lambda_n = \bs{\lambda}^c \cdot \bs{n}$ the normal component of the Lagrange multiplier (where we could use either $\bs{n}_1$ or $\bs{n}_2$, as at this stage the choice is arbitrary), and $\hat{\lambda}_n$ the augmented Lagrange multiplier given by $\hat{\lambda}_n = \lambda_n + \epsilon_n g_n =  p_n$, with  $\epsilon_n$  the augmentation (penalty) parameter in the normal direction.

\subsection{Variational formulation}

The solution of the boundary value problem makes the optimization problem~\eqref{eq:functional} stationary with respect to variations of all arguments. These three equations are obtained by taking the directional derivative along the directions of the variations:
\begin{align}
	\grad \Phi \left( \bs{u}, \bs{\lambda}^g, \bs{\lambda}^c \right) \cdot \delta \bs{u}         & = \sum_{i=1}^3 \left[ \int_{\Omega_i} \bs\varepsilon_i \left( \delta \bs u_i \right) :\bs\sigma_i \left(\bs u_i \right)  \, \dd{\Omega} - \int_{\Omega_i} \delta \bs u_i \cdot \bs b_i \, \dd{\Omega}  - \int_{\Gamma_i^t} \delta \bs u_i  \cdot \bar{\bs t}_i \, \dd{\Gamma}  \right]  \nonumber \\
	                                                                                             & \qquad + \int_{\Gamma^g} \bs{\lambda}^g \cdot \left( \delta \bs{u}_2 - \delta \bs{u}_3 \right) \, \dd{\Gamma} + \int_{\Gamma_c}   \big< \hat{\lambda}_n \big> \delta g \, \dd{\Gamma} = 0,  \label{eq:stationary_a}                                                                               \\
	\grad \Phi \left( \bs{u}, \bs{\lambda}^g, \bs{\lambda}^c \right) \cdot \delta \bs{\lambda}^g & = \int_{\Gamma_{g}} \delta \bs{\lambda}^g \cdot\left(\bs{u}_2-\bs{u}_3 \right) \, \dd{\Gamma} = 0, \label{eq:stationary_b}                                                                                                                                                                        \\
	\grad \Phi \left( \bs{u}, \bs{\lambda}^g, \bs{\lambda}^c \right) \cdot \delta \bs{\lambda}^c & = \int_{\Gamma_c} \frac{1}{\epsilon_n} \left[ \big< \hat{\lambda}_n \big> - \lambda_n \right] \delta \lambda_n \, \dd{\Gamma} = 0 \label{eq:stationary_c},
\end{align}\label{eq:lag-final}%
for all admissible variations $\delta \bs{u}$, $\delta \bs{\lambda}^g$, and $\delta \bs{\lambda}^c$. For the variations of the displacement, we define the vector-valued function space
\begin{equation}
	\bs{\mathcal{V}}\left( \Omega \right) = \left\{  \delta \bs u \in \left[ L^2 \left( \Omega \right) \right]^d, \,  \left. \delta \bs{u} \right|_{\Omega_i} \in     \left[ \mathcal{H}^1(\Omega_i) \right]^d, \, \left. \delta \bs{u} \right |_{\Gamma_i^u} =  \bs 0 \right\},
\end{equation}
where $\mathcal{H}^1 \left( \Omega_i \right)$ denotes the first-order Sobolev space on $\Omega_i$.
The primal field is chosen from the set
\begin{equation}
	\bs{\mathcal{U}} \left( \Omega \right) = \left\{  \bs u \in \left[ L^2 \left( \Omega \right)   \right]^d, \,  \left. \bs{u} \right|_{\Omega_i} \in     \left[ \mathcal{H}^1(\Omega_i) \right]^d, \, \left. \bs{u} \right |_{\Gamma_i^u} =  \bar{\bs u} \right\}.
\end{equation}
Lagrange multipliers $\bs\lambda$ and their variations $\delta \bs{\lambda}$ are taken from a fractional Sobolev space, \textit{i.e.}, $\bs{\lambda}^a  \in \bs{\Lambda} \left( \Gamma^a \right) \equiv  \left[ \mathcal{H}^{-1/2} \left( \Gamma_a \right) \right]^d,$ with $a = g,c$.

Multiple-point constraints (MPCs) are also investigated in this work to enforce both perfect bonding and contact. In such case, the enforcement works by writing simple constraint equations that describe the relationship between slave degrees of freedom (DOFs) as a function of the master DOFs. As there is no need for Lagrange multipliers, the rightmost two terms in~\eqref{eq:functional} disappear.

Next we discuss the discretization of Equations~\eqref{eq:stationary_a} through~\eqref{eq:stationary_c}. We start with the enriched formulation used in elements where constraints are to be enforced.

\subsection{The finite-dimensional interface-enriched generalized finite element formulation}
\label{sec:enriched}

It is not straightforward to couple meshes or simulate contact when the discretizations of the domains are non-matching, \textit{i.e.}, when there are hanging nodes in the former or contacting nodes do not occupy the same location in space in the latter (no node-to-node contact).
To alleviate the burden associated with the lack of mesh conformity, we employ an enrichment scheme inspired by the Interface-enriched Generalized Finite Element Method (IGFEM)~\cite{igfem,Aragon:2020a}, whereby enriched nodes are created along material interfaces to resolve weak discontinuities (those in the field gradient).
Consequently, all domains are discretized into non-overlapping finite elements such that $\Omega_i^h = \cup_i e_i$, $e_i \cap e_j = \emptyset \, (\forall i \neq j)$, with the entire computational domain given by $\cup_i \Omega_i^h = \Omega^h \approx \Omega$.
In other to solve numerically the finite-dimensional form of the equations in~\S\ref{eq:lag-final}, the trial solution and the weight function are chosen from the interface-enriched generalized finite element space
\begin{equation}
	\mathcal{S}_e = \Big\{ \left. \bs u^h  \right| \,   \bs u^h(\bs x)=\underbrace{\sum_{i\in \iota_h} N_i \left( \bs x \right) \bs u_i}_\text{std.~FEM} + \underbrace{\sum_{i\in \iota_e} s_i \psi_i(\bs x)\bs \alpha_i}_\text{enrichment}, \quad \bs{u}_i, \bs{\alpha}_i \in \mathbb{R}^d \Big\}.
	\label{eq:IGFEM_space}
\end{equation}
In~\eqref{eq:IGFEM_space} the first term represents the standard finite element component and the second term the enrichment. In the former, $\iota_h$ is the index set of all standard nodes in the mesh, $N_i$ is the $i$th standard Lagrange interpolation function associated with degrees of freedom $\bs u_i$ (which physically represent the nodal displacement at standard node $\bs{x}_i$). In the enrichment term, $\iota_e$ is the index set of enriched nodes, $\psi_i$ is the enrichment function associated with enriched DOFs $\bs \alpha_i$. Finally, $s_i$ is a scaling factor that is used to improve the condition number of the system matrix after assembly. This factor, which was studied thoroughly in a recent publication~\cite{Aragon:2020a}, is required for a robust implementation that handles interfaces getting arbitrarily close to standard mesh nodes.

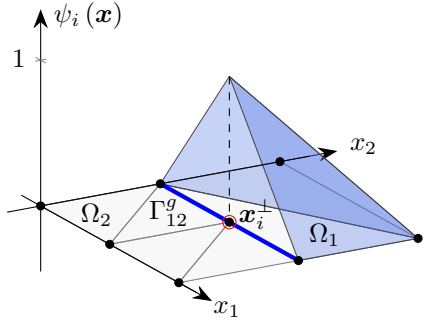
\begin{figure}
	\centering
	\input{enrich_func.tex}
	\caption{Enrichment function $\psi_i$ associated with enriched node $\enriched{x}{i}$, which coincides with a non-conforming standard node $\bs{x}_i$. The support of $\psi_i$ comprises integration elements only in $\Omega_1$; \protect\tikz{\protect\draw [red, thick] (0, 0) circle (2 pt);} represents an enriched node and  \protect\tikz{\protect\draw[fill = black, black] (0,0) circle (1.5pt);} a standard mesh node. }
	\label{fig:enrichedfunc}
\end{figure}

The enrichment function $\psi_i$ is constructed with the aid of Lagrange shape functions in integration elements (see enrichment $\psi_i$ corresponding to enriched node $\enriched{x}{i}$ in Figure~\ref{fig:enrichedfunc}), which not only form the support $\psi_i$, but, as the name implies, are also used for the numerical quadrature of the local stiffness and force arrays. Contrary to the original IGFEM formulation, where the support of an enrichment comprises integration elements at both sides of a material interface, here the support comprises only integration elements on one side of the non-conforming interface or contact surface. In our implementation, a computational geometric engine creates enriched nodes along the non-conforming interfaces or along contact surfaces, so that each enriched node corresponds to a standard mesh node on the surface of the opposite domain. As shown in Figure~\ref{fig:coupling_schematic}, the location of mismatching mesh nodes can be directly determined along the interface $\Gamma^g$ in mesh coupling problems, whereas for contact problems the closest node projection~\cite{Wriggers2006, Aragon:2013e} to an element edge is used to determine the location of enriched nodes at either side of contact surfaces.

\subsection{Stiffness matrix contributions}
\label{ssec:structural}

In this section we derive the discrete expressions for the stiffness contributions given by the first term in~\eqref{eq:stationary_a}. The assembly of the local stiffness matrix $\bs{k}_i$ and force vector $\bs{f}_i$  for finite elements that do not contain enriched nodes follows standard procedures. For enriched integration elements, and with reference to \cite{igfem, defem}, the local arrays  can be expressed as
\begingroup
\renewcommand*{\arraystretch}{1.2}
\begin{equation}
	\label{eq:igfem_stiff}
	\bs{k}_i =  \left[\begin{array}{ll}
			\bs k_{u u}                & \bs k_{u \alpha}      \\
			\bs k_{u \alpha}^\intercal & \bs k_{\alpha \alpha}
		\end{array}\right],  \qquad \bs{f}_i = \left[\begin{array}{c}
			\bs f_{u}      \\
			\bs f_{\alpha} \\
		\end{array}\right],
\end{equation}
with
\begin{alignat}{2}
	\bs{k}_{u u}      & =\int_{\Omega_{i}} \bs{B}_{u}^{\intercal} \bs{D} \bs{B}_{u} \, \dd{\Omega},
	                  & \bs{k}_{\alpha \alpha}                                                                                                                                     & =\int_{\Omega_{i}} \bs{B}_{\alpha}^{\intercal} \bs{D} \bs{B}_{\alpha} \, \dd{\Omega},  \label{eq:igfem_uu_aa}                                                        \\
	\bs{k}_{u \alpha} & =\int_{\Omega_{i}} \bs{B}_{u}^{\intercal} \bs{D} \bs{B}_{\alpha}  \, \dd{\Omega} =\bs{k}_{u \alpha}^{\intercal},
	                  & \bs{f}_{u}                                                                                                                                                 & =\int_{\Omega_{i}} \bs{N}_{u}^{\intercal} \bs{b}  \, \dd{\Omega} +\int_{\partial \Omega_{i}} \bs{N}_{u}^{\intercal} \bar{\bs t} \, \dd{\Gamma},                      \\
	\bs{f}_{\alpha}   & =\int_{\Omega_{i}} \bs{N}_{\alpha}^{\intercal} \bs{b}  \, \dd{\Omega} +\int_{\partial \Omega_{i}} \bs{N}_{\alpha}^{\intercal} \bar{\bs t}  \, \dd{\Gamma}, &                                                                                                                                                 & \label{eq:igfem_a}
\end{alignat}
where $\bs{D}$ is the constitutive matrix, and $\bs B_u$ and $\bs B_\alpha $ are the strain-displacement matrices of shape and enrichment functions, respectively~\cite{vandenBoom2018, defem-3d}. Finally, the global stiffness matrix $\bs K$ and global force vector $\bs F$ are obtained by standard assembly of their local counterparts. Denoting  by $\assembly$ the standard finite element assembly operator, global arrays are therefore
\begin{alignat}{2}
	\bs K          & =\assembly_i \bs k_i,                                                         & \bs{F}  =\assembly_i \bs f_i, \label{eq:assembled_arrays_a} \\
	\intertext{which, by making explicit the contributions of standard and enriched components, can be expressed as}
	\boldsymbol{K} & = \begin{bmatrix}
		\boldsymbol{K}_{u u}                  & \boldsymbol{K}_{u \alpha}      \\
		\boldsymbol{K}_{u \alpha}^{\intercal} & \boldsymbol{K}_{\alpha \alpha}
	\end{bmatrix},\qquad
	               & \boldsymbol{F} =\begin{bmatrix}
		\boldsymbol{F}_{u} \\
		\boldsymbol{F}_{\alpha}
	\end{bmatrix}.     \label{eq:assembled_arrays_b}
\end{alignat}
\endgroup

\subsection{Constraint enforcement}
\label{sec:constraint}

The enriched space alone does not ensure continuity across non-conforming interfaces and does not avoid interpenetration of contact surfaces. To properly resolve the field at mesh coupling and contact interfaces, constraints are imposed between enriched and standard DOFs.
Although the use of discrete MPCs and the augmented Lagrange method (ALM) is well established, their application in conjunction with this enriched framework is different and both methods are therefore thoroughly discussed in this section.

\subsubsection*{Multiple-point constraint method}

Conventionally, multiple-point constraints (MPCs) are used in a ``master and slave'' situation to enforce some sort of compatibility between nodes, \textit{e.g.}, for enforcing continuity along an interface or to enforce periodicity at opposite sides of a domain; an MPC can be simply regarded as an approach to create a ``tie'' among nodes. This method can be used in a single-pass or in a two-pass approach, which means that either side is selected as the master surface, or that either side serves as the master to its opposite side, respectively. In both approaches, the constraint is expressed as a linear combination of displacement vectors as in
\begin{equation}
	\bs u_s - \sum_{i\in \iota_m}N_i(\bs x)\bs u_i = \bs g,
	\label{eq:stdmpc}
\end{equation}
where $\bs u_s$ is the displacement of the slave node $\bs{x}_s$, $\iota_m$ is the index set of the slave's master nodes, and $\bs g$ is the gap vector. In mesh coupling problems, as the objective is to ensure continuity across an interface, the homogeneous MPC is used, \textit{i.e.}, $\bs g = \bs 0$. This constraint method is simple and straightforward to implement. However, as already discussed in the introduction, continuity cannot be ensured in a single-pass approach, and the system may become over-constrained in a two-pass approach~\cite{Haikal:2010}. In the following, we provide an alternative formulation of MPCs combined with the enriched approach outlined in the previous section for handling mesh coupling and contact problems without slipping and separation. In the context of IGFEM, MPCs have been employed in immersed domain problems~\cite{vandenBoom2018} and in the enforcement of Bloch-Floquet periodic boundary conditions~\cite{vandenboom2021}. Note, however, that for general contact problems it is preferred to adopt ALM (this will be explained in detail in the next section). For the cases where MPCs work, the objective is to ensure $C^0$-continuity across non-conforming interfaces or contact surfaces in a given direction.

For a standard hanging node $\bs{x}_i$, an enriched node $\enriched{x}{i}$ is created at the same location although placed in the adjacent element (see Figure~\ref{fig:enrichedfunc}). The constraint equation enforces that the displacement of standard and enriched nodes, $\bs u_i$ and $\enriched{u}{i}$, respectively, have to be equal. Referring back to~\eqref{eq:IGFEM_space}, this condition is expressed as
\begin{equation}
	\label{eq:enrichedmpc}
	\enriched{u}{i} = \sum_{j \in \indexset }N_j(\enriched{x}{i}) \bs u_j + s_i \bs \alpha_i = \bs u_i,
\end{equation}
where $\smash{\indexset} \subset \iota_h$ is the subset of mesh nodes with standard shape functions whose supports intersect the support of the enrichment function.
Mathematically, if the support of a standard shape function is defined as $\omega_i = \left\{ \left. \bs{x} \right| \, N_i \left( \bs{x} \right) \neq 0 \right\}$, and similarly for an enriched function $ \smash{\omega_i^\perp} = \left\{ \left. \bs{x} \right| \, \psi_i \left( \bs{x} \right) \neq 0 \right\}$, then
$\smash{\indexset} = \left\{ \left. i \in \iota_h \right| \, \omega_{i} \cap \smash{\omega_i^\perp} \neq \emptyset \right\} $.
Note that in Equation~\eqref{eq:enrichedmpc} we use $\psi_i \left( \smash{\enriched{x}{i}} \right) = 1$.

A similar constraint is enforced for every hanging standard node and enriched node pair at both sides of the non-conforming interface, making the proposed procedure actually correspond to a two-pass approach. However, when compared to just using Equation~\eqref{eq:stdmpc}, enriched DOFs render the interface response less stiff.
These constraints can be written in matrix form for all enriched nodes in the system as
\begin{equation}
	\bs U = \bs T \bar{\bs U},
\end{equation}
where $\bs U = \begin{bmatrix} \bs{u}_1  \ldots & \bs{\alpha}_1  \ldots \end{bmatrix}^\intercal $ is the vector containing all standard and enriched DOFs, $\bar{\bs U}$ is the vector of independent (unconstrained) DOFs, and $\bs T$ is a transformation matrix storing coefficients $N_i$ and $s_i$ in Equation~\eqref{eq:enrichedmpc}. Following standard procedures for MPCs~\cite{vandenBoom2018}, the unconstrained system $\bs K \bs U = \bs F $, where $\bs K $ and $\bs F$ were given in~\eqref{eq:assembled_arrays_b},
is transformed into $\bar{\bs K}\bar{\bs U} = \bar{\bs F}$, where $\bar{\bs K} = \bs T^{\intercal} \bs K \bs T$, $\bar{\bs F} = \bs T^{\intercal} \bs F$, and the transformation matrix $\bs T$ ties both standard and enriched nodes.

Notice that only a few additional enriched DOFs need to be added, and applying the MPC method between master and slave DOFs is straightforward. Numerical examples for coupling non-conforming discretizations will be shown in~\S~\ref{sec:numerical} to illustrate the accuracy and robustness of the approach.
\subsubsection*{Augmented Lagrange method}
\label{ssec:augmented_lagrange}

\begin{figure}
	\centering
	\def\svgwidth{4.5cm}
	\input{enriched_contact.pdf_tex}
	\caption{Schematic for the enriched discretization with ALM.  The locations of enriched nodes are detected using the projections of mismatching mesh nodes  with the normal vector (\textit{e.g.}, $\bs n_i^{+}$ for enriched node $\bs{x}_i^+$).
	The symbols \protect\tikz{\protect\draw [red, thick] (0, 0) circle (1.5pt);} represent enriched nodes and \protect\tikz{\protect\draw [fill=black, black] (0, 0) circle (1.5pt);} the standard mesh nodes. }
	\label{fig:alm}
\end{figure}
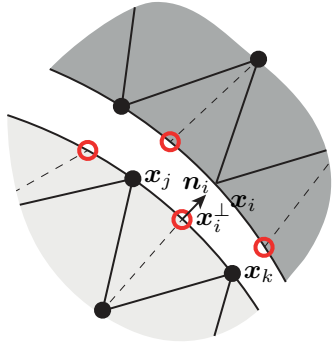

For general contact problems where relative slip and separation occur, it is not convenient to use MPCs to enforce constraints because the contact status needs to be detected during the nonlinear iterative contact step. The augmented Lagrange method can be regarded as a combination of penalty and Lagrange multiplier methods, whereby the inequality-constrained minimization contact problem is transformed into an unconstrained saddle point problem. Compared with pure penalty and Lagrange multiplier methods, this approach decreases the conditioning number of the stiffness matrix and is able to satisfy constraints exactly with finite penalty parameters~\cite{Simo1992}. In this section we obtain the discrete form corresponding to the weak form~\eqref{eq:stationary_a}-\eqref{eq:stationary_c}.
When considering Lagrange multipliers, we seek a solution to the system  $\hat{\bs K} \hat{\bs U} = \hat{\bs F}$ with
\begin{equation}
\begingroup
\renewcommand*{\arraystretch}{1.2}
\hat{\boldsymbol{K}}=\begin{bmatrix}
	\hat{ \boldsymbol{K}}_{u u}                 & \hat{\boldsymbol{K}}_{u \alpha}          & \hat{\bs K}_{u \lambda}      \\
	\hat{\boldsymbol{K}}_{u \alpha}^{\intercal} & \hat{\boldsymbol{K}}_{\alpha \alpha}     & \hat{\bs K}_{\alpha\lambda } \\
	\hat{\bs K}_{u \lambda}^{\intercal}         & \hat{\bs K}_{\alpha \lambda}^{\intercal} & \bs 0                        \\
\end{bmatrix}, \qquad  \hat{\bs{F}} =  \begin{bmatrix}
	\hat{\boldsymbol{F}}_{u} \\ \hat{\boldsymbol{F}}_{\alpha} \\ \hat{\bs F}_{\lambda}
\end{bmatrix}.
\label{eq:submatrix_mpc}
\endgroup
\end{equation}
where $\hat{\boldsymbol{U}} =   \begin{bmatrix}\boldsymbol{u} & \boldsymbol{\alpha} &\boldsymbol{\lambda}\end{bmatrix}^\intercal$ is the vector of unknowns in terms of standard DOFs, enriched DOFs, and Lagrange multipliers. The solution is found incrementally by making use of a generalized Newton loop~\cite{Curnier:54198, Karl2005}, \textit{i.e.}, $\hat{\bs K}  \Delta\hat{\bs U} = \Delta\hat{\bs F}$, where $\hat{\bs K}$ is mostly linear (we assume linear kinematics) and the nonlinear components arise due to contact. In the above expressions, an  explicitly differentiates this system from that used in the MPC method. Notice that a hat is also used for submatrix components, implying that the original matrices are modified by applying coupling terms.
In other words, the system arrays~\eqref{eq:assembled_arrays_b} are augmented with the contributions of the Lagrange multipliers, which are obtained by assembling the contributions of $n_\lambda$ constraints:
\begin{equation}
	\hat{\bs K} = \bs{K} +  \assembly_{i=1}^{n_\lambda} \bs k_i^c, \qquad \hat{\bs{F}}  = \bs{F} + \assembly_{i=1}^{n_\lambda} \bs f_i^c.
	\label{eq:systemAll}
\end{equation}
The explicit expressions of the Lagrange multiplier contributions $\bs k_i^c$ and $\bs f_i^c$ are given next.

\subsubsection*{Contact contribution}

Figure \ref{fig:alm} shows three standard mesh nodes $\bs{x}_i$, $\bs{x}_j$, and $\bs{x}_k$.  Without loss of generality and with reference to mesh node $\bs{x}_i$, we denote by $\enriched{x}{i}$ the position of the corresponding enriched node on the opposite surface. The position of $\enriched{x}{i}$ is determined by projecting $\bs{x}_i$ using the unit normal vector $\bs n_i$. The gap function $g_{n,i}$ along $\bs n_i$ and between $\bs{x}_i$ and $\enriched{x}{i}$ is expressed as
\begin{equation} \label{eq:gap_ith_node}
	g_{n,i} = \left( \bs x_i - \enriched{x}{i} \right) \cdot \bs n_i =  \left( \bs u_i - \enriched{u}{i} \right) \cdot \bs n_i +g_{0,i},
\end{equation}
where $g_{0,i} = \left( \bs{X}_i - \enriched{X}{i} \right) \cdot \bs{n}_i$ is the initial gap, $\bs{X}_i$ and $\enriched{X}{i}$ represent the initial positions of standard and enriched nodes, respectively, and $\bs{u}_i = \bs{x}_i - \bs{X}_i$ and  $\enriched{u}{i} = \enriched{x}{i} - \enriched{X}{i}$ their corresponding displacements.
Referring back to Figure~\ref{fig:alm}, the standard FEM shape functions attached to nodes $\bs{x}_j$ and $\bs{x}_k$ are the only ones that contribute to the displacement field at the location of enriched node $ \enriched{x}{i}$. By using~\eqref{eq:IGFEM_space} to express its displacement, the gap reads
\begin{equation}
	g_{n,i} = \left( \bs u_i - N_j( \enriched{x}{i}) \bs u_j - N_k(\enriched{x}{i}) \bs u_k - s_i \psi_i \left( \enriched{x}{i}\right)\bs \alpha_i   \right) \cdot \bs n_i + g_{0, i} =  \left( \bs{N}_i \otimes \bs I  \right) \enriched{U}{i}  \cdot \bs{n}_i + g_{0,i} ,
	\label{eq:enriched_gap}
\end{equation}
where
$ \otimes $ denotes the Kronecker product,
$\bs N_i =\begin{bmatrix}
		1 & - N_j(\enriched{x}{i}) & -N_k(\enriched{x}{i}) & - s_i\psi_i \left( \enriched{x}{i}\right) \end{bmatrix}$ with $\psi_i \left( \enriched{x}{i} \right) = 1$, $\enriched{U}{i} = \begin{bmatrix} \bs{u}_i & \bs{u}_j & \bs{u}_k & \bs{\alpha}_i \end{bmatrix}^\intercal$, and $\bs I$ is the $d \times d$ identity matrix.
Since the initial gap is constant during the analysis, the variation of~\eqref{eq:enriched_gap} is given by
\begin{equation}
	\delta g_{n,i} = \left( \delta \bs u_i - N_j( \enriched{x}{i}) \delta \bs u_j - N_k(\enriched{x}{i}) \delta \bs u_k - s_i \delta \bs \alpha_i \right) \cdot \bs n_i = \left( \bs{N}_i \otimes \bs I  \right) \delta\enriched{U}{i}  \cdot \bs{n}_i.
	\label{eq:enriched_vari_gap}
\end{equation}

The weak form of the contact contribution given by~\eqref{eq:stationary_c} is approximated by a summation over the $n_\lambda$ active contact nodes. Substituting $g_{n,i}$ and $\delta g_{n,i}$ with \eqref{eq:enriched_gap} and \eqref{eq:enriched_vari_gap}, respectively, the total contact contribution reads
\begin{equation}
	\int_{\Gamma_c}(\hat{\lambda}_n \delta g_n + \delta \lambda_n g_n) \, \dd{\Gamma} \approx \sum_{i=1}^{n_\lambda}\hat{\lambda}_{n,i} \left[  \left( \bs{N}_i \otimes \bs I  \right) \delta\enriched{U}{i}  \cdot \bs{n}_i  \right] + \sum_{i=1}^{n_\lambda}\delta\lambda_{n,i}\left[ \left( \bs{N}_i \otimes \bs I  \right) \enriched{U}{i}  \cdot \bs{n}_i + g_{0,i}  \right].
	\label{eq:weak_contact}
\end{equation}
Noteworthy, in the discretized right hand side of~\eqref{eq:weak_contact} the Lagrange multipliers represent the force acting on the enriched nodes. By introducing $\delta\hat{\bs U}_i = \begin{bmatrix}
		\delta\enriched{U}{i} & \delta \lambda_{n,i} \end{bmatrix} ^\intercal$, \eqref{eq:weak_contact} can be expressed as:
\begin{equation} \label{eq:contact_discrete}
	\int_{\Gamma_c}( \hat{\lambda}_n \delta g_n+\delta\lambda_n g_n ) \mathrm{d}\Gamma \approx \sum_{i=1}^{n_\lambda} \delta{\hat{\bs U} }_i^\intercal\bs G_i^{c}
\end{equation}
with
\begin{equation}
	\label{eq:computeC}
	\bs G_i^{c} = \begin{bmatrix}
		\lambda_{n,i} \bs C_i + \epsilon_{n, i}\bs{C}_i \bs C_i^\intercal \enriched{U}{i} & \bs C_i^\intercal\enriched{U}{i} + g_{0,i}
	\end{bmatrix}^\intercal~~\text{and}~~\bs{C}_i = \bs{N}_i^\intercal \otimes \bs{n}_i .
\end{equation}
Equation~\eqref{eq:contact_discrete} can be solved iteratively, and in this work we use the generalized Newton method~\cite{Curnier:54198, Karl2005}, where Lagrange multipliers are updated together with the primal field in the same iterative loop. In this method, which converges faster than Uzawa's algorithm, linearization of the contact contribution is required, leading to
\begin{equation}
	\frac{\partial }{\partial \hat{\bs{U}}_i} \left[ \sum_{i=1}^{n_\lambda} \delta\hat{\bs {U}}_i^\intercal\bs G_i^{c} \right] \Delta \hat{\bs U}_i = \sum_{i=1}^{n_\lambda} \delta\hat{\bs U}_i^\intercal \bs k_i^{c}\Delta \hat{\bs U}_i,
	\label{eq:contact_linearization}
\end{equation}
where $\Delta \hat{\bs U}_i = \begin{bmatrix} \Delta\enriched{U}{i} & \Delta\lambda_{n, i} \end{bmatrix}^\intercal$
and
\begingroup
\renewcommand*{\arraystretch}{1.2}
\begin{equation}
	\bs k_i^{c} = \begin{bmatrix}
		 & \epsilon_{n,i}\bs C_i \bs C_i^\intercal & \bs C_i \\
		 & \bs C_i^\intercal                       & 0
	\end{bmatrix}, \quad
	\bs f_i^{c} = -\begin{bmatrix}
		 & {\hat{\lambda}_{n, i} \bs C_i } \\
		 & {g}_{n,i}
	\end{bmatrix}.
	\label{eq:calculation Kc}
\end{equation}
\endgroup
Here $\bs k_i^c$ and $\bs f_i^c$ refer to the contact contribution of mesh node $i$ to the stiffness matrix and the load vector in a generalized Newton loop.

It is worth noting that this formulation and its solution procedure are similar to those of NTN contact (we refer the reader to \cite{Wriggers2006}). The only thing we needed to modify is the $\bs C_i$ vector, which is based on the vector $\bs N_i$ of enriched node $\enriched{x}{i}$.

\subsubsection*{Inactive constraints}\label{inactive}

In an inactive state of contact, \emph{i.e.}, when $\hat{\lambda}_n > 0$, Equation~\eqref{eq:stationary_c} is approximated by
\begin{equation}
	\int_{\Gamma_c} -\frac{1}{\epsilon_n}\lambda_n \delta \lambda_n \,\dd{\Gamma} \approx -\sum_{i=1}^{n_i}\frac{1}{\epsilon_{n, i}}\lambda_{n,i}\delta\lambda_{n, i},
\end{equation}
where $n_i$ is the number of inactive constraints, and the incremental equation is therefore expressed as
\begin{equation}
	\frac{\partial }{\partial \hat{\bs{U}}_i} \left[ - \sum_{i=1}^{n_i}\frac{1}{\epsilon_{n, i}}\lambda_{n,i}\delta\lambda_{n, i}  \right] \Delta \hat{\bs U}_i  = -\sum_{i=1}^{n_i}\frac{1}{\epsilon_{n, i}}\Delta\lambda_{n,i}\delta\lambda_{n, i}.
\end{equation}
For the inactive constraints, the contributions
\begin{equation}
	\bs k_i^c = \begin{bmatrix}
		 & \bs 0 & \bs 0                                     \\
		 & \bs 0 & \displaystyle{-\frac{1}{\epsilon_{n, i}}}
	\end{bmatrix} \quad \text{and} \quad
	\bs f_i^c=\left[
		\begin{array}{c}
			\bs 0 \\ \displaystyle{\frac{\lambda_{n, i}}{\epsilon_{n, i}}}
		\end{array}\right]
\end{equation}
are considered in~\eqref{eq:systemAll}, which practically implies the update of the Lagrage multiplier as described in Algorithm~\ref{alg:enrichedcontact}. Similar to the active case previously described, the solution increments are solved using the generalized Newton method. It is worth noticing that, compared to the standard way of applying constraints, only the enriched part described in \S~\ref{ssec:structural} needs to be added.

\section{Implementation }
In this section we discuss the implementation of the method in a displacement-based FEM framework.
Since the calculation of the element local arrays considering enrichment functions has been detailed elsewhere~\cite{defem,igfem}, here we only focus on the implementation of the constraints to enforce non-conforming mesh coupling and contact. Because the use of MPCs for coupling non-conforming meshes is standard~\cite[pp.~325--340]{fem2014}, we only provide the detailed pseudo-code for handling contact problems (where the coupling using LMs is also included).

\subsection{Coupling of non-conforming meshes}

For each non-conforming node, an enriched node with the same coordinates is created on the other side of the non-conforming interface as shown in Figure~\ref{fig:enrichedfunc}. The element that contains this enriched node is then split into integration elements~\cite{defem-3d}. An ordered tree data structure is recommended to store the associations among integration elements and their mesh parent elements. Integration elements are used to perform the numerical quadrature of elements' local stiffness and force arrays given by~\eqref{eq:igfem_stiff}-\eqref{eq:igfem_a}. The assembly of these contributions into the corresponding global counterparts follows standard procedures. MPCs or LMs are then used to enforce continuity constraints among non-conforming mesh nodes and their enriched slaves; while the application of MPCs is done according to standard procedures, the use of LMs can be regarded as a simplified version of the contact problem without a convergence test (more on this below).

\subsection{Contact}

We use a generalized Newton method to solve the nonlinear contact load increment following the procedure outlined in Algorithm~\ref{alg:enrichedcontact} which is now described. At each load increment we create enriched nodes using the closest projection method~\cite{Wriggers2006, Aragon:2013e} to determined their location (refer to Figure~\ref{fig:alm}). Integration elements are created afterwards, similarly to the procedure just discussed for the coupling of non-conforming meshes. Here, as the relative displacement between contacting bodies can be neglected, we only determine the locations of enriched nodes at the beginning of each contact step, in analogy with the active set strategy in NTS contact~\cite{Wriggers2006}.
For simplicity, we first detect enriched nodes without checking their gap in the normal direction; we then compute the gap and, if it is larger than zero, these nodes make no contribution to the stiffness or internal force calculation based on the inactive constraint formulation described in \S~\ref{inactive}. This approach helps the convergence within a step in the case where nodes are not in contact at the start of a step, but come in contact during the iteration.

Once the number of enriched nodes is determined, the number of required Lagrange multipliers is also determined (the cardinality of set $\mc{N}$ is denoted $\left| \mc{N} \right|$ in Algorithm~\ref{alg:enrichedcontact}). The total number of DOFs, which includes the DOFs for displacement and Lagrange multipliers in the generalized Newton method, is used to initialize the global arrays. Notice that in the displacement and Lagrange multiplier vectors, only the elements that correspond to active constraints are kept (and zeros are added for newly added enriched DOFs and multipliers). The unconstrained global arrays are then assembled considering also the enriched contributions~\cite{defem}.

The following loop over the index set of enriched nodes adds the contribution of contact constraints to the global arrays.
Augmented Lagrange multipliers are calculated based on the gap and penalty parameter. Note that contributions are added only if there is contact, \textit{i.e.}, $\hat{\lambda}_{n,i} \leq 0$.

After the solution of the system of equations and the update of the primal (displacement) and dual (Lagrange multiplier) fields, a convergence check is made. Whenever the norm of both vectors is lower than a user-specified tolerance, the analysis continues to the next contact step.
As mentioned above, LMs can be used for the mesh coupling problem, which can be regarded as a contact analysis where the enriched positions (mismatching nodes) are known at the beginning of the analysis; the solution to this problem can then be obtained directly in one iteration.

\begin{algorithm}[!ht]
	\begin{footnotesize}
		\caption{Generalized Newton pseudo-code for solving a contact step}\label{alg:enrichedcontact}
		\begin{algorithmic}[0]
			\Require{Solution {$\bs{U}$} and Lagrange multiplier $\bs\lambda$ vectors from previous step, sets of standard finite nodes $\mathcal{N}$ and elements $\mathcal{E}$, ordered tree $\mathcal{H}$, contact surfaces $\mathcal{C}$, boundary conditions $\mathcal{B}$, material properties $\mathcal{P}$, and penalty parameter $\epsilon$}
			\Statex
			\Function{solveContactStep}{$ \bs{U}, \bs{\lambda}, \mathcal{N}, \mathcal{E},\mathcal{H}, \mathcal{C}, \mathcal{B}, \mathcal{P}, \epsilon$}
			\Let{$\mc{N}_e = \left\{ \bigcup_i \enriched{x}{i} \right\} $}{findProjections$\left(\bs{U}_h, \mc{N}, \mc{C}\right)$} \Comment{create enriched nodes}
			\Let{$n_\lambda$}{$\left| \mc{N}_e \right|$} \Comment{get number of Lagrange multipliers}
			\Let{$\mathcal{H}$}{updateHierarchy$\left(\mathcal{N}_e, \mathcal{H}
					\right)$} \Comment{create/update element hierarchy}

			\Let{$n_d$}{$d \cross \left( \left| \mc{N} \right|  +  \left| \mc{N}_e \right|  \right) + n_\lambda$} \Comment{get total  DOFs, $d \equiv$ DOFs per node}

			\Let{$ \left\{ \bs U , \bs{\lambda} \right\} $}{copy($ \bs U_{\text{old}} , \bs{\lambda}_{\text{old}})$} \Comment{copy solution vector of previous step}

			\For{$k=1,2,\ldots$}
			\Comment{start generalized Newton loop}

			\Let{$\left\{ \bs K, \bs F \right\}$}{$\left\{ \bs{0}_{n_d \times n_d}, \bs{0}_{n_d\times 1} \right\}$} \Comment{initialize global arrays}



			\Let{$ \left\{  \bs K, \bs F \right\}$}{assemble($\bs{U},\mc{N} \cup \mc{N}_e, \mc{E} \cup \mc{H}, \mathcal{B},\mathcal{P}$)} \Comment{assemble global arrays~\cite{defem}}

			\For{$ i=1, \ldots, \left| \mc{N}_e \right| $} \Comment{loop over index set of enriched nodes}
			\Let{$\left\{ \bs{x}_i, \enriched{x}{i} \right\} $}{getContactPair($i, \mc{N}_e, \mc{N}$)} \Comment{get enriched segment}
			\Let{$ \bar{ \bs a}_{i}$}{getSegment($\enriched{x}{i} , \mc{H}$)} \Comment{get enriched node}
			\Let{$\bs n_{i}$}{getNormal($\bar{\bs{a}}_{i}$)} \Comment{get normal vector to segment}
			\Let{$g_{n, i}$}{gap($\bs{x}, \enriched{x}{i}, \bs n_{i}$)} \Comment{calculate normal gap for node pair}
			\Let{$\hat{\lambda}_{n, i}$}{$\lambda_{n, i} + \epsilon_{n, i} g_{n, i} $} \Comment{update augmented Lagrange multipler}

			\Let{$\bs C_i$}{$ \bs{N}_i^\intercal \otimes \bs n_{i}$} \Comment{compute $\bs C_i$ matrix}

			\If{$\hat{\lambda}_{n, i} \leq 0$} \Comment{in case of contact}

			\Let{$\bs k^c_i$}{$\begin{bmatrix}
						 & \epsilon_{n, i}\bs C_i \bs C_i^\intercal & \bs C_i \\
						 & \bs C_i^\intercal                        & 0\end{bmatrix}$} \Comment{compute $\bs{k}^c_i$}

			\Let{$\bs K$}{$\bs K+ \bs k^c_{i}$} \Comment{assemble $\bs k^c_i$ to global stiffness matrix}
			\Let{$\bs F$}{$\bs F -\begin{bmatrix}\hat{\bs \lambda}_{n, i} \bs C_i & g_{n,i}\end{bmatrix}^\intercal$} \Comment{update global residual vector}
			\ElsIf{$\hat{\lambda}_{n, i} > 0$}
			\Let{$\Delta\lambda_{n, i}$}{$-\lambda_{n, i}$} \Comment{set the increment as $-\lambda_{n, i}$}

			\EndIf

			\EndFor

			\Let{$ \smash{\begin{bmatrix}
							\Delta \bs{U} & \Delta \bs \lambda
						\end{bmatrix}^\intercal} $}{solve$\left(\bs{K}, \bs{F}\right)$} \Comment{compute increment in solution}
			\Let{$\bs{U}$}{$\bs{U}+\Delta\bs{U}$} \Comment{update displacements}
			\Let{$\bs{\lambda}$}{$\bs{\lambda}+\Delta\bs{\lambda}$} \Comment{update Lagrange multipliers}
			\If{$\lVert \Delta\bs U \rVert< \text{tol}$ and $\lVert \Delta\bs{\lambda} \rVert <\text{tol}$}
			break
			\EndIf
			\Statex
			\Return $\bs U, \mathcal{\bs\lambda}$
			\EndFor
			\EndFunction
		\end{algorithmic}
	\end{footnotesize}
\end{algorithm}

\section{Numerical examples}
\label{sec:numerical}

The accuracy and robustness of the proposed method is now demonstrated by means of numerical examples. Homogeneous linearly elastic materials and plane strain conditions are chosen. For convenience, no units are adopted so results are valid for any consistent unit system. Constant strain triangular elements are used with one point quadrature rule for both standard and integration elements.

\subsection{Contact patch test}
\label{sec:contactpatch}

\begin{figure}
	\centering
	\input{contact_middle.tex}
	\caption{Geometry and boundary conditions of the contact patch test. The top block is placed on the middle of the substrate.}
	\label{fig:contact_patch}
\end{figure}
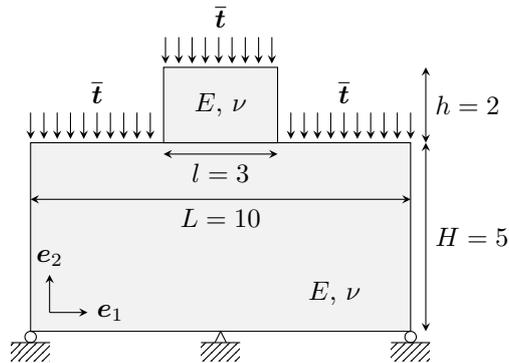

The commonly used contact patch test of Taylor and Papodopoulos~\cite{Taylor:1991} is used to study our method's ability to correctly transfer contact tractions. The problem consists of an elastic substrate with a rectangular punch as illustrated in Figure~\ref{fig:contact_patch}. Both the punch and the substrate are subject to a uniformly distributed vertical unit traction $\bar{\bs t}$ along their top edges. The substrate and the punch have the same Young's modulus $E = 10$ and Poisson's ratio $\nu = 0.3$.
The problem is setup so that the punch comes into contact with the substrate because of the applied pressure. Since the same material is used for the punch and substrate, and the pressure is uniformly distributed on the top surfaces, the substrate should experience a constant state of strain and stress.

The problem is then solved using the following discretization methods:
\begin{enumerate*}[\itshape i)]
	\item[(a)] Standard FEM using a single-pass MPC;
	\item[(b)] Standard FEM with a two-pass MPC;
	\item[(c)] Our enriched method using a two-pass MPC; and
	\item[(d)] Our enriched method using ALM.
\end{enumerate*}
For the single-pass MPC example, we defined the top surface of the substrate as master and the lower surface of the top block as slave (switching master and slave surfaces leads to an unconstrained system). Finally, the same strategy is used to integrate the applied pressure in all cases.

\begin{figure}
	\centering
	\begin{subfigure}[c]{0.48\textwidth}
		\centering
		\includegraphics[scale=0.25]{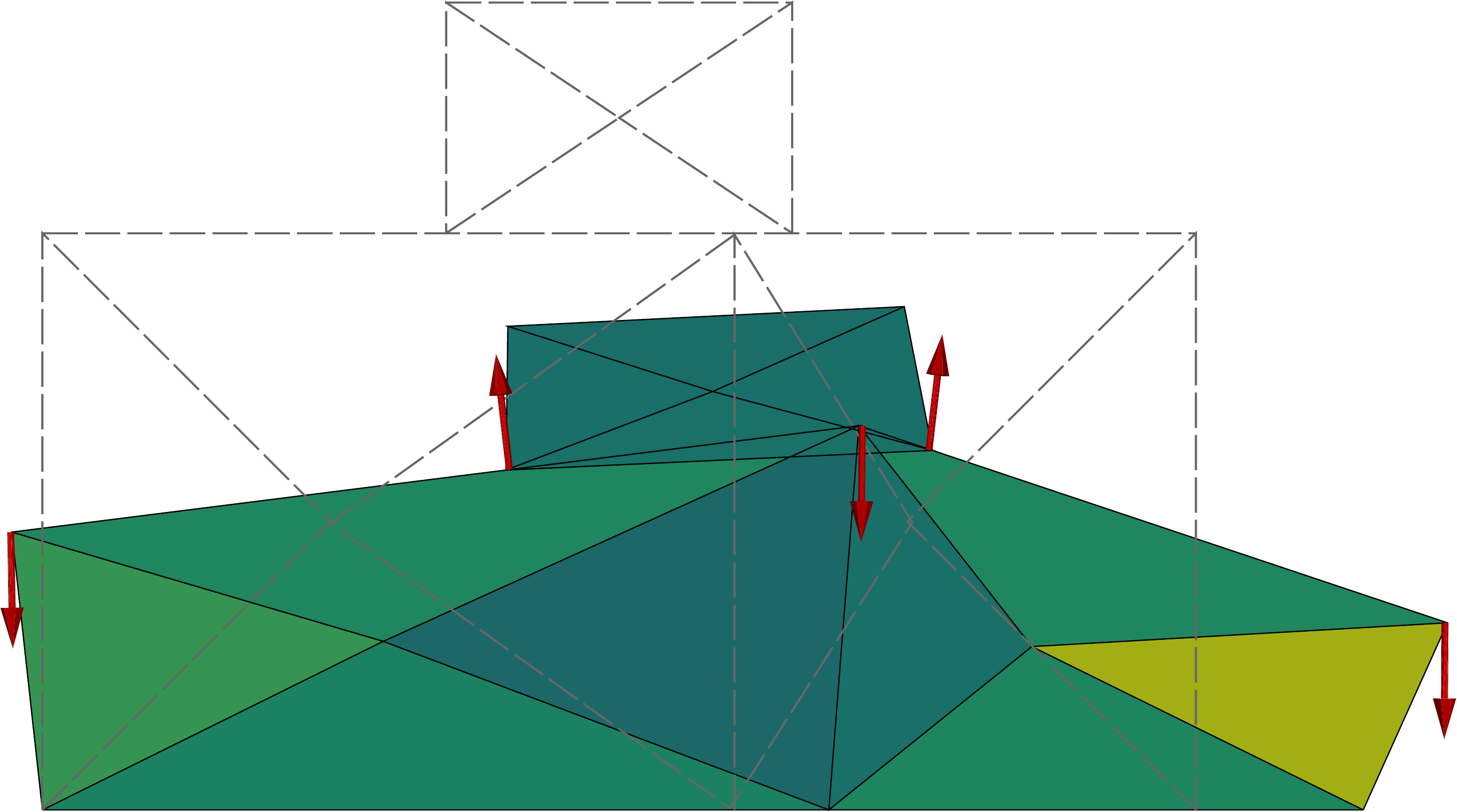}
		\caption{}
		\label{fig:deformed_contact_patch_c}
	\end{subfigure}
	\begin{subfigure}[c]{0.48\textwidth}
		\centering
		\includegraphics[scale = 0.25]{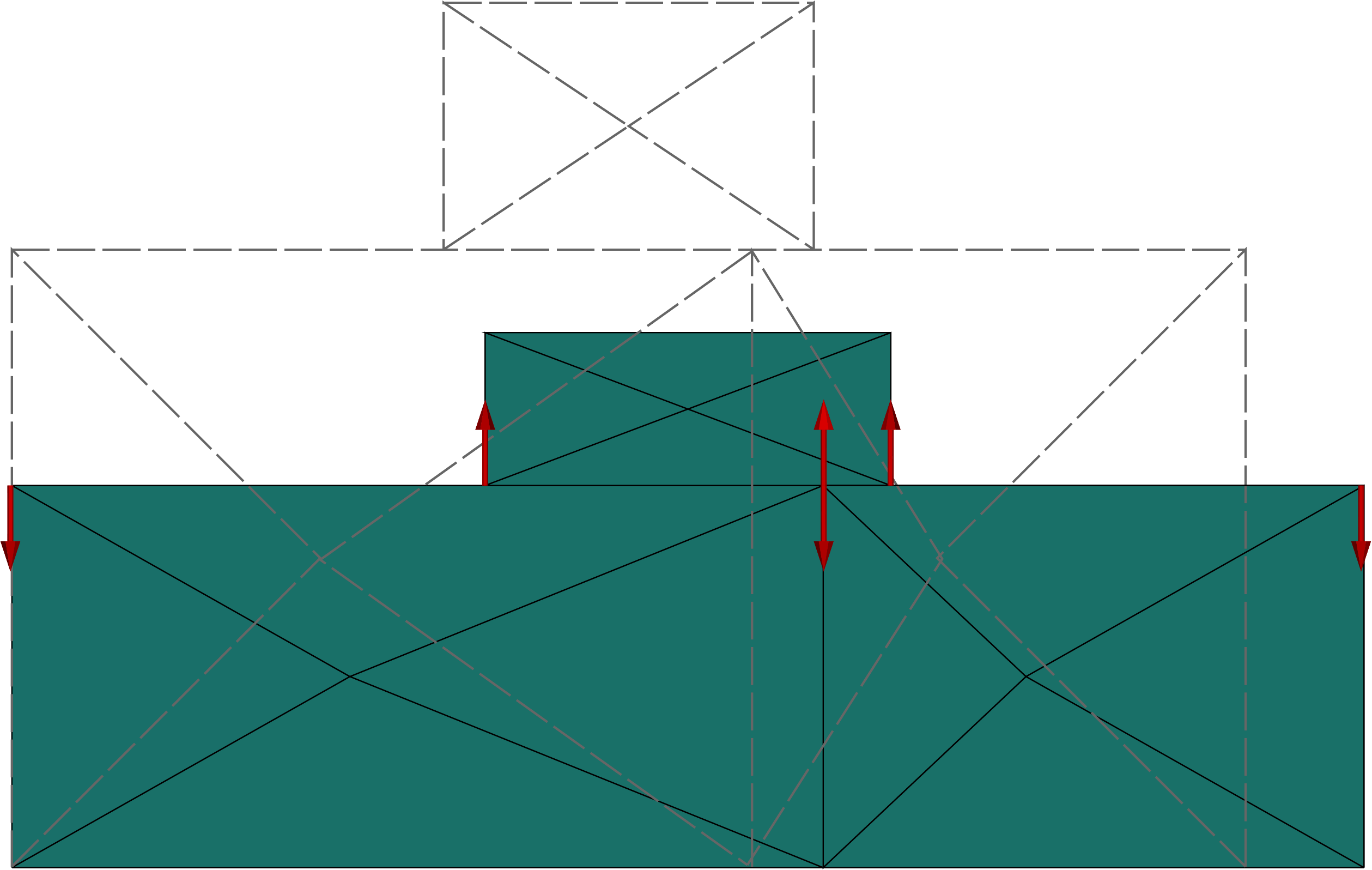}
		\begin{tikzpicture}
			\node at (0,1.2) {\includegraphics[height =2.4cm]{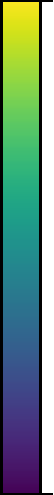}};
			\node at (0, 2.8) {$\sigma_{22}$};
			\node at (0.4, 2.4) {$-2.0$};
			\node at (0.4, 1.2) {$-1.0$};
			\node at (0.4, 0.0) {$-0.0$};
		\end{tikzpicture}
		\caption{}
		\label{fig:deformed_contact_patch_d}
	\end{subfigure}
	\\
	\begin{subfigure}[c]{0.48\textwidth}
		\centering
		\includegraphics[scale=0.25]{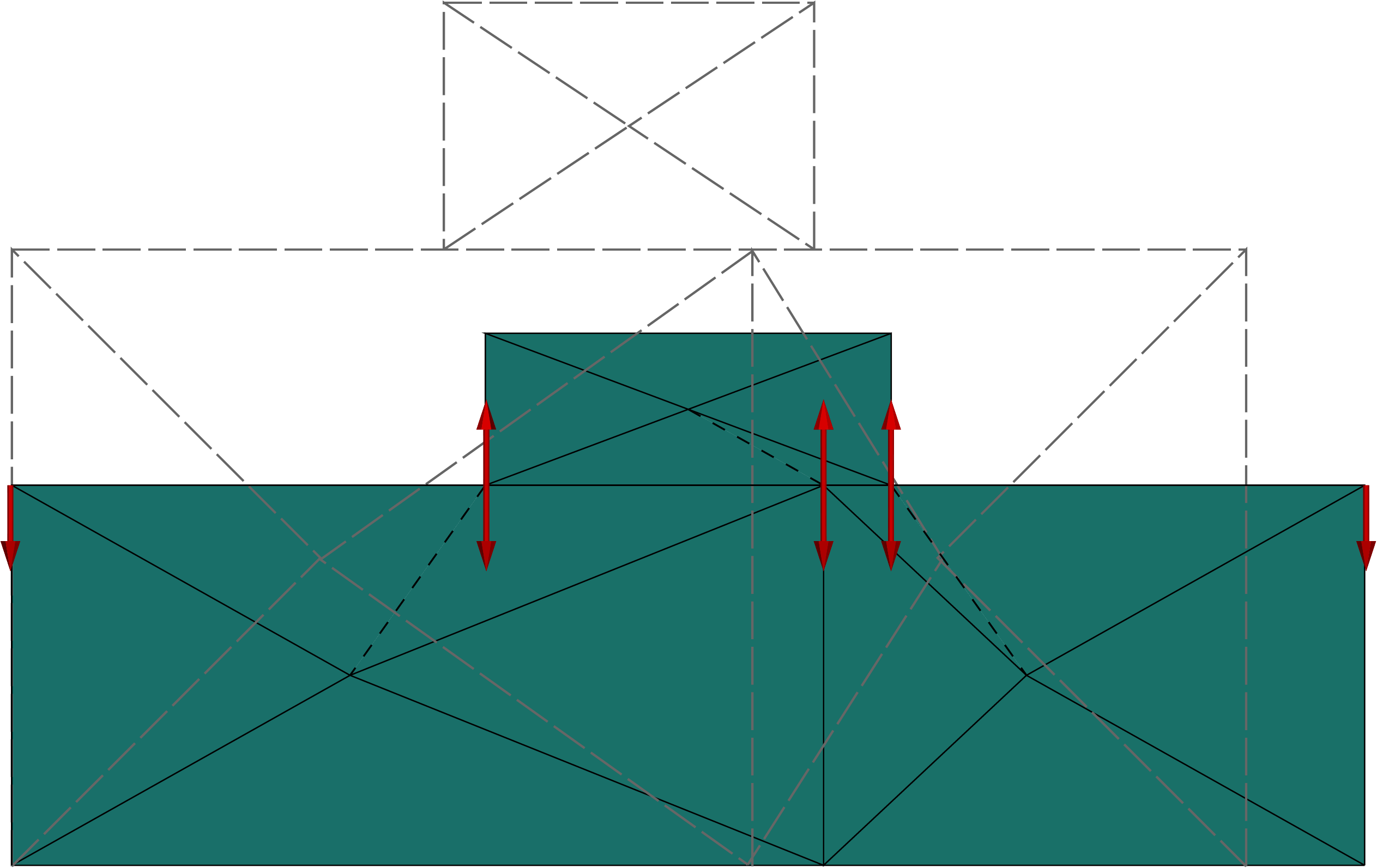}
		\caption{}
		\label{fig:deformed_contact_patch_e}
	\end{subfigure}
	\begin{subfigure}[c]{0.48\textwidth}
		\centering
		\includegraphics[scale=0.25]{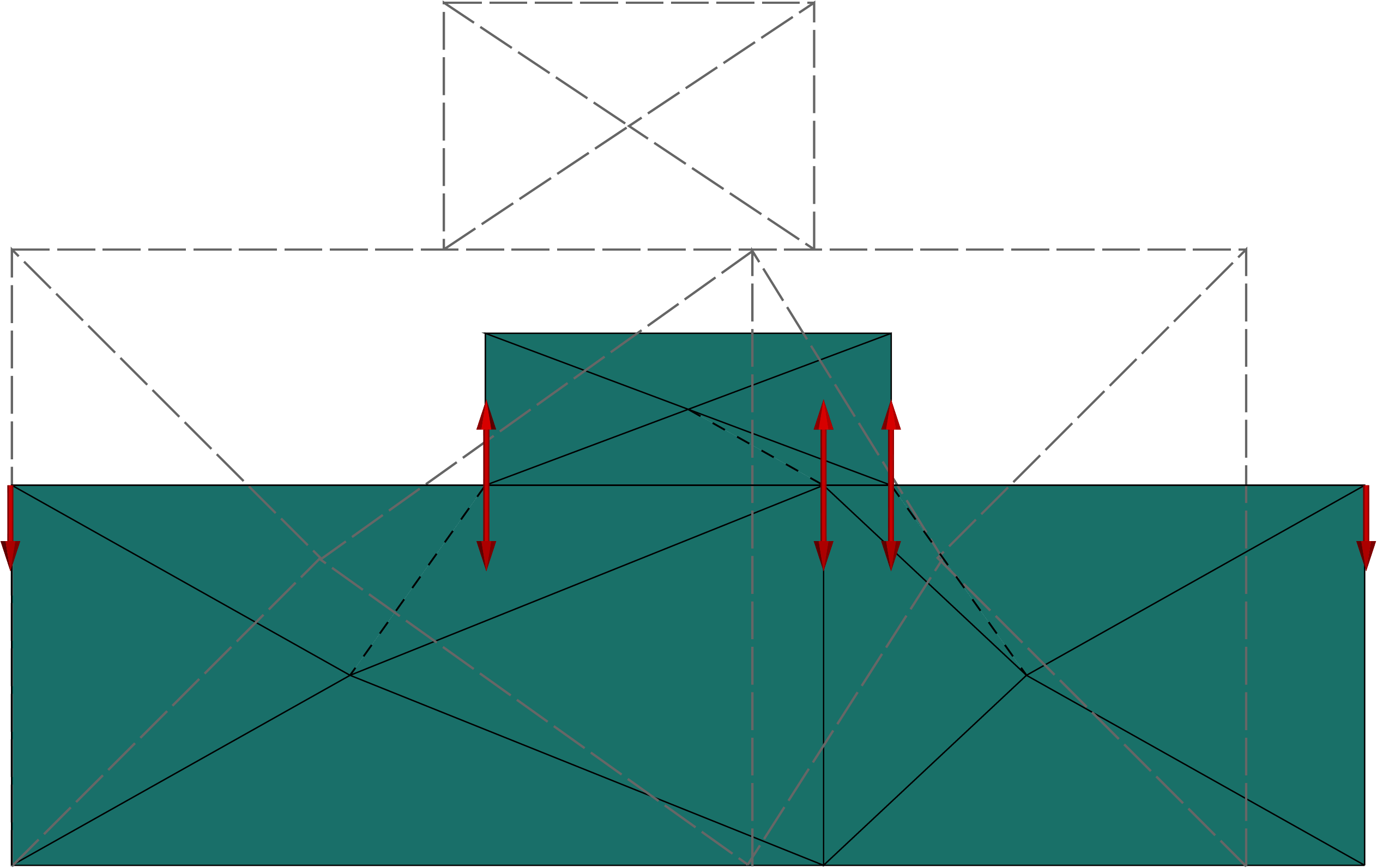}
		\begin{tikzpicture}
			\node at (0,1.2) {\includegraphics[height =2.4cm]{colorbar.pdf}};
			\node at (0, 2.8) {$\sigma_{22}$};
			\node at (0.4, 2.4) {$-2.0$};
			\node at (0.4, 1.2) {$-1.0$};
			\node at (0.4, 0.0) {$-0.0$};
		\end{tikzpicture}
		\caption{}
		\label{fig:deformed_contact_patch_f}
	\end{subfigure}
	\caption{Deformed configurations ($4\times$ magnification) for the contact patch test showing the element stress in the $\bs{e}_2$ direction and the contact tractions plotted at nodes with red arrows: Standard FEM with single-~(a) and two-pass~(b) MPC; Enriched approach with a two-pass MPC~(c) and ALM~(d).}
	\label{fig:deformed_contact_patch}
\end{figure}

Figure~\ref{fig:deformed_contact_patch} shows the stress field on the deformed configuration for all methods.
The single-pass MPC method (panel~(a)) is unable to ensure continuity and results in a non-constant stress field; notice also the interpenetration between the substrate and the punch. The two-pass method (panel~(b)) ensures continuity along the contact boundary and passes the patch test. The results obtained with our enriched formulation (two-pass MPC in panel~(c) and ALM in panel~(d)) show that the method ensures $C^0$-continuity and passes the patch test. In addition, exact integration of the applied pressure is readily possible because of the presence of integration elements with an enriched node at the location where the applied pressure is discontinuous.

\subsection{Convergence study}

\begin{figure}
	\centering
	\begin{subfigure}[c]{0.45\textwidth}
		\centering
		\input{plate_hole_schematic.pdf_tex}
		\caption{}\label{fig:platehole_geo}
	\end{subfigure}
	\begin{subfigure}[c]{0.25\textwidth}
		\centering
		\includegraphics[width=3.15cm]{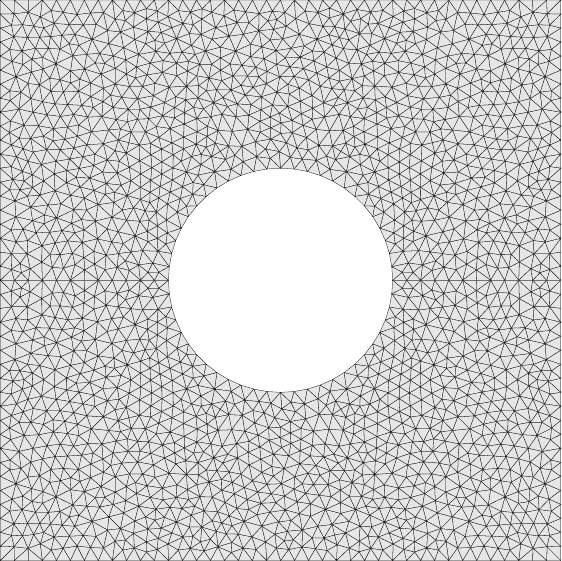}
		\caption{}\label{fig:platehole_conform}
	\end{subfigure}
	\begin{subfigure}[c]{0.25\textwidth}
		\centering
		\includegraphics[width=3.15cm]{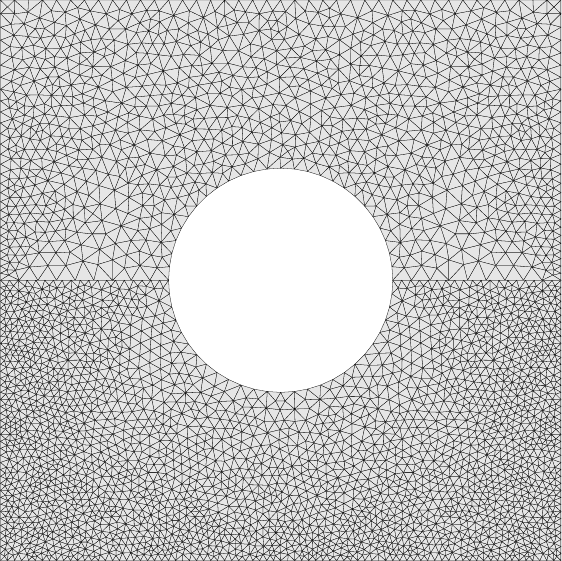}
		\caption{}\label{fig:platehole_nonconform}
	\end{subfigure}
	\caption{(a)~Schematic of an infinite plate with a circular hole subject to a uniform traction at $x_1 = \pm \infty$. The square region indicates the computational domain used in the convergence study. The problem is analyzed using the discretizations in panels (b-c) which represent a typical conforming mesh used for standard FEM~(b), and a typical mesh with a non-conforming interface~(c). Convergence results are shown for the cases of horizontal and vertical interfaces (in the latter case, the mesh is rotated).}
	\label{fig:platehole_meshes}
\end{figure}

The convergence of the proposed method is investigated by means of the classical problem of a circular hole in an infinite plate, for which the exact solution can be found in References~\cite{Kirsch1898, book:Rock}. As shown in the schematic of Figure~\ref{fig:platehole_geo}, a square computational domain of size $L=20$ with a centered whole of radius $r=4$ is chosen. The material properties of the plate are $E=10$ and $\nu = 0.3$. On the boundary of the square domain we prescribe the exact displacement field corresponding to the uniform far-field traction $\bar{\bs t} = \pm \sigma_\infty \bs e_1$, with $\sigma_\infty = 1$.

Figure~\ref{fig:platehole_meshes} shows the two discretizations for this problem: a standard conforming FE mesh in panel~(a) and a mesh composed of two parts that are non-conforming along the coupling interface in panel~(b), where the ratio between the  the element sizes in the top and bottom domains is equal to 2 and is kept constant with mesh refinement. Four different analysis approaches are compared:
\begin{enumerate*}[\itshape i)]
	\item Standard FEM using conforming meshes; \item Two-pass MPC on non-conforming discretizations; \item Our enriched method using a two-pass MPC; and \item Our enriched method using LM.
\end{enumerate*} Since the single-pass MPC cannot ensure continuity along the coupling interface, as demonstrated in the previous example, it has been discarded in this analysis.

Convergence is studied via the standard, $\mathcal{L}^2$, and energy, $\mathcal{E}$, norms of the error defined as
\begin{align}
	\label{eq:l2norm}
	\| \epsilon \|_{\mathcal{L}^{2}} & \equiv \frac{ \left\Vert \bs u - \bs u^h \right\Vert_{\mathcal{L}^2 \left( \Omega \right)} }{\left\Vert \bs u \right\Vert_{\mathcal{L}^2 \left( \Omega \right)}} = \frac{\sqrt{ \sum_{e \in \Omega^h} \int_{e} \|\bs u - \bs u^h \|^2 \, \dd{\Omega} }}{\sqrt{ \sum_{e \in \Omega^h} \int_e \|\bs u\|^2 \, \dd{\Omega} }},                                                                                                                             \\
	\label{eq:energynorm}
	\| \epsilon \|_{\mathcal{E}}     & \equiv \frac{ \left\Vert \bs u - \bs u^h \right\Vert_{\mathcal{E} \left( \Omega \right)} }{\left\Vert \bs u \right\Vert_{\mathcal{E} \left( \Omega \right)}}=\frac{\sqrt{\sum_{e \in \Omega^h} \int_{e} \left( \bs \varepsilon - \bs \varepsilon^h \right)^\intercal  \bs D \left( \bs \varepsilon -\bs \varepsilon^h \right) \,\dd{\Omega} }}{\sqrt{ \sum_{e \in \Omega^h} \int_{e}\bs \varepsilon ^\intercal  \bs D \bs \varepsilon \,\dd{\Omega}}},
\end{align}
where quantities with and without the superscript $h$ refer to approximate and exact solutions, respectively.

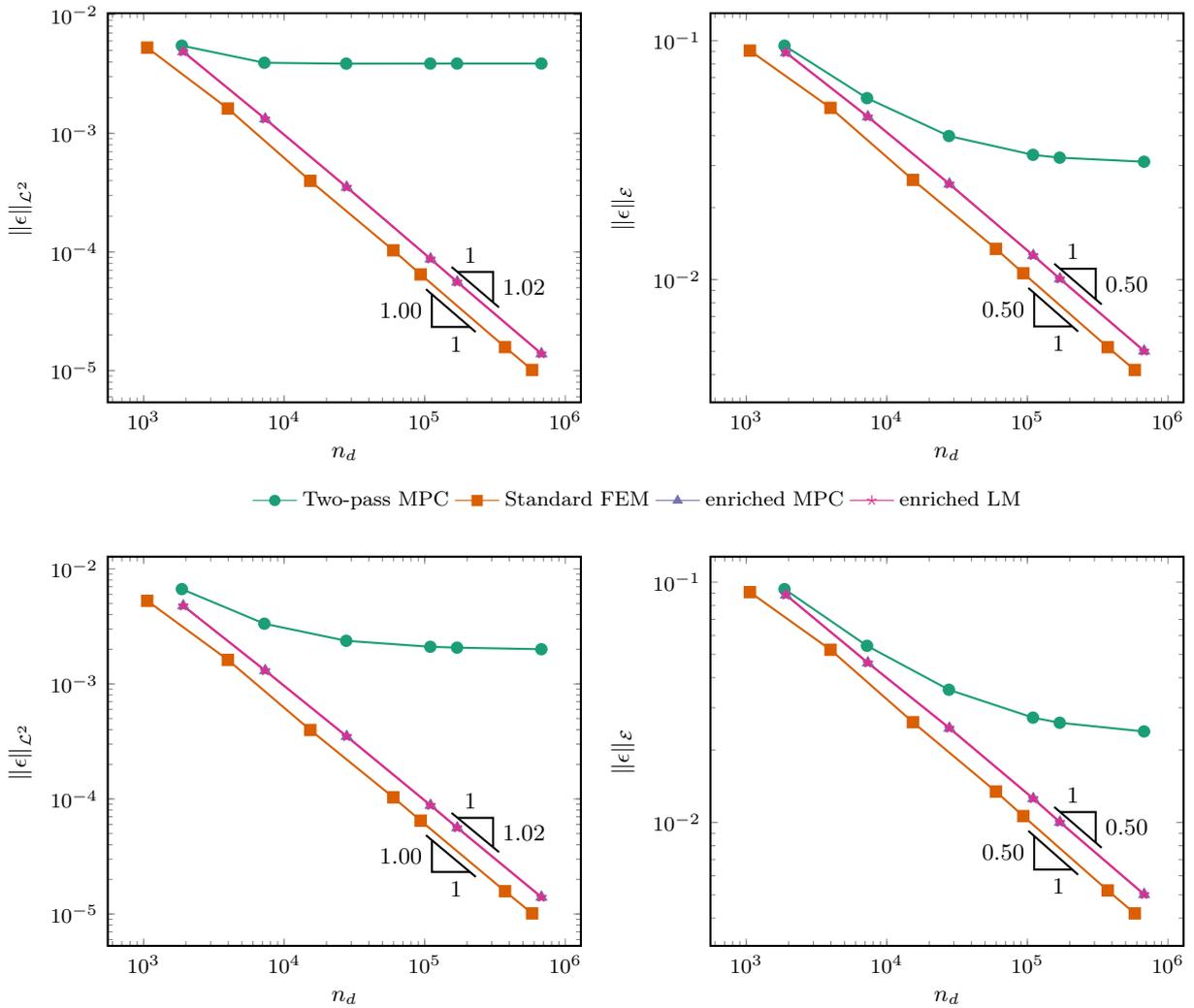
\begin{figure}
	\centering
	\input{plate_hole_convergence_single_legend.tex}
	\caption{Convergence results for the horizontal (top row) and vertical (bottom row) non-conforming interface. The figures show the error in the $\mathcal{L}^2$ norm (left column) and the energy norm (right column) as a function of the total number of degrees of freedom $n_d$. The curves for enriched methods with MPC and LM overlap.}
	\label{fig:convergence}
\end{figure}

The plots in the top row of Figure~\ref{fig:convergence} show the convergence results with the configuration shown in Figure~\ref{fig:platehole_meshes} (horizontal interface).
While the two-pass MPC method does not converge due to locking originated from an overconstrained interface~\cite{Haikal:2010,babuskaFEM}, the curves of the two enriched approaches overlap and achieve the same rate of convergence as that of standard FEM with conforming discretizations: about $1$ for the $\mathcal{L}^2$ norm and $0.5$ for the energy norm.
Similar results, reported in the second row of Figure~\ref{fig:convergence}, are obtained when the non-conforming mesh is rotated $90^\circ$, resulting in a vertical interface.

\begin{figure}
	\centering
	\begin{subfigure}[c]{1.\textwidth}
		\centering
		\hspace{1.7cm}
		\begin{tikzpicture}
			\node at (0, 0)	{\includegraphics[width=5.5cm]{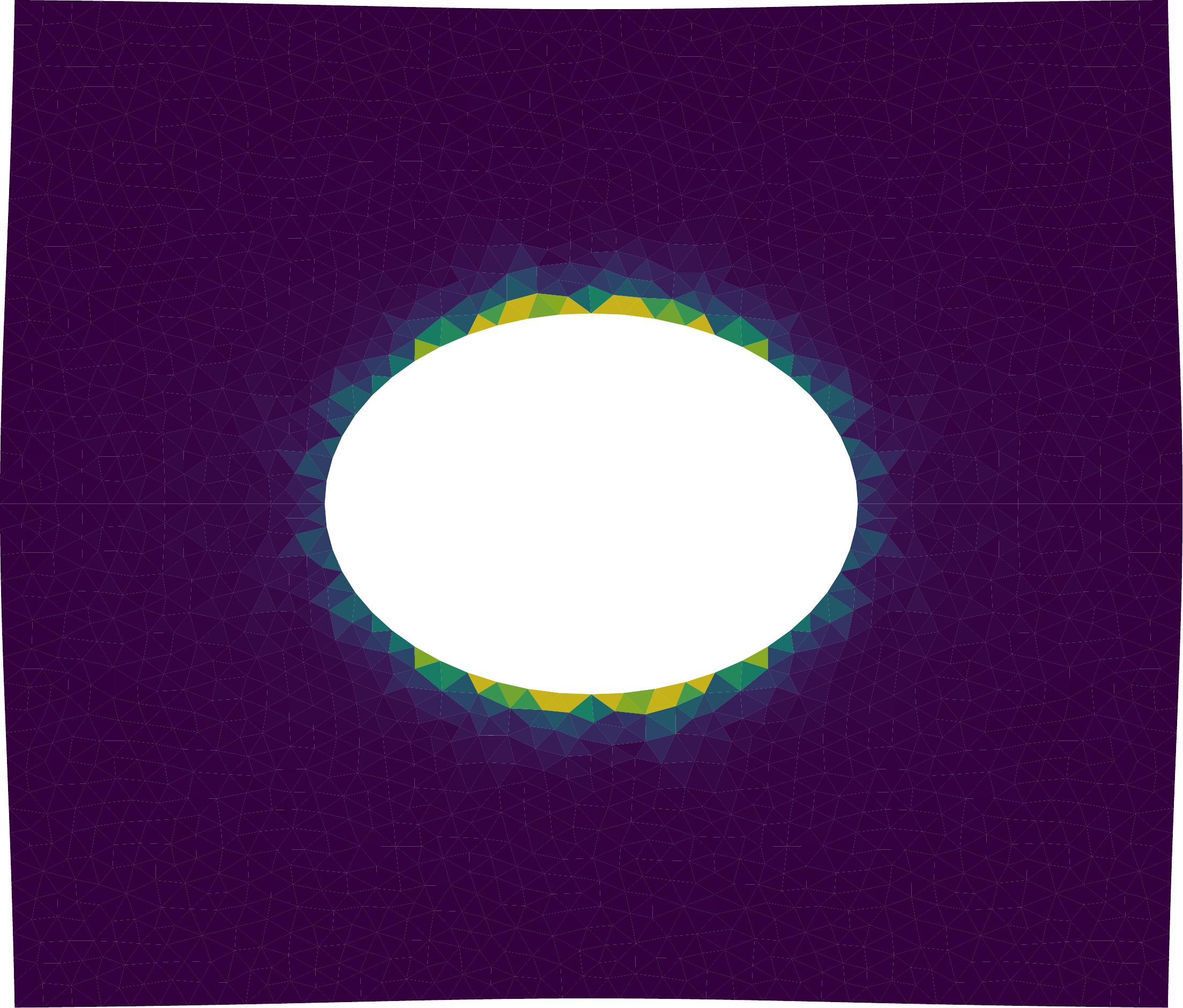}};
			\node at (3.5,0) {\includegraphics[height =2.4cm]{colorbar.pdf}};
			\node at (3.5, 1.6) {$\|e\|_{\mathcal{E}}$};
			\node at (4.2, 1.2) {$10^{-2}$};
			\node at (4.0, -1.2) {$0$};
		\end{tikzpicture}
		\caption{}
	\end{subfigure}
	\begin{subfigure}[c]{0.32\textwidth}
		\includegraphics[width=1.\linewidth]{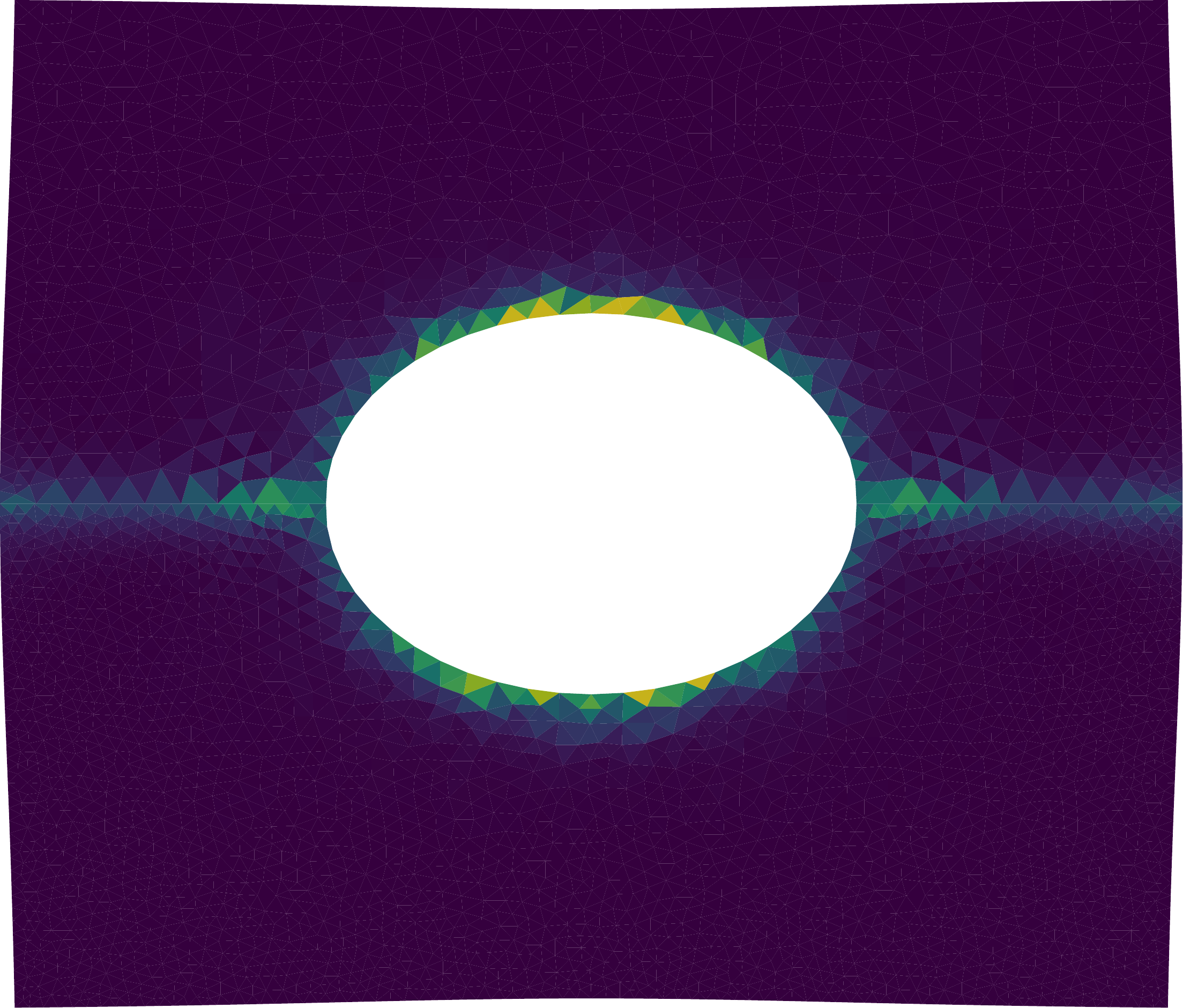}
		\caption{}
	\end{subfigure}
	\begin{subfigure}[c]{0.32\textwidth}
		\centering
		\includegraphics[width=1.\linewidth]{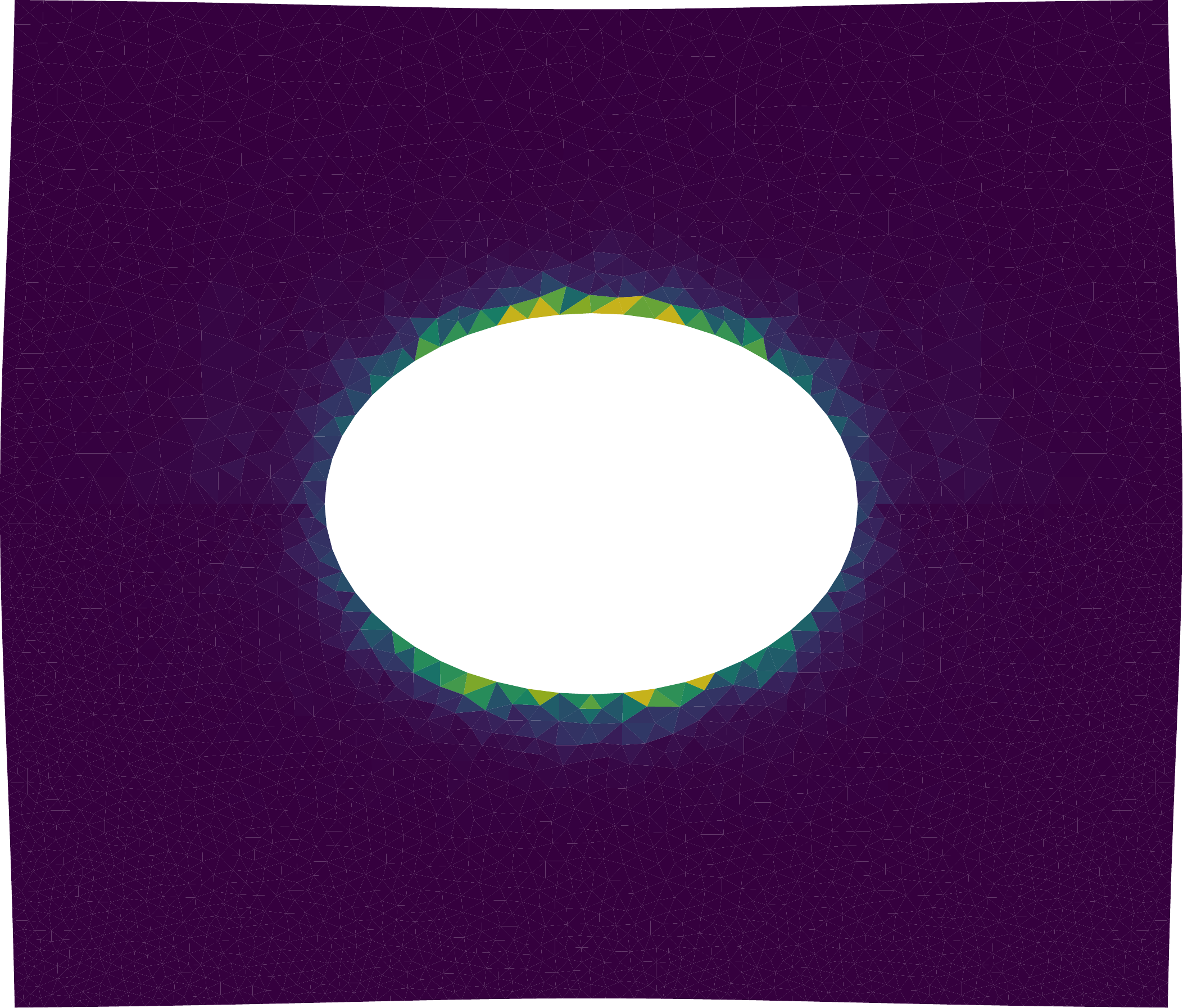}
		\caption{}
	\end{subfigure}
	\begin{subfigure}[c]{0.32\textwidth}
		\includegraphics[width=1.\linewidth]{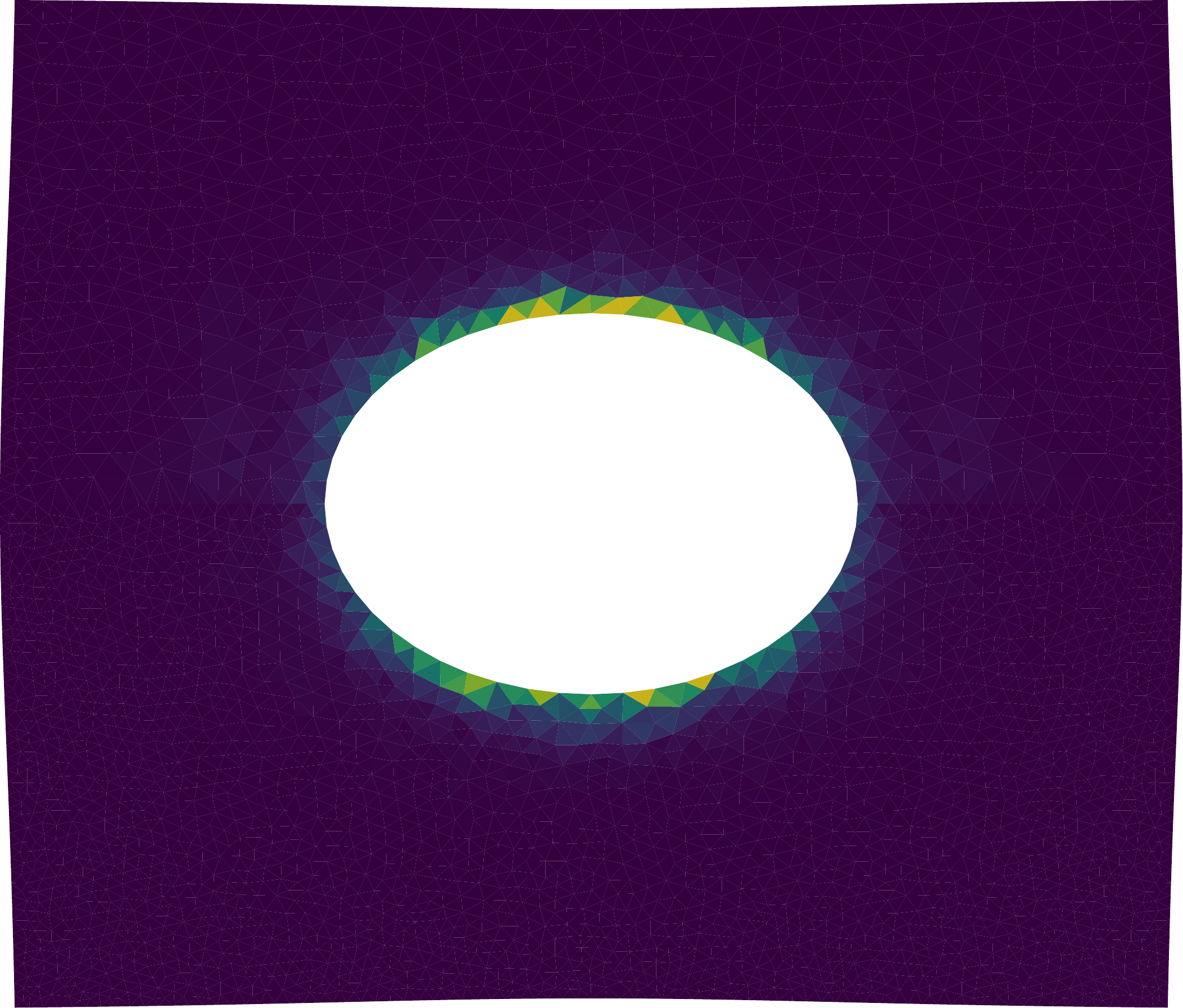}
		\caption{}
	\end{subfigure}

	\begin{subfigure}[c]{0.32\textwidth}
		\centering
		\includegraphics[width=1.\linewidth]{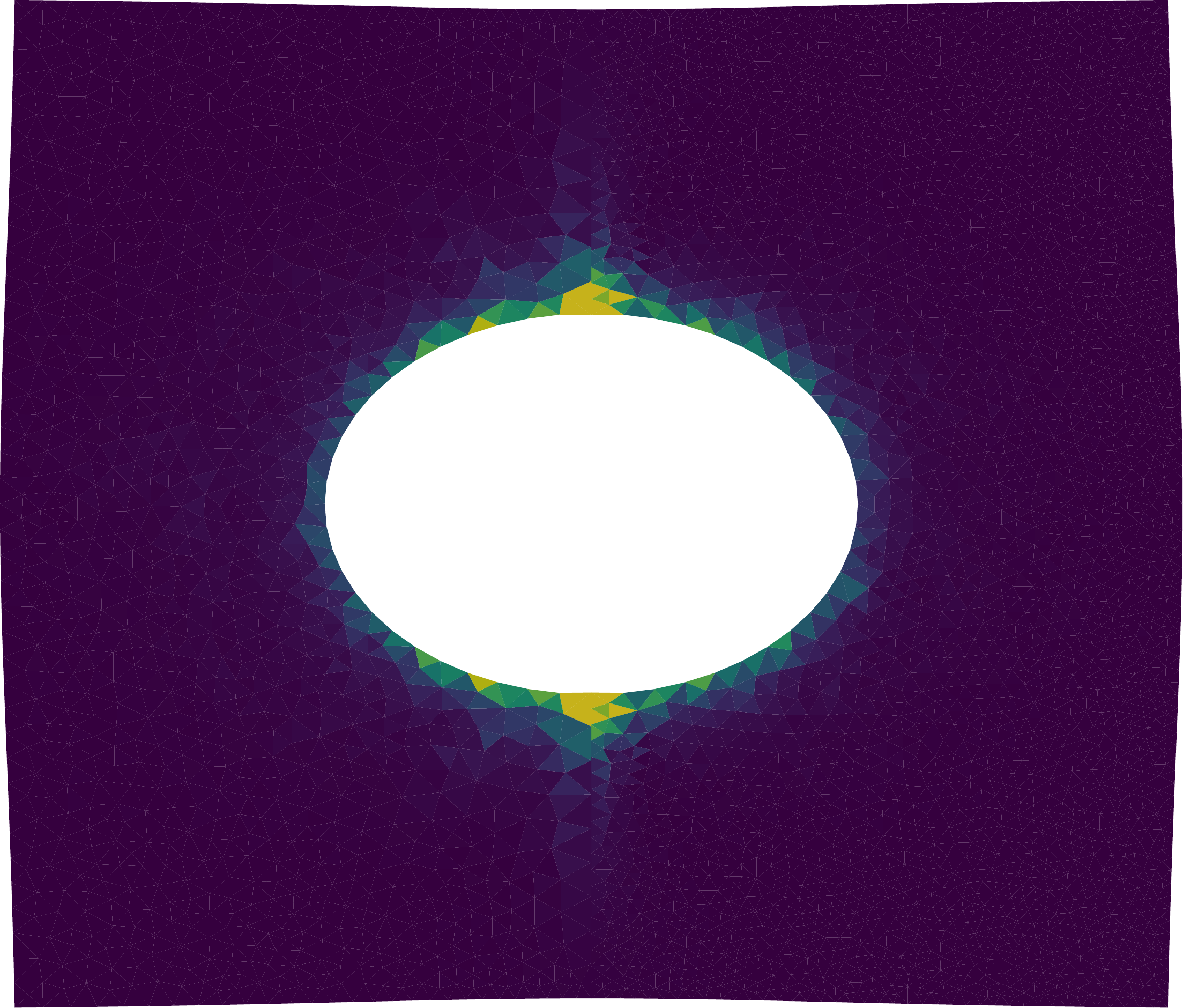}
		\caption{}
	\end{subfigure}
	\begin{subfigure}[c]{0.32\textwidth}
		\includegraphics[width=1.\linewidth]{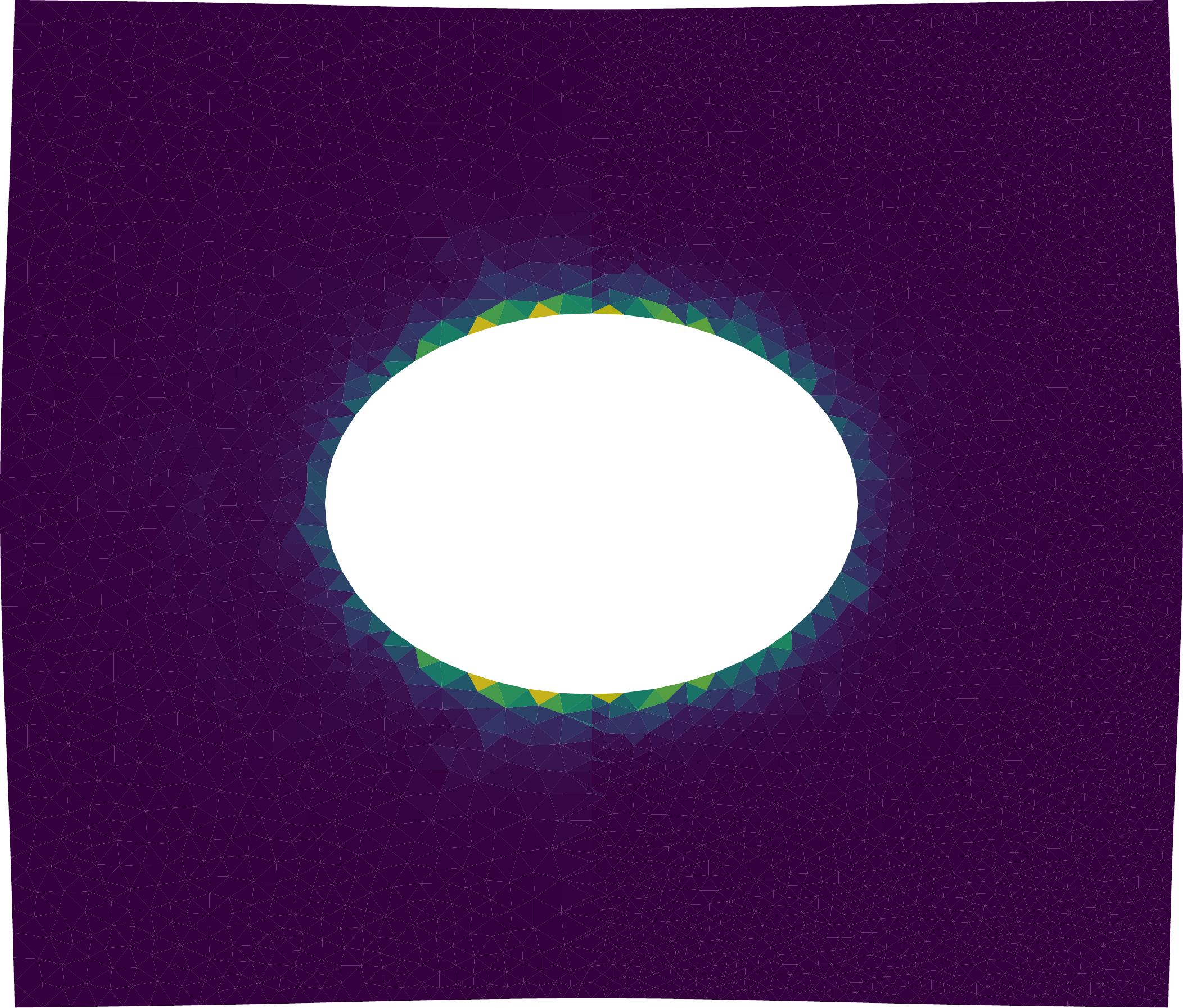}
		\caption{}
	\end{subfigure}
	\begin{subfigure}[c]{0.32\textwidth}
		\includegraphics[width=1.\linewidth]{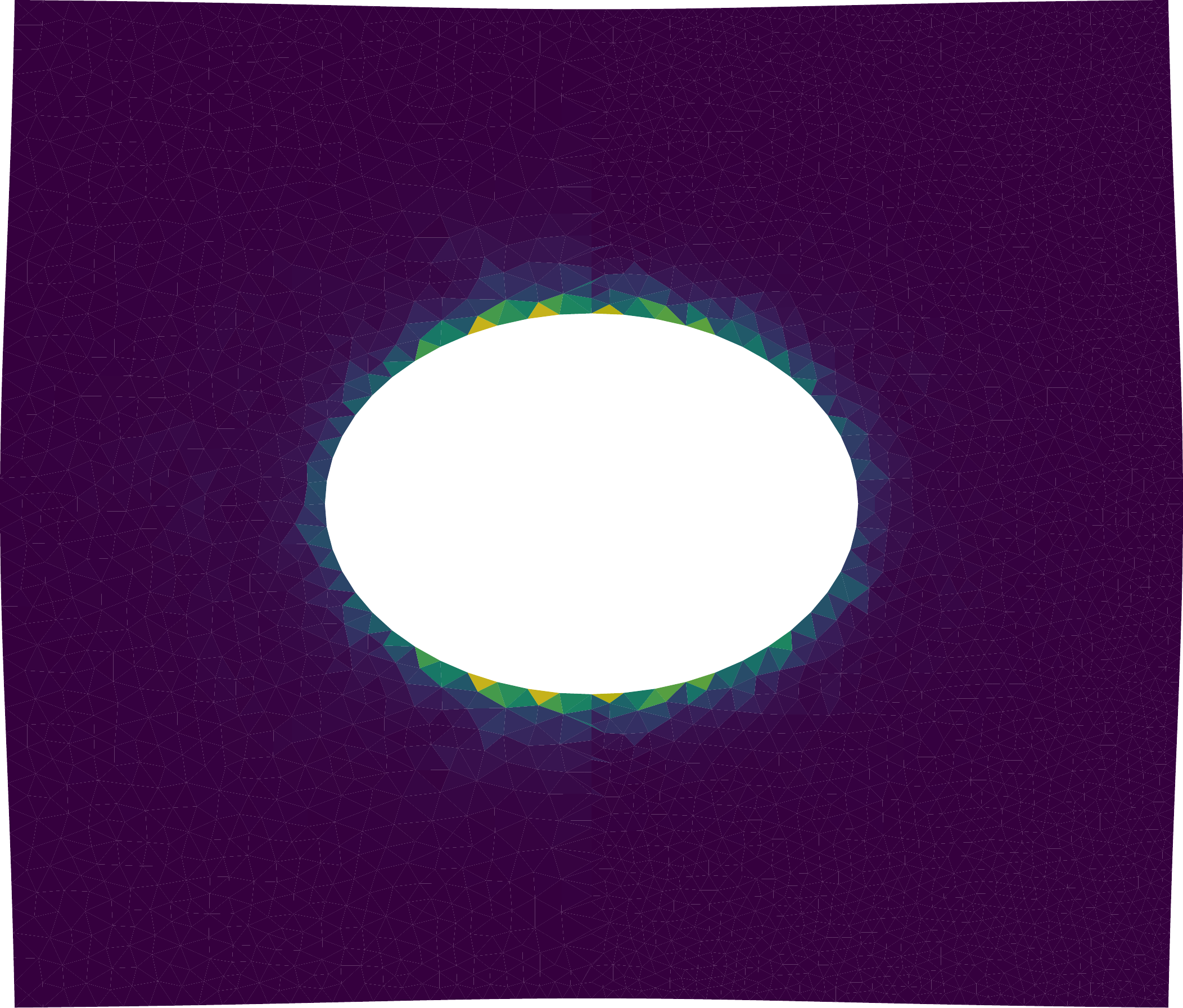}
		\caption{}
	\end{subfigure}
	\caption{Contour plot of the error in energy norm, averaged per element: Standard FEM~(a); Horizontal non-conforming interface with two-pass MPC~(b), enriched FEM with MPC~(c), and enriched FEM with LM~(d); Vertical non-conforming interface with two-pass MPC~(e), enriched FEM with MPC~(f), and enriched FEM with LM~(g).}
	\label{fig:convergence_error}
\end{figure}

The energy norm of the error corresponding to the four methods using the mesh with horizontal and vertical non-conforming interfaces are shown in Figure~\ref{fig:convergence_error}, plotted per element (average value). The results corresponding to the enriched methods and standard FEM are in good agreement, whereas those obtained by the two-pass MPC (panel~(b)) show some clear differences along the coupling interface.

This example shows that the performance of the proposed method for mesh coupling is basically identical to that of the standard FEM on conforming meshes. Because the same convergence rates are obtained, it can be concluded that the LBB condition is fulfilled. In contrast to the two-pass MPC method, our method avoids interface locking because enriching the primal field gives more kinematic freedom to the interface. The enrichment therefore enables an accurate representation of the mechanical behavior at the coupling interface.

\subsection{Stability} \label{sec:stability}

Following our previous work on enriched FEM~\cite{vandenBoom2018,defem-3d,Aragon:2020a}, we investigate the stability of the proposed method. Using the same problem geometry of the contact patch test used in \S~\ref{sec:contactpatch},  we examine the influence of punch location and mesh size on the condition number of the system matrix. We compute the condition number of the global system matrix $\bs{K}$ as
\begin{equation}
	\cond \left( \bs{K} \right) \equiv \kappa \left( \bs{K} \right) = \frac{\lambda_\text{max}}{\lambda_\text{min}},\quad \lambda_\text{min} \neq 0,
	\label{eq:condition}
\end{equation}
where  $\lambda_\text{max}$ and $\lambda_\text{min}$ denote, respectively, the highest and lowest (non-zero) eigenvalues of the system matrix.
No Dirichlet boundary conditions are enforced on the system and therefore we discard the lowest six eigenvalues, which correspond to the rigid body modes of both blocks.

We investigate the condition number of the matrices with MPC and ALM. For each approach three variations of the enriched method are compared: \begin{enumerate*}
	\item[$\mathrm{ns}$)]~The enriched method without scaling enrichment functions, \textit{i.e.}, $s_i = 1$ in Equation~\eqref{eq:IGFEM_space}; \item[$\mathrm{os}$)]~The enriched method with the optimal scaling proposed in Reference~\cite{Aragon:2020a}, \textit{i.e.}, $s_i = \sqrt{2 \zeta \left( 1- \zeta \right) }$, where $0 \leq \zeta \leq 1$ denotes the (relative) location of the enriched node in the finite element side that contains it; and \item[$\mathrm{pc}$)]~The enriched method without scaling, but using a diagonal preconditioner such that $\bs K_{\mathrm{pc}} = \bs \Delta \bs K \bs \Delta$, where $\Delta_{ij} =\delta_{ij}/\sqrt{K_{ij}}$ is a diagonal matrix with $\delta_{ij}$ denoting the Kronecker delta.
\end{enumerate*}

\subsubsection*{Effect of punch location}

\begin{figure}
	\begin{subfigure}[b]{0.48\textwidth}
		\centering
		\input{contact_condition2.tex}
		\caption{}
		\label{fig:punch_location_a}
	\end{subfigure}
	\begin{subfigure}[b]{0.48\textwidth}
		\centering
		\includegraphics[height=3.5cm]{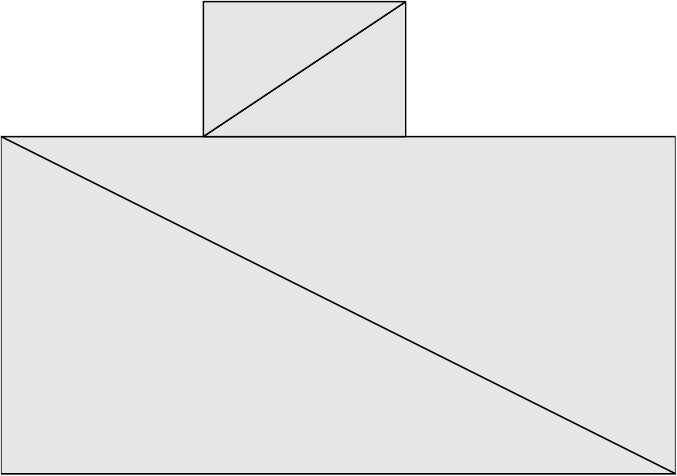}
		\caption{}
		\label{fig:punch_location_b}
	\end{subfigure}
	\caption{(a)~Moving punch problem schematic; (b)~Finite element mesh, where substrate and punch are discretized with two triangular elements. The punch location $x_1$ measures the horizontal distance of the punch from the origin of the coordinate system.}
	\label{fig:punch_location}
\end{figure}

In the first test, the influence of the punch location on the condition number as it moves on the substrate is evaluated (see Figure~\ref{fig:punch_location_a}). Both substrate and punch are discretized with two triangular elements (see Figure~\ref{fig:punch_location_b}). Their material properties are, respectively, $E_1=10, E_2=\num{10000}$ and $\nu_1=\nu_2=0.3$.

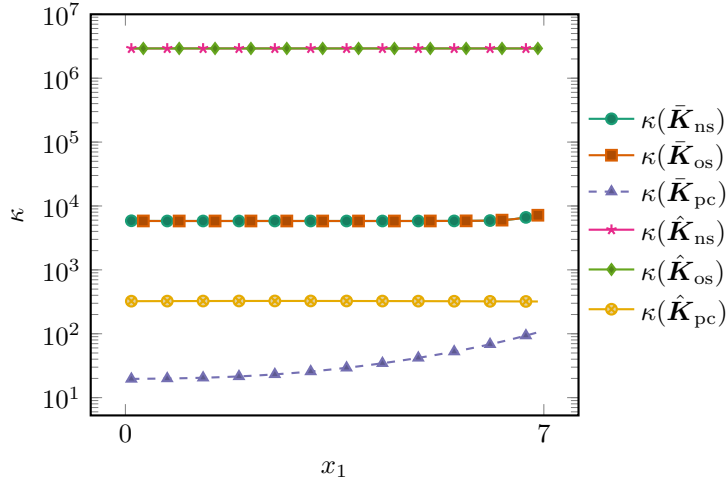
\begin{figure}
	\centering
	\input{contact_condition_moving.tex}
	\caption{Condition number as a function of punch location $x_1$.}
	\label{fig:condition_moving}
\end{figure}

Figure~\ref{fig:condition_moving} shows the condition number of the stiffness matrix as a function of the punch location. For MPC, the condition number for the unscaled enriched method (labelled $\kappa ( \bar{\bs K}_{\mathrm{ns}} ) $) rises slightly when the punch approaches the sides of the substrate. However, when using the optimal scaling proposed in Reference~\cite{Aragon:2020a} (labelled $\kappa ( \bar{\bs K}_{\mathrm{os}}) $), the condition number is the same as that of unscaled method, showing that the ineffectiveness of the scaling factor demonstrated in the one-dimensional example in Appendix~\ref{appendix} holds also in this case. Applying the diagonal preconditioner (labelled $\kappa ( \bar{\bs K}_{\mathrm{pc}})$) improves the condition number. Overall, the condition number remains bounded as enriched nodes are placed arbitrarily close to standard finite element nodes when dealing with MPCs.

For ALM, the condition number of the original stiffness matrix without enrichment scaling (labelled $\kappa ( \hat{\bs K}_{\mathrm{ns}} ) $) also overlaps the one that uses optimal enrichment scaling (labelled $\kappa ( \hat{\bs K}_{\mathrm{os}} )$). However, with the diagonal preconditioner (labelled $\kappa ( \hat{\bs K}_{\mathrm{pc}} )$, the condition number improves significantly, but is still higher than that of the MPC method.

\subsubsection*{Effect of mesh size}

\begin{figure}
	\begin{subfigure}[b]{0.32\textwidth}
		\centering
		\input{contact_condition1}
		\caption{}
		\label{fig:fixed_punch_a}
	\end{subfigure}
	\hfill
	\begin{subfigure}[b]{0.32\textwidth}
		\centering
		\includegraphics[height=3.5cm]{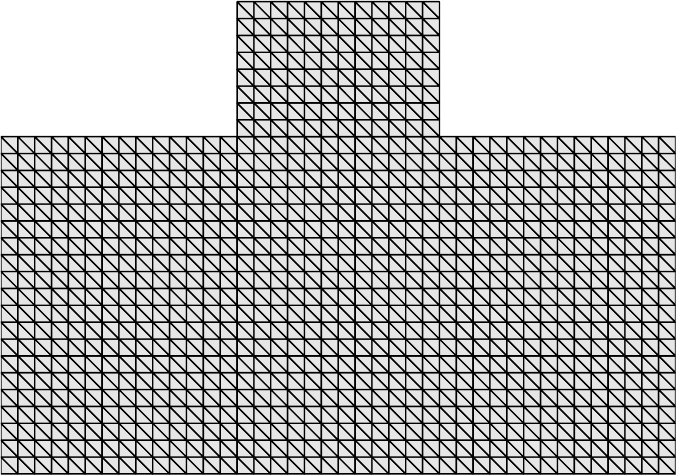}
		\caption{}
		\label{fig:fixed_punch_b}
	\end{subfigure}
	\hfill
	\begin{subfigure}[b]{0.32\textwidth}
		\centering
		\includegraphics[height=3.5cm]{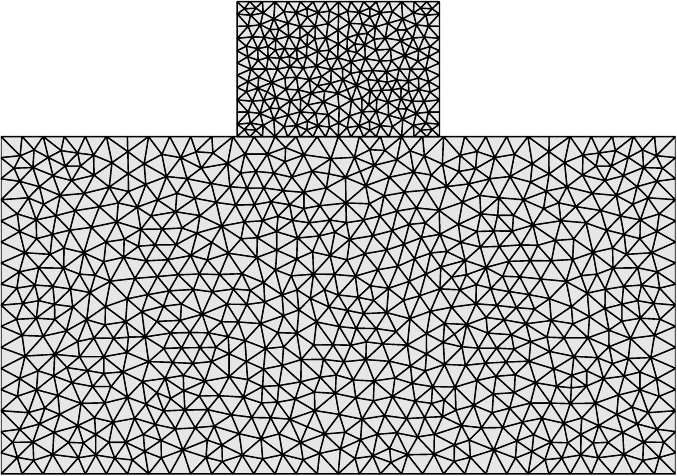}
		\caption{}
		\label{fig:fixed_punch_c}
	\end{subfigure}
	\hfill
	\caption{(a) Fixed punch schematic; (b) Standard FEM conforming node-to-node contact mesh used used as reference; and (c) Non-conforming mesh used for the enriched methods.}
	\label{fig:fixed_punch}
\end{figure}

In this second test, which is illustrated schematically in Figure~\ref{fig:fixed_punch_a}, we study the influence of mesh refinement with a fixed punch. The material properties are the same as those used in the first test. The results of our enriched approach are compared to those of standard FEM using conforming node-to-node contact discretizations, as shown in Figure~\ref{fig:fixed_punch_b}. Figure~\ref{fig:fixed_punch_c} shows a typical finite element discretization used for all other results.

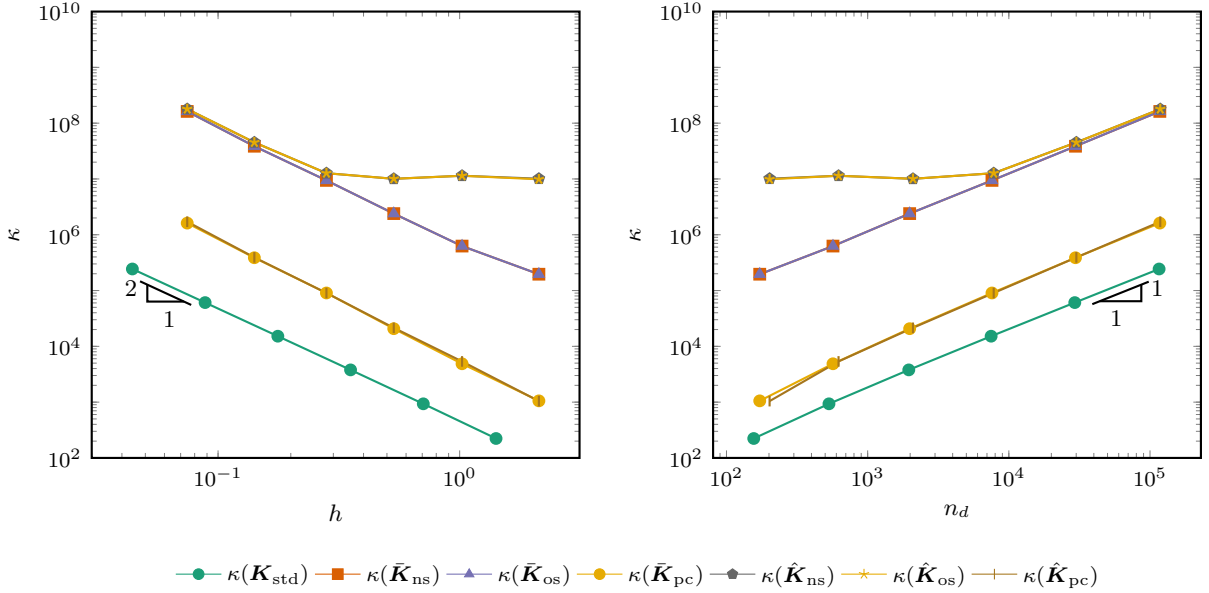
\begin{figure}
	\centering
	\input{contact_condition_single_legend.tex}
	\caption{Condition number of the entire stiffness matrix as a function of mesh size $h$ (left) and total number of DOFs~(right). }
	\label{fig:condition_refine}
\end{figure}

The results in Figure~\ref{fig:condition_refine} show the condition number as a function of mesh size (left) and total number of DOFs (right). The reference curve (labeled $\kappa ( \bs K_{\mathrm{std}} )$) is computed using the conforming mesh shown in Figure~\ref{fig:fixed_punch_b}.
It is well known that the condition number in standard FEM scales as $\mathcal{O}\left( h^{-2}\right)$ with mesh size $h$ and $\mathcal{O}\left( n_d \right)$ with the total number of DOFs $n_d$. For MPC, the condition number of the original enriched system matrix indeed deteriorates with mesh refinement (curve $\kappa(\bar{\bs{K}}_{\mathrm{ns}})$). Also in this case, the optimal scaling proposed in Reference~\cite{Aragon:2020a} (curve $\kappa(\bar{\bs{K}}_{\mathrm{os}})$) has no effect on the conditioning. However, applying the simple diagonal preconditioner improves the condition number significantly (curve $\kappa(\bar{\bs{K}}_{\mathrm{pc}})$). Noteworthy, the condition number of the enriched system matrices increases at the same rate as that of standard FEM with node-to-node contact. Therefore, the enriched method constrained using MPCs with a simple preconditioner is as stable as standard FEM.

For ALM, the condition numbers are generally worse than those obtained with MPCs. The condition number also deteriorates with mesh refinement, and the magnitude with and without scaling is also the same (curves $\kappa(\hat{\bs{K}}_{\mathrm{ns}})$ and $\kappa(\hat{\bs{K}}_{\mathrm{os}})$, respectively). We find that the condition number of the enriched system matrix with a simple pre-conditioner (curve $\kappa(\hat{\bs{K}}_{\mathrm{pc}})$) is close to that of the preconditioned matrix using MPCs and grows at the same rate as FEM with node-to-node contact. 

\subsection{Hertzian contact problem}

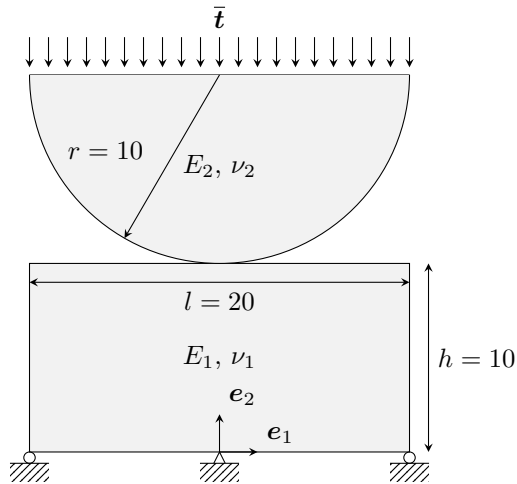
\begin{figure}
	\centering
	\input{hertz_geo.tex}
	\caption{Geometry of the Hertzian contact example.}
	\label{fig:hertz_geo}
\end{figure}

Figure~\ref{fig:hertz_geo} illustrates a Hertzian contact problem, where a semi-circular punch with radius $r=10$ and material properties $E_2 =\num{7e5}$ and $\nu_2 = 0.3$ is subject to a uniformly distributed load $\bar{\bs{t}}=-25\bs{e}_2$ and pushed against a substrate with material properties $E_1 = \num{7000}$ and $\nu_1=0.3$. The substrate has length $l = 20$ and height $h= 10$ and is simply supported along the bottom edge. The solution to this problem is given in Reference~\cite{Wriggers2016}:
\begin{equation}
	p_{n}= \frac{4 r \bar{t}}{\pi b^{2}} \sqrt{b^{2}-x_1^{2}},   \qquad \
	b= 2 \sqrt{\frac{2 r^{2} \bar{t}}{\pi E^{*}}}, \qquad \
	\frac{1}{E^{*}}=  \frac{1}{2} \left(\frac{1-\nu_{1}^{2}}{E_{1}}+\frac{1-\nu_{2}^{2}}{E_{2}}\right),
\end{equation}
where $\bar{t}$ is the magnitude of applied traction, $p_n$ denotes the contact pressure, $b$ the contact area, and $E^{*}$ the effective stiffness.

\begin{figure}
	\begin{subfigure}[b]{0.4\textwidth}
		\centering
		\includegraphics[width = \textwidth]{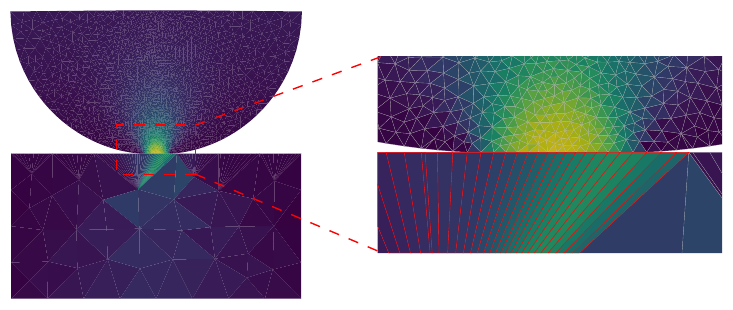}
		\caption{}
	\end{subfigure}
	\begin{subfigure}[b]{0.18\textwidth}
		\centering
		\begin{tikzpicture}
			\node at (6,0.0){\includegraphics[height=2.4cm]{colorbar.pdf}};
			\node at (6.0, 1.6) {$\sigma_{22}$};
			\node at (6.55, 1.2) {$-350$};
			\node at (6.3, -1.1) {$0$};
		\end{tikzpicture}
	\end{subfigure}
	\begin{subfigure}[b]{0.4\textwidth}
		\includegraphics[width = \textwidth]{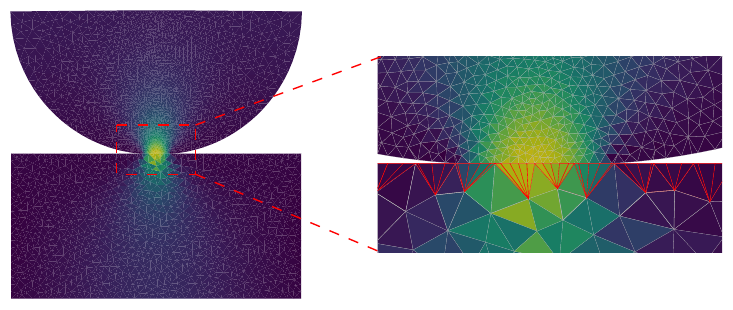}
		\caption{}
	\end{subfigure}
	\caption{Stress component $\sigma_{22}$ for different discretization designs (integration elements are drawn with red edges): (a)~substrate mesh much coarser than punch mesh, (b)~similar mesh sizes for punch and substrate.}\label{fig:hertz_stress}
\end{figure}

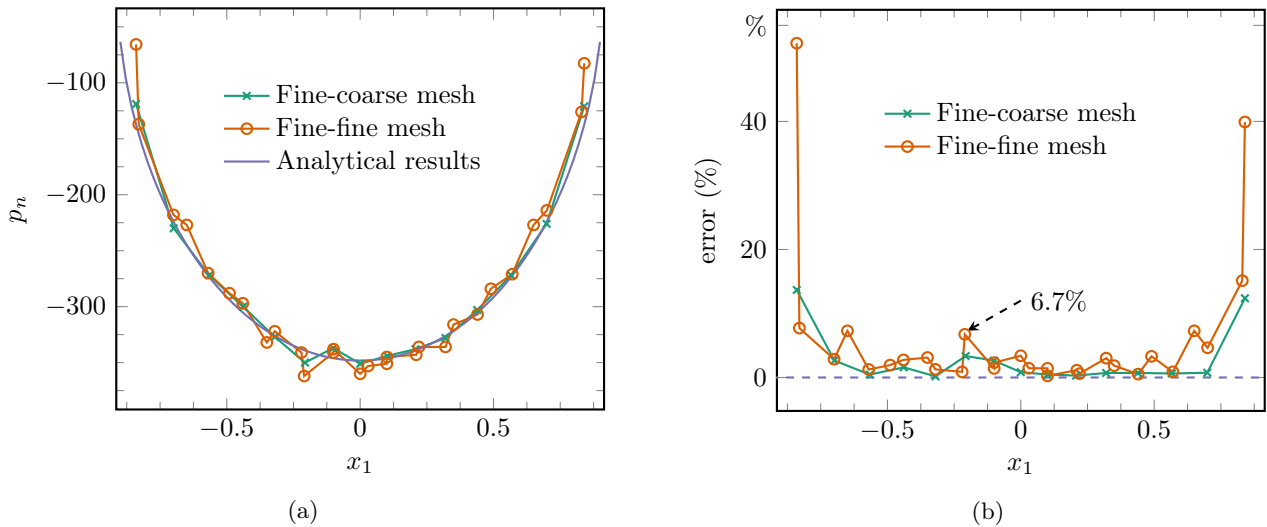
\begin{figure}
	\centering
	\begin{subfigure}[position]{0.47\textwidth}
		\input{hertz_traction.tex}
		\caption{}\label{fig:hertz_traction_pressure}
	\end{subfigure}
	\hfill
	\begin{subfigure}[position]{0.47\textwidth}
		\input{hertz_traction_error.tex}
		\caption{}	\label{fig:hertz_traction_error}
	\end{subfigure}
	\caption{Traction along the contact interface in the Hertzian contact problem: Comparison of the numerical profiles obtained with the discretizations reported in Figure~\ref{fig:hertz_stress} to the analytical solution~(a), and corresponding relative error with respect to the analytical solution~(b).}
	\label{fig:hertz_traction}
\end{figure}

The load is applied in twenty load increments of equal magnitude. We used the enriched method with ALM, and convergence (with a tolerance $\norm{\Delta \bs U}/\norm{\bs U}< 10^{-5}$) was reached within four iterations per step.
Two different meshes are used in this numerical test: for the first one, the substrate discretization is coarser than that of the punch, while in the second one punch and substrate are discretized with elements of similar sizes.
In both cases, Figure~\ref{fig:hertz_stress} shows the presence of enriched nodes also far from the contact area. Enriched nodes are actually added also far from the contact area for convenience (in practice, all standard nodes from one contact surface are projected to the other and vice versa). Enriched nodes that do not come into contact with the corresponding standard node will be regarded as inactive in the calculation of stiffness matrix and force vector components.
The results are reported in terms of stress distribution $\sigma_{22}$ in Figure~\ref{fig:hertz_stress} and contact pressure profile in Figure~\ref{fig:hertz_traction}. The stress field shows a typical Hertzian contact distribution. Figure~\ref{fig:hertz_traction_error} shows the contact pressure relative error, computed as $e = \left| p_n-p_n^h \right| / \left|  p_n\right|$, where $p_n^h$ is the pressure obtained numerically.
We find that the steep pressure gradients in elements that transition from contact to no contact are responsible of yielding inaccurate contact tractions.
The error in the interior region is within 7\% for both discretizations studied. Therefore, in this region the numerical contact pressure profile approximates the analytical solution closely.

\section{Discussion}
\label{sec:conclusion}

Compared to traditional contact and coupling formulations, the proposed method has a number of advantages: As the enriched formulation essentially transforms the problem into a node-to-node discretization, it is possible to utilize the most straightforward coupling and contact techniques. This also facilitates the implementation in existing standard displacement-based finite element packages. Furthermore, as the tractions are properly transferred and over-constrained locking is avoided, no contact stabilization techniques are required. This important intrinsic property of the formulation, which was first noticed while comparing DE-FEM with X/GFEM in Reference~\cite{defem}, relates also to the recovery of smooth tractions in immersed analysis using IGFEM and DE-FEM~\cite{ramos2015, vandenBoom2018}.
Noteworthy, our traction profiles are on par with those obtained by the virtual element method (VEM)~\cite{Wriggers2016}, whereby a non-conforming problem is transformed into a node-to-node VEM-conforming discretization; in our procedure, however,  formulation and computer implementation are much simpler than VEM.
It is important to note that the original element's shape functions are kept intact, and that enrichments are only nonzero in the elements along the contact boundary (thus the partition of unity property in these elements is lost). Because enrichment functions are local by construction and vanish at the original mesh nodes, the Kronecker delta property on those nodes is retained, and all standard DOFs preserve their physical interpretation. As a result, post-processing is only required to compute the solution at enriched nodes, and because every enriched node is matched to a standard node, obtaining the latter's DOFs further reduces post-processing.

It has been acknowledged in previous works that interface- and discontinuity-enriched formulations may have difficulties in properly reconstructing field gradients (strains and thus stresses). This issue stems from the construction of the enriched finite element space, which may use sliver integration elements that degrade the accuracy of field gradients (as in standard FEM). While the issue is more pronounced for material interfaces~\cite{Soghrati:2017,Nagarajan:2018}, it has been shown recently that the issue is negligible near Dirichlet boundaries~\cite{vandenBoom2018} and near traction-free cracks~\cite{defem-3d}. In the context of contact and mesh coupling problems, our numerical experiments indicate that this issue is not present. As enriched nodes are only placed along the contact/coupling interfaces, we conjecture that the presence of sliver integration elements does not adversely affect the gradient field accuracy and thus the method can properly recover strains and stresses.
It is worth noting that there is work that aims at improving the accuracy of recovered gradient fields from enriched FEMs (and in fact unfitted FEMs in general)~\cite{Zhang:2021}.

Although in this work we considered linearized kinematics and frictionless contact, the extension of the current enriched framework to more advanced problems such as frictional contact, contact in 3D, and contact in large deformation is relatively straightforward. The only drawbacks we see at the moment are related to the non-symmetric global stiffness matrix stemming from the frictional contact formulation, the necessity of a more efficient way of contact detection in a three-dimensional implementation, and the possible need of smoothing techniques to achieve better convergence properties in large deformation problems.
Finally, the proposed method could also be applied to high-order approximations, albeit high-order enriched functions would be needed to properly describe curved edges (assuming that geometry is described nonlinearly).

\section*{Acknowledgements}
The authors would like to thank Dr. Vladislav A. Yastrebov for his insightful comments during the preparation of this manuscript. Dongyu Liu gratefully acknowledges the financial support from the China Scholarship Council (CSC) under contract 201606250027.

\appendix
\section{An analytical 1D problem to investigate the role of the optimal scaling factor}\label{appendix}
\begin{figure}[h!]
	\centering
	\input{analytic.pdf_tex}
	\caption{One-dimensional coupling problem.}
	\label{fig:1d}
\end{figure}
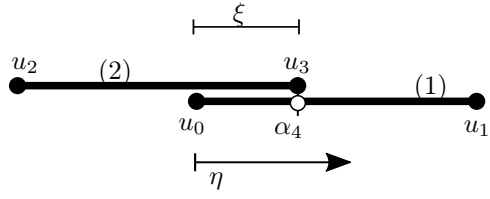

A simple 1D problem is proposed to investigate the ineffectiveness of the optimal scaling factor in the results reported in \S~\ref{sec:stability}.
Figure~\ref{fig:1d} shows the problem, consisting of a 1D rod (element~1) that is coupled somewhere along its length at location $\xi$ to another 1D rod (element~2) through an enriched node.

For simplicity, both elements have unit lengths, cross sections, and stiffnesses. The shape functions of both elements are
\begin{equation}
	N_0 = 1-\eta \;\;,\;\; N_1=\eta.
\end{equation}
The enrichment function below, endowed with scaling factor $s$, is defined in element~1:
\begin{equation}
	s\psi =  \left\{
	\begin{array}{ll}
		\frac{s}{\xi} \eta                  & 0\leq \eta\leq \xi \\
		\frac{s}{1-\xi} \left(1-\eta\right) & \xi\leq \eta\leq 1 \\
	\end{array}
	\right..
\end{equation}

The element matrix for element~1, including the enrichment function contribution, is computed as
\begin{equation}
	\begin{aligned}
		\boldsymbol{k}_1 & = \int_{0}^{\xi} \begin{bmatrix} \bs{B}_{u}^\intercal \\ \bs{B}_{\alpha}^\intercal \end{bmatrix} \bs{D} \begin{bmatrix}
			\bs{B}_{u} & \bs{B}_{\alpha}
		\end{bmatrix} \, \dd{\eta} + \int_{\xi}^{1} \begin{bmatrix} \bs{B}_{u}^\intercal \\ \bs{B}_{\alpha}^\intercal \end{bmatrix} \bs{D} \begin{bmatrix}
			\bs{B}_{u} & \bs{B}_{\alpha}
		\end{bmatrix} \, \dd{\eta} \\
		                 & = \xi\begin{bmatrix}
			-1 \\
			1  \\
			\frac{s}{\xi}
		\end{bmatrix}\begin{bmatrix}
			-1 & 1 & \frac{s}{\xi}
		\end{bmatrix} + \left(1-\xi\right)\begin{bmatrix}
			-1 \\
			1  \\
			-\frac{s}{1-\xi}
		\end{bmatrix}\begin{bmatrix}
			-1 & 1 & -\frac{s}{1-\xi}
		\end{bmatrix}                                                    \\
		                 & = \begin{bmatrix}
			1  & -1 & 0                                   \\
			-1 & 1  & 0                                   \\
			0  & 0  & \frac{s^2}{1-\xi} + \frac{s^2}{\xi}
		\end{bmatrix},
	\end{aligned}
\end{equation}
while that for element 2 reads
\begin{equation}
	\boldsymbol{k}_2 = \int_{0}^{1} \bs{B}_{u}^\intercal \bs{D}
	\bs{B}_{u}  \, \dd{\eta} = \begin{bmatrix}
		-1 \\
		1  \\
	\end{bmatrix}\begin{bmatrix}
		-1 & 1
	\end{bmatrix} = \begin{bmatrix}
		1  & -1 \\
		-1 & 1
	\end{bmatrix}.
\end{equation}
The assembled stiffness matrix of the entire system is then
\begin{equation}
	\boldsymbol{K} = \begin{bmatrix}
		1  & -1 & 0  & 0  & 0                                   \\
		-1 & 1  & 0  & 0  & 0                                   \\
		0  & 0  & 1  & -1 & 0                                   \\
		0  & 0  & -1 & 1  & 0                                   \\
		0  & 0  & 0  & 0  & \frac{s^2}{1-\xi} + \frac{s^2}{\xi}
	\end{bmatrix},
\end{equation}
which clearly shows that standard and enriched elements are decoupled. Coupling is achieved by means of the multiple-point constraint relation
\begin{equation}
	u\left(\xi\right) = N_0\left(\xi\right)u_0 + N_1\left(\xi\right)u_1 + s \Psi\left(\xi\right)\alpha_4 = u_3
\end{equation}
from which, since $\psi\left(\xi\right)=1$,
\begin{equation}
	\alpha_4 = \frac{1}{s}u_3 - \frac{N_0\left(\xi\right)}{s}u_0 -\frac{N_1\left(\xi\right)}{s}u_1.
\end{equation}
Considering that
\begin{equation}
	\boldsymbol{U} = \boldsymbol{T}\bar{\boldsymbol{U}} = \begin{bmatrix}
		u_0 \\
		u_1 \\
		u_2 \\
		u_3 \\
		\alpha_4
	\end{bmatrix}=\begin{bmatrix}
		1                 & 0               & 0 & 0           \\
		0                 & 1               & 0 & 0           \\
		0                 & 0               & 1 & 0           \\
		0                 & 0               & 0 & 1           \\
		- \frac{1-\xi}{s} & - \frac{\xi}{s} & 0 & \frac{1}{s}
	\end{bmatrix}\begin{bmatrix}
		u_0 \\
		u_1 \\
		u_2 \\
		u_3 \\
	\end{bmatrix},
\end{equation}
the transformation matrix $\boldsymbol{T}$ for the multiple-point constraint is therefore
\begin{equation}
	\boldsymbol{T} = \begin{bmatrix}
		1                 & 0               & 0 & 0           \\
		0                 & 1               & 0 & 0           \\
		0                 & 0               & 1 & 0           \\
		0                 & 0               & 0 & 1           \\
		- \frac{1-\xi}{s} & - \frac{\xi}{s} & 0 & \frac{1}{s}
	\end{bmatrix},
\end{equation}
from which one obtains the modified stiffness matrix
\begin{equation}
	\bar{\boldsymbol{K}} = \boldsymbol{T}^\intercal\boldsymbol{K}\boldsymbol{T}= \begin{bmatrix}
		\frac{1}{\xi}  & 0                & 0  & -\frac{1}{\xi}                                        \\
		0              & -\frac{1}{\xi-1} & 0  & \frac{1}{\xi-1}                                       \\
		0              & 0                & 1  & -1                                                    \\
		-\frac{1}{\xi} & \frac{1}{\xi-1}  & -1 & \frac{\xi\left(\xi-1\right)-1}{\xi\left(\xi-1\right)} \\
	\end{bmatrix}.
\end{equation}

Note that the scaling factor $s$ does not appear in the modified stiffness matrix $\bar{\boldsymbol{K}}$. The scaling factor, therefore, does not have any influence on eigenvalues and condition number of the system matrix. It can be concluded that scaling of enrichment functions is ineffective towards the improvement of the condition number of enriched coupling and contact problems using MPCs.

\bibliographystyle{unsrt}
\bibliography{bibliography}

\end{document}

%% file: figures/potatos_contact_nonmatch.pdf_tex
\begingroup%
  \makeatletter%
  \providecommand\color[2][]{%
    \errmessage{(Inkscape) Color is used for the text in Inkscape, but the package 'color.sty' is not loaded}%
    \renewcommand\color[2][]{}%
  }%
  \providecommand\transparent[1]{%
    \errmessage{(Inkscape) Transparency is used (non-zero) for the text in Inkscape, but the package 'transparent.sty' is not loaded}%
    \renewcommand\transparent[1]{}%
  }%
  \providecommand\rotatebox[2]{#2}%
  \ifx\svgwidth\undefined%
    \setlength{\unitlength}{180.31156041bp}%
    \ifx\svgscale\undefined%
      \relax%
    \else%
      \setlength{\unitlength}{\unitlength * \real{\svgscale}}%
    \fi%
  \else%
    \setlength{\unitlength}{\svgwidth}%
  \fi%
  \global\let\svgwidth\undefined%
  \global\let\svgscale\undefined%
  \makeatother%
  \begin{picture}(1,0.64319751)%
    \put(0,0){\includegraphics[width=\unitlength,page=1]{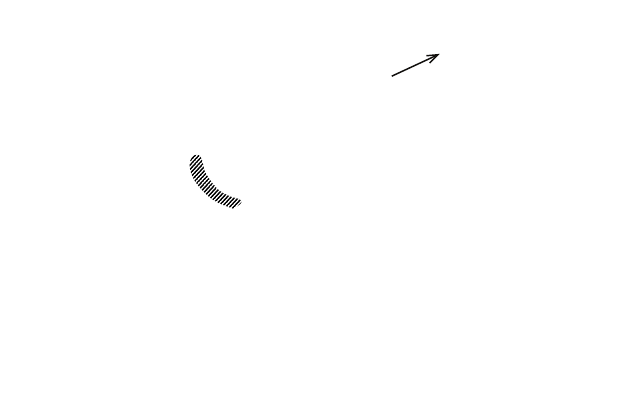}}%
    \put(0.8156557,0.23792593){\color[rgb]{0,0,0}\makebox(0,0)[lb]{\smash{$\bs n_2$ }}}%
    \put(0,0){\includegraphics[width=\unitlength,page=2]{potatos_contact_nonmatch.pdf}}%
    \put(0.48669994,0.53060895){\color[rgb]{0,0,0}\makebox(0,0)[lb]{\smash{$\Gamma^t_1$}}}%
    \put(0.37983372,0.49024104){\color[rgb]{0,0,0}\makebox(0,0)[lb]{\smash{$\Omega_1$}}}%
    \put(0,0){\includegraphics[width=\unitlength,page=3]{potatos_contact_nonmatch.pdf}}%
    \put(0.54138479,0.30098061){\color[rgb]{0,0,0}\makebox(0,0)[lb]{\smash{$\Omega_2$}}}%
    \put(0.6048943,0.63058142){\color[rgb]{0,0,0}\makebox(0,0)[lb]{\smash{$\bs{\bar{t}}_1$}}}%
    \put(0.70903991,0.5521693){\color[rgb]{0,0,0}\makebox(0,0)[lb]{\smash{$\bs n_1$ }}}%
    \put(0,0){\includegraphics[width=\unitlength,page=4]{potatos_contact_nonmatch.pdf}}%
    \put(0.56648417,0.41479772){\color[rgb]{0,0,0}\makebox(0,0)[lt]{\begin{minipage}{0.09041633\unitlength}\raggedright $\Gamma^c$\end{minipage}}}%
    \put(0,0){\includegraphics[width=\unitlength,page=5]{potatos_contact_nonmatch.pdf}}%
    \put(0.30408743,0.14866835){\color[rgb]{0,0,0}\makebox(0,0)[lb]{\smash{$\bs n_3$ }}}%
    \put(0.56872043,0.11212752){\color[rgb]{0,0,0}\makebox(0,0)[lb]{\smash{$\Omega_3$}}}%
    \put(0.34588428,0.36859909){\color[rgb]{0,0,0}\makebox(0,0)[lb]{\smash{$\Gamma^u_1$}}}%
    \put(0.53156138,0.0454606){\color[rgb]{0,0,0}\makebox(0,0)[lb]{\smash{$\Gamma^u_3$}}}%
    \put(0.78417921,0.15076606){\color[rgb]{0,0,0}\makebox(0,0)[lb]{\smash{$\bs{\bar{t}}_3$}}}%
    \put(0.45483561,0.23261307){\color[rgb]{0,0,0}\makebox(0,0)[lt]{\begin{minipage}{0.09041633\unitlength}\raggedright $\Gamma^g$\end{minipage}}}%
  \end{picture}%
\endgroup%

%% file: figures/enrich_func.tex
\begin{tikzpicture}[scale = 1.0]
	\newcommand \wen {1};

	\begin{axis}[
		colormap/blackwhite,
		axis lines*=left,
		xmin=-0.1,
		xmax= 4.5,
		ymin=-0.1,
		ymax= 4.5,
		zmin=-0.3,
		zmax=1.2,
		axis lines=center,
		enlargelimits=true,
		z axis line style={-{Stealth[scale=1.5]}},
		ztick={0, 1},
		zticklabels={$0$, $1$},
		view={60}{25}, 
		y axis line style={-{Stealth[scale=1.5]}},
		x axis line style={-{Stealth[scale=1.5]}},
		xtick=\empty,
		ytick=\empty,
		scale=1.0
		]

		\pgfplotsset{meshstyle/.style={patch,patch type=triangle,color=gray!10,faceted color=gray,fill opacity=0.5}}

		\coordinate (a) at (0.0, 0.0, 0);
		\coordinate (b) at (2.0, 0.0, 0);
		\coordinate (c) at (4.0, 0.0, 0);
		\coordinate (d) at (0.0, 2.0, 0);
		\coordinate (e) at (2.0, 2.0, 0);
		\coordinate (f) at (4.0, 2.0, 0.0);
		\coordinate (g) at (0.0, 4.0, 0.0);
		\coordinate (h) at (4.0, 4.0, 0.0);

		\pgfplotsset{psistyle/.style={patch, patch type=triangle, color=RoyalBlue, ultra thin, fill opacity=0.3}}
		\pgfplotsset{crackstyle/.style={patch, patch type=triangle, color=BrickRed, faceted color=BrickRed, ultra thin, fill opacity=0.25}}

		\pgfmathsetmacro{\wenOne}{1.0};

		\addplot3[meshstyle] coordinates {(0, 0, 0) (2, 0, 0) (0, 2, 0)};
		\addplot3[meshstyle] coordinates {(0,2,0) (2.0, 2.0, 0) (2.0, 0, 0.0)};
		\addplot3[meshstyle] coordinates {(2,2,0) (4.0, 0.0, 0) (2.0, 0, 0.0)};
		\addplot3[meshstyle] coordinates {(4,0,0) (4.0, 2.0, 0) (2.0, 2.0, 0.0)};
		\addplot3[meshstyle] coordinates {(0,2,0) (4.0, 2.0, 0) (4.0, 4.0, 0.0)};
		\addplot3[meshstyle] coordinates {(4,4.,0) (0.0, 4.0, 0) (0.0, 2.0, 0.0)};
		\addplot3[psistyle] coordinates {(2,2, 1.0) (0., 2.0, 0.0) (4.0, 4.0, 0.0)};
		\addplot3[psistyle] coordinates {(2,2, 1.0) (4.0, 2.0, 0.0) (4.0, 4.0, 0.0)};
		\draw [dashed] (2.0, 2.0, \wenOne) -- (2.0, 2.0, 0);
		\draw [ultra thick, blue] (0, 2, 0) -- (4, 2, 0);

		\draw [fill = black] (a) circle [radius=1.5pt];
		\draw [fill = black] (b) circle [radius=1.5pt];
		\draw [fill = black] (c) circle [radius=1.5pt];
		\draw [fill = black] (d) circle [radius=1.5pt];
		\draw [fill = black] (e) circle [radius=1.5pt];
		\draw [fill = black] (f) circle [radius=1.5pt];
		\draw [fill = black] (g) circle [radius=1.5pt];
		\draw [fill = black] (h) circle [radius=1.5pt];

		\draw [line width=0.0, black] (0, 0, 0) -- (4.5, 0, 0) node[anchor=center, pos=1.2]{{$x_1$}};
		\draw [line width=0.0, black] (0, 0, 0) -- (0, 4.5, 0) node[anchor=center, pos=1.2]{{$x_2$}};
		\node at (0, 0.8, 1.25) {$\psi_i \left( \bs x \right)$};
		\draw [red] (e) circle [radius = 2.5pt];
		\node [right, yshift=3pt] at (e) {$\enriched{x}{i}$};
		\node [above left] at (2.3, 1.3, -0.) {$\Gamma^g_{12}$};
		\node at (0.75, 0.5, 0.0) {$\Omega_2$};
		\node at (3.2, 2.9, 0.0) {$\Omega_1$};

	\end{axis}
\end{tikzpicture}

%% file: figures/enriched_contact.pdf_tex
\begingroup%
  \makeatletter%
  \providecommand\color[2][]{%
    \errmessage{(Inkscape) Color is used for the text in Inkscape, but the package 'color.sty' is not loaded}%
    \renewcommand\color[2][]{}%
  }%
  \providecommand\transparent[1]{%
    \errmessage{(Inkscape) Transparency is used (non-zero) for the text in Inkscape, but the package 'transparent.sty' is not loaded}%
    \renewcommand\transparent[1]{}%
  }%
  \providecommand\rotatebox[2]{#2}%
  \newcommand*\fsize{\dimexpr\f@size pt\relax}%
  \newcommand*\lineheight[1]{\fontsize{\fsize}{#1\fsize}\selectfont}%
  \ifx\svgwidth\undefined%
    \setlength{\unitlength}{84.62999725bp}%
    \ifx\svgscale\undefined%
      \relax%
    \else%
      \setlength{\unitlength}{\unitlength * \real{\svgscale}}%
    \fi%
  \else%
    \setlength{\unitlength}{\svgwidth}%
  \fi%
  \global\let\svgwidth\undefined%
  \global\let\svgscale\undefined%
  \makeatother%
  \begin{picture}(1,1.06049872)%
    \lineheight{1}%
    \setlength\tabcolsep{0pt}%
    \put(0,0){\includegraphics[width=\unitlength,page=1]{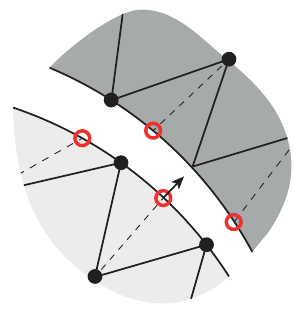}}%
    \put(0.55662573,0.47736699){\makebox(0,0)[lt]{\lineheight{1.25}\smash{\begin{tabular}[t]{l}$\bs{n}_i$\end{tabular}}}}%
    \put(0.59638481,0.36944973){\makebox(0,0)[lt]{\lineheight{1.25}\smash{\begin{tabular}[t]{l}$\enriched{x}{i}$\end{tabular}}}}%
    \put(0.69638481,0.4244973){\makebox(0,0)[lt]{\lineheight{1.25}\smash{\begin{tabular}[t]{l}$\bs{x}_{i}$\end{tabular}}}}%
    \put(0.44492199,0.49819308){\makebox(0,0)[lt]{\lineheight{1.25}\smash{\begin{tabular}[t]{l}$\bs{x}_j$\end{tabular}}}}%
    \put(0.73648783,0.21420032){\makebox(0,0)[lt]{\lineheight{1.25}\smash{\begin{tabular}[t]{l}$\bs{x}_k$\end{tabular}}}}%
  \end{picture}%
\endgroup%

%% file: figures/contact_middle.tex
\begin{tikzpicture}[scale=0.5]
\tikzstyle{ground}=[fill,pattern=north east lines,draw=none,minimum width=0.5cm,minimum height=0.2cm]
\draw [fill=gray!10] (0,0) --(10,0) -- (10,5) -- (0, 5) -- cycle;
\draw [fill=gray!10] (3.5, 5) -- (6.5, 5) -- (6.5, 7) -- (3.5, 7) -- cycle;

\draw [-stealth] (0.5, 0.5) -- (0.5, 1.5);
\draw [-stealth] (0.5, 0.5) -- (1.5, 0.5);
\node [right] at (1.5, 0.5) {$\bs{e}_1$};
\node [above] at (0.5, 1.5) {$\bs{e}_2$};

\draw [stealth-stealth] (10.4, 0) -- (10.4, 5);
\node [right] at (10.4, 2.5) {$H = 5$};
\draw [stealth-stealth] (10.4, 5) -- (10.4, 7);
\node [right] at (10.4, 6) {$h = 2$};

\draw [stealth -stealth] (0, 3.5) -- (10, 3.5);
\node [below] at (5, 3.5) {$L = 10$};

\draw [stealth - stealth] (3.5, 4.7) -- (6.5, 4.7);
\node [below] at (5, 4.7) {$l = 3$};

\coordinate (A) at (0, 0);
\coordinate (B) at (5, 0);
\coordinate (C) at (10, 0);
    
\node (ground0) at (A) [ground, yshift = -0.15cm, anchor = north] {};
\draw (ground0.north west) -- (ground0.north east);
\draw (0, -0.15) circle  (0.15cm);
\node (ground1) at (B) [ground, yshift = -0.15cm, anchor=north] {};
\draw  (ground1.north west) -- (ground1.north east);
\draw (4.85, -0.3) -- (5, 0) -- (5.15, -0.3);
\node (ground2) at (C) [ground, yshift = -0.15cm, anchor=north] {};
\draw  (ground2.north west) -- (ground2.north east);
\draw (10, -0.15) circle  (0.15cm);
\foreach \i in {0,...,9}
{
	\draw [-stealth] (0.35*\i, 5.8) -- (0.35*\i, 5.1);
	\draw [-stealth] (6.85 + 0.35*\i, 5.8) -- (6.85+ 0.35*\i, 5.1);
}
\foreach \i in {0,...,8}
{
	\draw [-stealth] (3.55+0.35*\i, 7.8) -- (3.55+0.35*\i, 7.1);
}
\node [above] at (1.75, 5.8) {$\overline{\bs t}$};
\node [above] at (8.25, 5.8) {$\overline{\bs t}$};
\node [above] at (5., 7.8) {$\overline{\bs t}$};

\node at (8, 1) {$E$, $\nu$};
\node at (5, 6) {$E$, $\nu$};
\end{tikzpicture}

%% file: figures/plate_hole_schematic.pdf_tex
\begingroup%
  \makeatletter%
  \providecommand\color[2][]{%
    \errmessage{(Inkscape) Color is used for the text in Inkscape, but the package 'color.sty' is not loaded}%
    \renewcommand\color[2][]{}%
  }%
  \providecommand\transparent[1]{%
    \errmessage{(Inkscape) Transparency is used (non-zero) for the text in Inkscape, but the package 'transparent.sty' is not loaded}%
    \renewcommand\transparent[1]{}%
  }%
  \providecommand\rotatebox[2]{#2}%
  \newcommand*\fsize{\dimexpr\f@size pt\relax}%
  \newcommand*\lineheight[1]{\fontsize{\fsize}{#1\fsize}\selectfont}%
  \ifx\svgwidth\undefined%
    \setlength{\unitlength}{198.42501068bp}%
    \ifx\svgscale\undefined%
      \relax%
    \else%
      \setlength{\unitlength}{\unitlength * \real{\svgscale}}%
    \fi%
  \else%
    \setlength{\unitlength}{\svgwidth}%
  \fi%
  \global\let\svgwidth\undefined%
  \global\let\svgscale\undefined%
  \makeatother%
  \begin{picture}(1,0.78571747)%
    \lineheight{1}%
    \setlength\tabcolsep{0pt}%
    \put(0,0){\includegraphics[width=\unitlength,page=1]{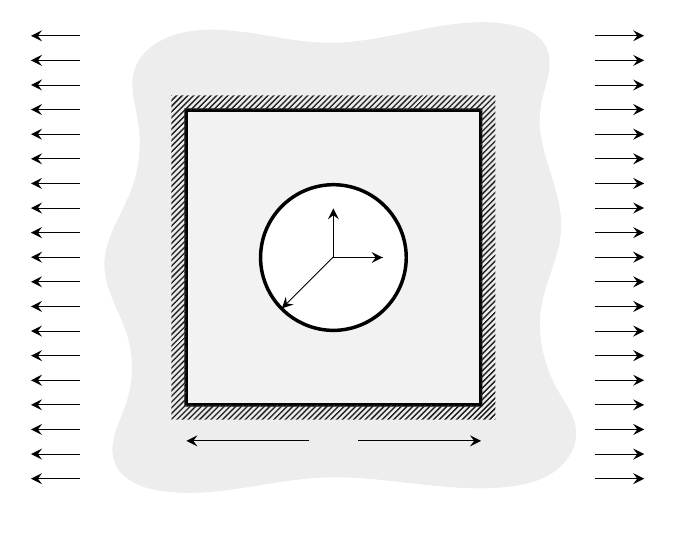}}%
    \put(0.45659669,0.34927477){\makebox(0,0)[lt]{\lineheight{1.25}\smash{\begin{tabular}[t]{l}$r$\end{tabular}}}}%
    \put(0.52525254,0.42488618){\makebox(0,0)[lt]{\lineheight{1.25}\smash{\begin{tabular}[t]{l}$\bs{e}_1$\end{tabular}}}}%
    \put(0.42631069,0.46463555){\makebox(0,0)[lt]{\lineheight{1.25}\smash{\begin{tabular}[t]{l}$\bs{e}_2$\end{tabular}}}}%
    \put(0.46714061,0.12884472){\makebox(0,0)[lt]{\lineheight{1.25}\smash{\begin{tabular}[t]{l}$L$\end{tabular}}}}%
    \put(0.94850079,0.68914805){\makebox(0,0)[lt]{\lineheight{1.25}\smash{\begin{tabular}[t]{l}$\sigma_{\infty}$\end{tabular}}}}%
    \put(0.12469112,0.68914805){\makebox(0,0)[lt]{\lineheight{1.25}\smash{\begin{tabular}[t]{l}$\sigma_{\infty}$\end{tabular}}}}%
    \put(0.29542353,0.22525681){\makebox(0,0)[lt]{\lineheight{1.25}\smash{\begin{tabular}[t]{l}$E, \nu$\end{tabular}}}}%
  \end{picture}%
\endgroup%

%% file: figures/plate_hole_convergence_single_legend.tex
  \begin{tikzpicture}
    \pgfplotsset{footnotesize,samples=10}
    \begin{groupplot}[cycle list/Dark2,
		cycle multiindex* list={
			mark list\nextlist
			Dark2\nextlist},
			group style = {group size = 2 by 2, horizontal sep = 50pt, vertical sep = 60pt}, width=8cm, xlabel=$n_d$, 	xmode =log, ymode = log, yminorticks=true, xminorticks=true, minor tick num=3,]

\pgfplotsset{cycle list shift=0}
			
\nextgroupplot[
	width=8cm,
	ylabel=$\| \epsilon \|_{\mathcal{L}^{2}}$,
	legend columns=4,
	minor tick num=3,
	legend style={
		color =black,
		draw=none,
		fill=none,
		legend cell align={left},
		legend to name = grouplegend1},
	thick=1pt
]

\addplot table [x index=1, y index =2, col sep=comma] {convergence/tp_plate_hole.dat};
\addlegendentry{Two-pass MPC}

\addplot table [x index=1, y index =2, col sep=comma] {convergence/std_plate_hole.dat}
coordinate [pos=0.75] (A)
coordinate [pos=0.85]  (B)
;\addlegendentry{Standard FEM}
\coordinate (A') at ($(A)!-3pt!90:(B)$);
\coordinate (B') at ($(B)!-3pt!270:(A)$);
\draw  (B') -| (A') node[pos=0.18,below]{\small $1$} node[pos=0.75,left]{\small $1.00$};
\draw [shorten <=-3pt,shorten >=-3pt] (A') -- (B');


\addplot table [x index=1, y index =2, col sep=comma] {convergence/nonmatch_plate_hole.dat}
coordinate [pos=0.75] (A)
coordinate [pos=0.85]  (B)
;
\coordinate (A') at ($(A)!3pt!90:(B)$);
\coordinate (B') at ($(B)!3pt!270:(A)$);
\draw  (A') -| (B') node[pos=0.18,above]{\small $1$} node[pos=0.75,right]{\small $1.02$};
\draw [shorten <=-3pt,shorten >=-3pt] (A') -- (B');\addlegendentry{enriched MPC}
5

\addplot table [x index=1, y index =2, col sep=comma] {convergence/lag_plate_hole.dat};
\addlegendentry{enriched LM}

\nextgroupplot[	
	ylabel=$\| \epsilon \|_{\mathcal{E}}$,
	thick=1pt
]

\addplot table [x index=1, y index =3, col sep=comma] {convergence/tp_plate_hole.dat};

\addplot table [x index=1, y index =3, col sep=comma] {convergence/std_plate_hole.dat}
coordinate [pos=0.75] (A)
coordinate [pos=0.85]  (B)
;
\coordinate (A') at ($(A)!-3pt!90:(B)$);
\coordinate (B') at ($(B)!-3pt!270:(A)$);
\draw  (B') -| (A') node[pos=0.18,below]{\small $1$} node[pos=0.75,left]{\small $0.50$};
\draw [shorten <=-3pt,shorten >=-3pt] (A') -- (B');

\addplot table [x index=1, y index =3, col sep=comma] {convergence/nonmatch_plate_hole.dat}
coordinate [pos=0.75] (A)
coordinate [pos=0.85]  (B)
;
\coordinate (A') at ($(A)!3pt!90:(B)$);
\coordinate (B') at ($(B)!3pt!270:(A)$);
\draw  (A') -| (B') node[pos=0.18,above]{\small $1$} node[pos=0.75,right]{\small $0.50$};
\draw [shorten <=-3pt,shorten >=-3pt] (A') -- (B');

\addplot table [x index=1, y index =3, col sep=comma] {convergence/lag_plate_hole.dat};

\nextgroupplot[
	ylabel=$\| \epsilon \|_{\mathcal{L}^{2}}$,
	thick=1pt
]

\addplot table [x index=1, y index =2, col sep=comma] {convergence/tp_plate_hole_vertical.dat};

\addplot table [x index=1, y index =2, col sep=comma] {convergence/std_plate_hole.dat}
coordinate [pos=0.75] (A)
coordinate [pos=0.85]  (B)
;
\coordinate (A') at ($(A)!-3pt!90:(B)$);
\coordinate (B') at ($(B)!-3pt!270:(A)$);
\draw  (B') -| (A') node[pos=0.18,below]{\small $1$} node[pos=0.75,left]{\small $1.00$};
\draw [shorten <=-3pt,shorten >=-3pt] (A') -- (B');

\addplot table [x index=1, y index =2, col sep=comma] {convergence/nonmatch_plate_hole_vertical.dat}
coordinate [pos=0.75] (A)
coordinate [pos=0.85]  (B)
;
\coordinate (A') at ($(A)!3pt!90:(B)$);
\coordinate (B') at ($(B)!3pt!270:(A)$);
\draw  (A') -| (B') node[pos=0.18,above]{\small $1$} node[pos=0.75,right]{\small $1.02$};
\draw [shorten <=-3pt,shorten >=-3pt] (A') -- (B');

\addplot table [x index=1, y index =2, col sep=comma] {convergence/lag_plate_hole_vertical.dat};

\nextgroupplot[
	ylabel=$\| \epsilon \|_{\mathcal{E}}$,
	thick=1pt
]

\addplot table [x index=1, y index =3, col sep=comma] {convergence/tp_plate_hole_vertical.dat};

\addplot table [x index=1, y index =3, col sep=comma] {convergence/std_plate_hole.dat}
coordinate [pos=0.75] (A)
coordinate [pos=0.85] (B)
;

\coordinate (A') at ($(A)!-3pt!90:(B)$);
\coordinate (B') at ($(B)!-3pt!270:(A)$);
\draw  (B') -| (A') node[pos=0.18,below]{\small $1$} node[pos=0.75,left]{\small $0.50$};
\draw [shorten <=-3pt,shorten >=-3pt] (A') -- (B');

\addplot table [x index=1, y index =3, col sep=comma] {convergence/nonmatch_plate_hole_vertical.dat}
coordinate [pos=0.75] (A)
coordinate [pos=0.85] (B)
;
\coordinate (A') at ($(A)!3pt!90:(B)$);
\coordinate (B') at ($(B)!3pt!270:(A)$);
\draw  (A') -| (B') node[pos=0.18,above]{\small $1$} node[pos=0.75,right]{\small $0.50$};
\draw [shorten <=-3pt,shorten >=-3pt] (A') -- (B');

\addplot table [x index=1, y index =3, col sep=comma] {convergence/lag_plate_hole_vertical.dat};

\end{groupplot}
\node at ($(group c1r1) + (4,-4.cm)$) {\ref{grouplegend1}}; 
\end{tikzpicture}

%% file: figures/contact_condition2.tex
\begin{tikzpicture}[scale=0.5]
\tikzstyle{ground}=[fill,pattern=north east lines,draw=none,minimum width=0.5cm,minimum height=0.2cm]
\draw [black,fill = gray!10] (0,0) --(10,0) -- (10,5) -- (0, 5) -- cycle;
\draw [black,fill = gray!10] (0, 5) -- (3, 5) -- (3, 7) -- (0, 7) -- cycle;

\draw [ -stealth] (3.1, 6) -- (6.9, 6);
\draw [, dashed] (7, 5) -- (10, 5) -- (10, 7) -- (7, 7) -- cycle;

\draw [stealth-stealth] (9.5, 0) -- (9.5, 5);
\node [left] at (9.5, 2.5) {$H = 5$};
\draw [stealth-stealth] (9.5, 5) -- (9.5, 7);
\node [left] at (9.5, 6) {$h = 2$};

\draw [stealth -stealth] (0, 3.5) -- (10, 3.5);
\node [below] at (5, 3.5) {$L = 10$};

\draw [stealth - stealth] (0, 4.7) -- (3, 4.7);
\node [below] at (1.75, 4.7) {$l = 3$};

\coordinate (A) at (0, 0);
\coordinate (B) at (5, 0);
\coordinate (C) at (10, 0);

	\draw [-stealth] (0, 0) -- (0, 2.0);
	\draw [-stealth] (0, 0) -- (2.0, 0);
	\node [right] at (0, 2.0) {$\bs{e}_2$};
	\node [above] at (2.0, 0) {$\bs{e}_1$};
\node at (5, 1) {$E_1$, $\nu_1$};
\node at (1.5, 6) {$E_2$, $\nu_2$};

\end{tikzpicture}

%% file: figures/contact_condition_moving.tex
\begin{tikzpicture}[baseline]
\begin{axis}[cycle list/Dark2,
		cycle multiindex* list={
			mark list\nextlist
			Dark2\nextlist},
	xlabel=$x_1$, 
	ylabel=$\kappa$,
	ymax = 1e7,
	ymode = log,
	legend columns=1,
	yminorticks=true,
	legend style={at={(1,0.5)},
		anchor=west,
		draw=none,
		fill=none,
		legend cell align={left},
		},
	xtick={0,7},
	width=8cm,
	thick=1pt]

\addplot +[mark repeat=6] table [x index=0, y index =2, col sep=comma] {condition/1e4/contact_mpc_condition_moving_none_all.dat};\addlegendentry{$\kappa(\bar{\bs{K}}_{\mathrm{ns}})$}
\addplot +[mark repeat=6,mark phase=3] table [x index=0, y index =2, col sep=comma] {condition/1e4/contact_mpc_condition_moving_scaled_all.dat};\addlegendentry{$\kappa(\bar{\bs{K}}_{\mathrm{os}})$ }
\addplot +[mark repeat=6, dashed] table [x index=0, y index =2, col sep=comma] {condition/1e4/contact_mpc_condition_moving_precond_all.dat};\addlegendentry{$\kappa(\bar{\bs{K}}_{\mathrm{pc}})$ }

\addplot +[mark repeat=6] table [x index=0, y index =2, col sep=comma] {condition/1e4/contact_alm_condition_moving_none_all.dat};\addlegendentry{$\kappa(\hat{\bs{K}}_{\mathrm{ns}})$}
 \addplot +[mark repeat=6, mark phase=3] table [x index=0, y index =2, col sep=comma] {condition/1e4/contact_alm_condition_moving_scaled_all.dat};\addlegendentry{$\kappa(\hat{\bs{K}}_{\mathrm{os}})$}
 \addplot +[mark repeat=6] table [x index=0, y index =2, col sep=comma] {condition/1e4/contact_alm_condition_moving_precond_all.dat};\addlegendentry{$\kappa(\hat{\bs{K}}_{\mathrm{pc}})$}

\end{axis}
\end{tikzpicture}

%% file: contact_condition1.tex
\begin{tikzpicture}[scale=0.5]
\tikzstyle{ground}=[fill,pattern=north east lines,draw=none,minimum width=0.5cm,minimum height=0.2cm]
\draw [black,fill = gray!10] (0,0) --(10,0) -- (10,5) -- (0, 5) -- cycle;
\draw [black, fill = gray!10] (3.5, 5) -- (6.5, 5) -- (6.5, 7) -- (3.5, 7) -- cycle;


\draw [stealth-stealth] (9.5, 0) -- (9.5, 5);
\node [left] at (9.5, 2.5) {$H = 5$};
\draw [stealth-stealth] (7, 5) -- (7, 7);
\node [right] at (7, 6) {$h = 2$};

\draw [stealth -stealth] (0, 3.5) -- (10, 3.5);
\node [below] at (5, 3.5) {$L = 10$};

\draw [stealth - stealth] (3.5, 4.7) -- (6.5, 4.7);
\node [below] at (5, 4.7) {$l = 3$};
\draw [stealth - stealth] (0, 5.4) -- (3.5, 5.4);
\node [above] at (1.75, 5.4) {$x_1=3.5$};

\coordinate (A) at (0, 0);
\coordinate (B) at (5, 0);
\coordinate (C) at (10, 0);

	\draw [-stealth] (0, 0) -- (0, 2.0);
	\draw [-stealth] (0, 0) -- (2.0, 0);
	\node [right] at (0, 2.0) {$\bs{e}_2$};
	\node [above] at (2.0, 0) {$\bs{e}_1$};

\node at (5, 1) {$E_1$, $\nu_1$};
\node at (5, 6) {$E_2$, $\nu_2$};
\end{tikzpicture}

%% file: figures/contact_condition_single_legend.tex
  \begin{tikzpicture}
    \pgfplotsset{footnotesize,samples=10}
    \begin{groupplot}[cycle list/Dark2,
		cycle multiindex* list={
			mark list\nextlist
			Dark2\nextlist},
			group style = {group size = 2 by 1, horizontal sep = 50pt}]
			
\nextgroupplot[
	xlabel=$h$, 
	ylabel=$\kappa$,
	xmode =log,
	ymode = log,
	ymax = 1e10,
	ymin = 1e2,
	legend columns=1,
	yminorticks=true,
	xminorticks=true,
	legend columns=7,
	legend style={at={(1,0.5)},
		anchor=west,
		color = black,
		draw=none,
		fill=none,
		legend cell align={left},
		legend to name = grouplegend},
	ytick distance=10^1,
	ytick={1e2,1e3,1e4,1e5,1e6,1e7,1e8,1e9,1e10},
	yticklabels={$10^{2}$,,$10^{4}$,,$10^{6}$,,$10^{8}$,,$10^{10}$},
	height=7.5cm,
	width=8cm,
	thick=1pt
	]


\addplot table [x index=1, y index =4, col sep=comma] {condition/1e4/contact_std_condition_R_precond.dat}
coordinate [pos=0.85] (A)
coordinate [pos=0.95]  (B)
;
\coordinate (A') at ($(A)!3pt!90:(B)$);
\coordinate (B') at ($(B)!3pt!270:(A)$);
\draw  (B') |- (A') node[pos=0.8,below]{\small $1$} node[pos=0.1,left]{\small $2$};
\draw [shorten <=-3pt,shorten >=-3pt] (A') -- (B');
\addlegendentry{$\kappa({\bs{K}_{\mathrm{std}}})$}


\addplot table [x index=1, y index=5, col sep= comma] {condition/1e4/contact_enriched_condition_Rmpcmaster_none.dat};
\addlegendentry{$\kappa(\bar{\bs{K}}_{\mathrm{ns}})$}

\addplot table [x index=1, y index=5, col sep= comma] {condition/1e4/contact_enriched_condition_Rmpcmaster_scaled.dat};
\addlegendentry{$\kappa(\bar{\bs{K}}_{\mathrm{os}})$}

\pgfplotsset{cycle list shift=2}
\addplot table [x index=1, y index=5, col sep= comma] {condition/1e4/contact_enriched_condition_Rmpcmaster_precond.dat};
\addlegendentry{$\kappa(\bar{\bs{K}}_{\mathrm{pc}})$}

\pgfplotsset{cycle list shift=3}
\addplot table [x index=1, y index=5, col sep= comma] {condition/1e4/contact_alm_condition_refine_none_all.dat};
\addlegendentry{$\kappa(\hat{\bs{K}}_{\mathrm{ns}})$}

\pgfplotsset{cycle list shift=8}
\addplot table [x index=1, y index=5, col sep= comma] {condition/1e4/contact_alm_condition_refine_scaled_all.dat};
\addlegendentry{$\kappa(\hat{\bs{K}}_{\mathrm{os}})$}

\pgfplotsset{cycle list shift=0}
\addplot table [x index=1, y index=5, col sep= comma] {condition/1e4/contact_alm_condition_refine_precond_all.dat};
\addlegendentry{$\kappa(\hat{\bs{K}}_{\mathrm{pc}})$}

\nextgroupplot[
	xlabel=$n_d$, 
	ylabel=$\kappa$,
	xmode = log,
	ymode = log,
	ymax = 1e10,
	ymin = 1e2,
	yminorticks=true,
	xminorticks=true,
	ytick distance=10^1,
	ytick={1e2,1e3,1e4,1e5,1e6,1e7,1e8,1e9,1e10},
	yticklabels={$10^{2}$,,$10^{4}$,,$10^{6}$,,$10^{8}$,,$10^{10}$},
	height=7.5cm,
	width=8cm,
	thick=1pt
	]


\addplot table [x index=3, y index =4, col sep=comma] {condition/1e4/contact_std_condition_R_precond.dat}
coordinate [pos=0.85] (A)
coordinate [pos=0.95]  (B)
;
\coordinate (A') at ($(A)!3pt!270:(B)$);
\coordinate (B') at ($(B)!3pt!90:(A)$);
\draw  (B') |- (A') node[pos=0.8,below]{\small $1$} node[pos=0.1,right]{\small $1$};
\draw [shorten <=-3pt,shorten >=-3pt] (A') -- (B');

\addplot table [x index=2, y index=5, col sep= comma] {condition/1e4/contact_enriched_condition_Rmpcmaster_none.dat};

\addplot table [x index=2, y index=5, col sep= comma] {condition/1e4/contact_enriched_condition_Rmpcmaster_scaled.dat};

\pgfplotsset{cycle list shift=2}
\addplot table [x index=2, y index=5, col sep= comma] {condition/1e4/contact_enriched_condition_Rmpcmaster_precond.dat};

\pgfplotsset{cycle list shift=3}
\addplot table [x index=2, y index=5, col sep= comma] {condition/1e4/contact_alm_condition_refine_none_all.dat};

\pgfplotsset{cycle list shift=8}
\addplot table [x index=2, y index=5, col sep= comma] {condition/1e4/contact_alm_condition_refine_scaled_all.dat};

\pgfplotsset{cycle list shift=0}
\addplot table [x index=2, y index=5, col sep= comma] {condition/1e4/contact_alm_condition_refine_precond_all.dat};

    \end{groupplot}
    \node at ($(group c1r1) + (4,-4.5cm)$) {\ref{grouplegend}}; 
\end{tikzpicture}

%% file: figures/hertz_geo.tex
\begin{tikzpicture}[scale = 0.5]
\tikzstyle{ground}=[fill,pattern=north east lines,draw=none,minimum width=0.5cm,minimum height=0.2cm]
\draw [black, fill = gray!10] (0,0) --(10,0) -- (10,5) -- (0, 5) -- cycle;
\draw [black] (0,10) -- (10, 10);
\draw [fill=gray!10] (10, 10) arc (0:-180:5);
\node at (5, 2.5) {$E_1$, $\nu_1$};
\node [above] at (5, 7) {$E_2$, $\nu_2$};
\draw [-stealth] (5, 0.) --(5, 1);
\draw [-stealth] (5, 0) -- (6, 0);
\node [above right] at (6, 0) {$\bs{e}_1$};
\node [above right] at (5, 1) {$\bs{e}_2$};
\draw [stealth-stealth] (0, 4.5) -- (10, 4.5);
\node [below] at (5, 4.5) {$l=20$};
\draw [stealth-stealth] (10.5, 0) -- (10.5, 5);
\node [right] at (10.5, 2.5) {$h=10$};
\draw [-stealth] (5, 10) -- (2.5, {10-2.5*sqrt(3)});
\node at (2, 8) {$r =10$};
\draw (0, -0.15) circle  (0.15cm);
\node (ground1) at (0, 0) [ground, yshift = -0.15cm, anchor=north] {};
\draw  (ground1.north west) -- (ground1.north east);
\draw (4.85, -0.3) -- (5, 0) -- (5.15, -0.3);
\node (ground1) at (5, 0) [ground, yshift = -0.15cm, anchor=north] {};
\draw  (ground1.north west) -- (ground1.north east);
\draw (10, -0.15) circle  (0.15cm);
\node (ground1) at (10, 0) [ground, yshift = -0.15cm, anchor=north] {};
\draw  (ground1.north west) -- (ground1.north east);
\foreach \i in {0,...,20}
{
	\draw [-stealth] (0.5*\i, 11) -- (0.5*\i, 10.2);
}
\node [above] at (5, 11) {$\overline{\bs{t}}$};

\end{tikzpicture}

%% file: figures/hertz_traction.tex
\begin{tikzpicture}[baseline]
\begin{axis}[
xlabel=$x_1$, 
ylabel= $p_n$,
legend columns=1,
yminorticks=true,
xminorticks=true,
minor tick num=3,
xmin =-{sqrt(800*25/pi*(0.91/7e3+0.91/7e5))},
xmax = {sqrt(800*25/pi*(0.91/7e3+0.91/7e5))},
legend style={at={(0.2,0.7)},
		anchor=west,
		draw=none,
		fill=none,
		legend cell align={left},
		},
		cycle list/Dark2,
		cycle multiindex* list={
		mark list\nextlist
		Dark2\nextlist},
   		width=8cm,
		thick=1pt]

\addplot +[mark = x] table [x index=0, y index =1] {./hertz/traction_coarse.csv};\addlegendentry{Fine-coarse mesh}

\addplot +[mark = o] table [x index=0, y index =1] {./hertz/traction_fine.csv};\addlegendentry{Fine-fine mesh}
	
\addplot +[domain=-1:1, 
    samples=100, 
    mark = none] {-1/(20*(0.91*(1/7e3+1/7e5)))*sqrt(800*25/pi*(0.91*(1/7e3+1/7e5)) -x*x)};\addlegendentry{Analytical results}

\end{axis}
\end{tikzpicture}

%% file: figures/hertz_traction_error.tex
	\begin{tikzpicture}[baseline]
		\begin{axis}[
			xlabel=$x_1$, 
			ylabel= error ($\%$),
			legend columns=1,
			yminorticks=true,
			xminorticks=true,
			minor tick num=3,
			ytick = {0., 0.2, 0.4, 0.55},
			yticklabels = {$0$, $20$, $40$, $\%$ },
			xmin =-{sqrt(800*25/pi*(0.91/7e3+0.91/7e5))},
			xmax = {sqrt(800*25/pi*(0.91/7e3+0.91/7e5))},
			legend style={at={(0.2,0.7)},
				anchor=west,
				draw=none,
				fill=none,
				legend cell align={left},
			},
			cycle list/Dark2,
			cycle multiindex* list={
				mark list\nextlist
				Dark2\nextlist},
			width=8cm,
			thick=1pt]
			
			\addplot +[mark = x] table [x index=0, y index =2] {./hertz/traction_coarse.csv};\addlegendentry{Fine-coarse mesh}
			
			\addplot +[mark = o] table [x index=0, y index =2] {./hertz/traction_fine.csv};\addlegendentry{Fine-fine mesh}
			
			\addplot +[domain=-1:1, 
			samples=100, 
			mark = none, dashed] {0};
			
			\draw [-stealth, dashed] (0.0, 0.12) -- (-0.2, 0.073);
			\node [right] at (0.0, 0.12) {$6.7\%$};
			
		\end{axis}
	\end{tikzpicture}

%% file: figures/analytic.pdf_tex
\begingroup%
  \makeatletter%
  \providecommand\color[2][]{%
    \errmessage{(Inkscape) Color is used for the text in Inkscape, but the package 'color.sty' is not loaded}%
    \renewcommand\color[2][]{}%
  }%
  \providecommand\transparent[1]{%
    \errmessage{(Inkscape) Transparency is used (non-zero) for the text in Inkscape, but the package 'transparent.sty' is not loaded}%
    \renewcommand\transparent[1]{}%
  }%
  \providecommand\rotatebox[2]{#2}%
  \newcommand*\fsize{\dimexpr\f@size pt\relax}%
  \newcommand*\lineheight[1]{\fontsize{\fsize}{#1\fsize}\selectfont}%
  \ifx\svgwidth\undefined%
    \setlength{\unitlength}{200.69477516bp}%
    \ifx\svgscale\undefined%
      \relax%
    \else%
      \setlength{\unitlength}{\unitlength * \real{\svgscale}}%
    \fi%
  \else%
    \setlength{\unitlength}{\svgwidth}%
  \fi%
  \global\let\svgwidth\undefined%
  \global\let\svgscale\undefined%
  \makeatother%
  \begin{picture}(1,0.35167492)%
    \lineheight{1}%
    \setlength\tabcolsep{0pt}%
    \put(0,0){\includegraphics[width=\unitlength,page=1]{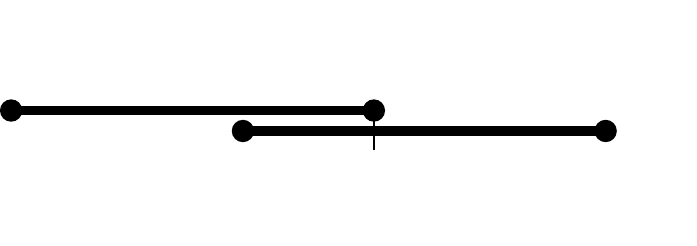}}%
    \put(0.31517284,0.10919025){\color[rgb]{0,0,0}\makebox(0,0)[lt]{\lineheight{1.25}\smash{\begin{tabular}[t]{l}$u_0$\end{tabular}}}}%
    \put(0.84506946,0.10642171){\color[rgb]{0,0,0}\makebox(0,0)[lt]{\lineheight{1.25}\smash{\begin{tabular}[t]{l}$u_1$\end{tabular}}}}%
    \put(0.00456443,0.2237286){\color[rgb]{0,0,0}\makebox(0,0)[lt]{\lineheight{1.25}\smash{\begin{tabular}[t]{l}$u_2$\end{tabular}}}}%
    \put(0.51030174,0.22372839){\color[rgb]{0,0,0}\makebox(0,0)[lt]{\lineheight{1.25}\smash{\begin{tabular}[t]{l}$u_3$\end{tabular}}}}%
    \put(0.49262736,0.1018174){\color[rgb]{0,0,0}\makebox(0,0)[lt]{\lineheight{1.25}\smash{\begin{tabular}[t]{l}$\alpha_4$\end{tabular}}}}%
    \put(0,0){\includegraphics[width=\unitlength,page=2]{analytic.pdf}}%
    \put(0.37203469,0.00878297){\color[rgb]{0,0,0}\makebox(0,0)[lt]{\lineheight{1.25}\smash{\begin{tabular}[t]{l}$\eta$\end{tabular}}}}%
    \put(0,0){\includegraphics[width=\unitlength,page=3]{analytic.pdf}}%
    \put(0.41592281,0.31333156){\color[rgb]{0,0,0}\makebox(0,0)[lt]{\lineheight{1.25}\smash{\begin{tabular}[t]{l}$\xi$\end{tabular}}}}%
    \put(0.7525512,0.17785589){\color[rgb]{0,0,0}\makebox(0,0)[lt]{\lineheight{1.25}\smash{\begin{tabular}[t]{l}(1)\end{tabular}}}}%
    \put(0.16554532,0.20522426){\color[rgb]{0,0,0}\makebox(0,0)[lt]{\lineheight{1.25}\smash{\begin{tabular}[t]{l}(2)\end{tabular}}}}%
  \end{picture}%
\endgroup%